\documentclass[12pt]{article}
\usepackage{amsmath}
\usepackage{amsfonts}
\usepackage{setspace}
\usepackage{graphicx}
\usepackage{comment}
\usepackage{placeins}
\usepackage{bm}
\usepackage{bbm}
\usepackage{parskip}
\usepackage[dvipsnames]{xcolor}

\usepackage[backend=biber, style=apa, natbib=true]{biblatex}
\usepackage{fullpage}
\usepackage{multirow}
\usepackage{booktabs}
\usepackage{subcaption}
\usepackage{lineno}

\usepackage{longtable}
\usepackage{array}
\usepackage{wrapfig}
\usepackage{float}
\usepackage{colortbl}
\usepackage{pdflscape}
\usepackage{tabu}
\usepackage{threeparttable}
\usepackage{threeparttablex}
\usepackage[normalem]{ulem}
\usepackage{makecell}
\usepackage{hyperref}
\hypersetup{
  colorlinks   = true, 
  urlcolor     = blue,
  linkcolor    = blue, 
  citecolor   = blue 
}

\newcommand{\X}{\textbf{X}}

\newcommand{\Var}{\mathbb{V}}
\newcommand{\Cov}{\text{Cov}}
\newcommand{\E}{\mathbb{E}}
\newcommand{\W}{\mathbb{W}}
\newcommand{\dotXt}{\dot{X}}
\newcommand{\ddotXt}{\ddot{X}}
\newcommand{\ddotRt}{\ddot{R}_{i}}
\newcommand{\dotRt}{\dot{R}_{i}}
\newcommand{\bbeta}{\beta}

\newcommand{\wdim}{\hat{\tau}_{wDIM}}

\newcommand{\M}{\textbf{M}}
\newcommand{\BSigma}{\mathbf{\Sigma}}

\newcommand\indep{\protect\mathpalette{\protect\independenT}{\perp}}
\def\independenT#1#2{\mathrel{\rlap{$#1#2$}\mkern5mu{#1#2}}}
\newtheorem{assumption}{Assumption}
\newcommand{\phibstar}{\overline{\phi}^{\, \star}}

\definecolor{neyman}{HTML}{000000}
\definecolor{separate}{HTML}{000000}

\expandafter\def\expandafter\normalsize\expandafter{%
    \normalsize%
    \setlength\abovedisplayskip{-0.5\baselineskip}%
    \setlength\belowdisplayskip{0pt}%
    \setlength\abovedisplayshortskip{-0.5\baselineskip}%
    \setlength\belowdisplayshortskip{0pt}%
}

\begin{document}

\title{\vspace{-1em}Inference with weights: Residualization produces short, valid intervals for varying estimands and varying resampling processes}
\date{\today\\
}

\author{\begin{tabular}{c c c}  
   Erin Hartman\thanks{The authors wish to thank Peng Ding, Avi Feller, Noah Greifer, Melody Huang, Sam Pimentel, Andy Shen, Yuki Shiraito, Soichiro Yamauchi, and the participants of the Berkeley Causal Lab and Polmeth XLII. The authors gratefully acknowledge student travel support from the Social Statistics (SSS), Government Statistics (GSS), and Survey Research Methods (SRMS) Sections of ASA to present this research. }  & Chad Hazlett & Arisa Sadeghpour \\
     UC Berkeley & UCLA & UC Berkeley 
\end{tabular}}

\parskip=0.5\baselineskip
\parindent=0pt

\maketitle

\begin{abstract}
Weighting is widely used in observational causal inference, yet basic questions remain about how to estimate the variance of weighted effect estimates. For example, when and how should sampling uncertainty in the weights be propagated into the final result? Should these weights reduce rather than inflate variance? Does the choice of weighted variance estimator depend on the target estimand and the assumed source of randomness? We show that a single, familiar estimator addresses these concerns across settings: robust standard errors from a weighted regression of the outcome on treatment, the centered balancing covariates, and their interaction with treatment. These residualized standard errors are valid and often substantially smaller than those from weighted regression on treatment alone, with reductions of up to 30\% when covariates are prognostic. We establish their justification under design-based, model-based, and superpopulation frameworks, for both ATE and ATT estimands, and for exact balancing, approximate balancing, and inverse propensity weights. For inference to the superpopulation, we employ a simple correction. We further show that M-estimation yields nearly identical results when constructed to answer the same questions, and that the key issue is not ``uncertainty in the weights'' given a sample, but uncertainty over the sample. Simulations and three empirical applications confirm our analyses.
\end{abstract}

\newpage
\doublespacing
\begin{refsection}

\section{Introduction}\label{sec:intro}
Weighting procedures of many kinds are popular in observational causal inference research. Researchers may construct weights from a modeled probability of treatment (the propensity score), by matching or stratification, or to achieve approximate or exact balance on (functions of) the covariates. One common practice for inference after weighting is to conceptualize the weights as a pre-processing step, then simply estimate (robust) standard errors from a weighted difference-in-means, or equivalently a weighted regression of outcome on treatment (see e.g. \citealp{hernan2020causal, reifeis_hudgens_2022}), similar to approaches sometimes advocated with matching \citep{ho2007matching, greifer2021matching}.

Throughout, we take the weights as given and assume they do their job: that the weighted difference-in-means they produce consistently estimates the causal target (see Section~\ref{subsec:weighting_methods}). Our subject is the variance of this estimate---how it should be conceived, and how it should be estimated. Will the weights make those standard errors inappropriately large? Does uncertainty in estimation of the weights need to be propagated into our standard errors, and if so, how?  In short, we argue that inference for such weighted comparisons should be performed using the heteroskedasticity-robust (HC0) standard error from a \emph{weighted outcome regression}. Specifically, we recommend the weighted regression of the outcome on the treatment while additionally including the (centered) balancing covariates and their interaction with treatment, i.e. a weighted version of the \citet{lin_agnostic_2013} estimator. We refer to these as the ``residualized'' standard errors. This estimator ``takes credit'' for the variance reduction the weights achieve under different resampling frameworks.

The point estimate produced by this weighted regression will be identical to that of the weighted difference-in-means (wDIM) when the weights achieve exact balance on the included covariates. Where balance is only approximate, the point estimate will differ slightly in finite samples. However, this may be beneficial to efficiency. In particular, the weighted outcome regression is numerically an augmented inverse propensity-weighted (AIPW) estimator. If the weights converge to the true inverse propensity weights (a stronger condition than the consistency we assume throughout), and the outcome model is additionally correctly specified, it attains the semiparametric efficiency bound (Section~\ref{sec:exact_v_approx}).

Under both the design-based and model-based sampling frameworks, we first show that the variance of sample-specific estimands is given by residualized standard errors. We then describe adjustments that make them suitable under the superpopulation framework, which allows additional sampling uncertainty in the distribution of the balancing covariates in the population. One lesson is that no notion of ``uncertainty in the weights'' beyond that driven by uncertainty in covariate distributions (in the superpopulation setting) is required for valid inference. Relatedly, M-estimation approaches \citep[e.g.][]{stefanski2002calculus, ross2024m} can be constructed to capture either superpopulation variances (by regarding sample moments of the covariates as uncertain), or tailored to the fixed sample (by taking observed sample moments as fixed), for any type of balancing weight, and for uncertainty in the weighted residualized regression estimate. Such M-estimation approaches closely reproduce the corresponding closed-form residualized standard error estimator we recommend.

The use of such residualized or linearized standard errors is not new, and has a particularly long history in count or mean estimation problems in the survey literature \citep[e.g.][]{deville1992calibration, breidt_opsomer_2017}.  Our aim is to demystify and justify this approach to inference with weighted comparisons. Conveniently, this approach produces valid confidence intervals regardless of the choice of estimand or resampling framework, and for weights from any source meeting mild conditions set out in Section~\ref{subsec:weighting_methods}, including entropy balancing, inverse propensity, and matching weights. The results emphasize that weights need not lead to inappropriately wider uncertainty. Rather, the common variance estimator does not naturally ``take credit'' for the stabilizing influence the weights have in each finite sample, while the residualized standard error does. We confirm that these residualized standard errors are smaller than those from non-residualizing estimators, while retaining nominal or conservative coverage, with reductions of  up to 30\% across our simulations and three empirical applications. In our empirical applications, we find that residualized standard errors can be even smaller than standard errors of unweighted estimates. The gains grow with the degree to which the weighted covariates are prognostic.

\section{Background}\label{sec:setup}
We focus on an observational setting, where researchers cannot control treatment assignment, and where it is important that they address potential bias induced by selection into treatment when estimating causal effects.  Weighting estimators---in which researchers construct a set of weights that adjust the data, or a subset of the data, to look ``representative'' of a specified target distribution in varying senses described below---are a popular method for covariate adjustment. Our objective is to clarify what are appropriate variance estimators after weighting.  To justify these choices, we find it necessary to clarify the target estimands and sources of sampling uncertainty relevant to researchers under different data-generating paradigms.  We introduce some common notation here, with details specific to the data-generating process outlined in subsequent sections.  We then introduce our primary weighted estimator for causal effects.

\subsection{Notation and assumptions}\label{sec:notation}

We consider a binary treatment $Z_i \in \{0, 1\}$ and observed outcome $Y_i$ with potential outcomes $Y_i(1)$ and $Y_i(0)$, and covariates $X_i$. We assume a single version of treatment and no interference, collectively referred to as the Stable Unit Treatment Value assumption (SUTVA), with consistency holding ($Y_i(z)=Y_i$ for unit $i$ taking treatment $Z_i=z$) \citep{rubin1980randomization}.  We will let $n$ denote the sample size of our realized data, with $n_t = \sum_i Z_i$ and $n_c = \sum_i (1 - Z_i)$.  

Finally, our analysis below takes causal quantities to be the target of
inference after weighting, and so we require conditional ignorability, or
selection on observables \citep{rubin1974estimating}:

\begin{equation*} 
\{Y(1),Y(0)\} \indep Z \mid \X \quad \quad \text{(conditional ignorability)}
\end{equation*}

We return in Section~\ref{subsec:weighting_methods} to what our results provide when this assumption, or the specification of the weights, is in doubt.

\subsection{Weighting Methods}\label{subsec:weighting_methods}
We briefly review the variety of covariate adjustment weights that researchers may employ, and to which our approach applies. 

\paragraph{Inverse propensity score weights} 
Inverse propensity score weights are justified on the basis that under conditional ignorability, one need not condition on $X_i$, but rather on a one-dimensional score: the propensity score, $\Pr(Z_i = 1 \mid X_i) = e(X_i)$ \citep{rosenbaum_central_1983}. This scalar $e(X_i)$ can then be used as a basis
for weighting. Among other options, the propensity score can be directly estimated, $\hat e(X_i)$ (e.g. by logit or probit model), then used to construct inverse propensity score weights (IPW). When targeting the average treatment effect (ATE) these weights are proportional to the reciprocal of the probability of receiving the assigned treatment status, i.e. $1/\hat{e}(X_i)$ for treated units and $1/(1-\hat{e}(X_i))$ for control units.  When targeting the average treatment effect on the treated (ATT), the treated units are left with uniform weights, while control units are weighted by their odds of being treated, $\hat{e}(X_i)/(1-\hat{e}(X_i))$. In practice, these weights are often stabilized in the ``H\'{a}jek'' style and scaled to sum to 1 in each treatment condition, as we do below, replacing the numerator with the unconditional probability of receiving the assigned treatment ($\Pr(Z=1)$ or $\Pr(Z=0)$). 

An important feature of the IPW is that they balance all functions of the covariates in expectation, sometimes referred to as the population balance property \citep[see e.g.][]{rosenbaum_central_1983, hirano2003efficient, benmichael2021balancingactcausalinference}. In our discussion below, this has the important implication that IPW thus falls into the category of  \textit{approximate} weighting approaches, those that do not produce exact finite-sample covariate balance, for which our motivating logic differs slightly from the case of exact balancing weights.

\paragraph{Matching methods}  Matching methods seek to find, for each unit, one or more units of the opposite treatment status that lie nearby in the covariate space by some chosen measure. Because units may be never-selected, multiply-selected, or (in some procedures) partially selected as comparators in the final (``matched'') sample, the results of matching procedures can be represented by weights \citep{greifer2021matching}.  For concreteness, we consider ``1-to-many'' matching with replacement, and that the matching weights have been normalized to sum to 1 within each treatment group, though the results should apply in general. Excepting cases of exact matching on categorical variables (equivalent to sub-classification/poststratification), matching typically results in approximate balance. 

\paragraph*{Weighting for balance} An alternative to IPW is constructing weights that target within-sample balance on chosen moments of the covariates. This approach can be motivated as a method of moments estimator for the inverse propensity score weights \citep[e.g.][]{cohn2023balancing}, or by other motivations rooted in functional form claims on the treatment or outcome model  \citep[e.g.][]{zhao2017entropy, hazlett2020kernel}, discussed below. Regardless of motivation, balancing weights solve an optimization problem of the form 

\begin{equation}
\min_{w} \sum_{i:\,Z_i=z} f(w_i)
\quad \text{such that} \quad
\Big|\sum_{i:\,Z_i=z} w_i \phi_{ik} - \phi^{\star}_{k}\Big| \le \xi_k
\ \ \text{for each } k,
\quad \text{and} \quad \sum_{i:\,Z_i=z} w_i = 1,
\end{equation}

where $z \in \{0,1\}$ indicates treatment status, $\phi_i$ corresponds to
features of $\X$ for which balance is targeted\footnote{We could write this $\phi(X_i)$ to make clear it is a function of $X$, but choose not to for notational simplicity.}, and $\phibstar$ is the \emph{target profile}---the feature means of the population to which arm $z$ is reweighted. The estimand is determined by the choice of $\phibstar$ and the arms reweighted to it: for the ATT, $\phibstar = n_t^{-1} \sum_{i:\,Z_i = 1} \phi_i$ and only the controls ($z = 0$) are reweighted; for the ATE, $\phibstar = n^{-1} \sum_{i} \phi_i$ and both arms are reweighted to it, sometimes referred to as the ``three-way balancing constraints'' in \citet{kallberg2023large}. The prevalent practice of ``mean balancing'' corresponds to
choosing $\bm{\phi} = \X$.

Common weighting methods correspond to different dispersion penalties $f(w)$\footnote{Sometimes this dispersion penalty is written as a deviation from starting weights or initial design weights.  For simplicity here we assume uniform starting weights within the treatment and control groups.}, with occasional variations on the balancing constraints and typically imposing an additional non-negativity constraint on the weights \citep{cohn2023balancing}. This includes linear weighting and poststratification \citep{deville1992calibration}, entropy balancing or raking \citep{hainmueller_entropy_2012}, and many other related methods.  
Here, $\xi = (\xi_1, \ldots, \xi_K)$ is a vector of imbalance tolerances, one per balanced feature. When all $\xi_k = 0$, we refer to these as exact balancing weights; when some $\xi_k > 0$ the weights achieve only approximate balance, typically desired to reduce dispersion and/or for feasibility when $\phi$ is high-dimensional \citep[e.g.][]{zubizarreta2015, hazlett2020kernel, hartman2024kpop}.

\paragraph{Effect Estimator.} Once the researcher has constructed the weights, treatment effects are estimated using a weighted difference-in-means:

\begin{equation}
    \wdim = \sum_{i} w_i Z_i Y_i - \sum_{i} w_i (1 - Z_i) Y_i = w_1^{\top}Y_1 - w_0^{\top}Y_0 \label{eqn:wtd_dim}
\end{equation}

\noindent where we have relied on the weights summing to one (within each treatment group), $w_1$ and $w_0$ represent vectors of weights and $Y_1$ and $Y_0$ vectors for outcomes for treated and control units respectively.

While researchers can directly estimate the quantity in Equation~\eqref{eqn:wtd_dim}, it is common to obtain the point estimate for the weighted difference-in-means using a bivariate weighted least squares regression of $Y$ on $Z$ with an intercept.  The coefficient on $Z$ will be numerically equivalent to Equation~\eqref{eqn:wtd_dim}.  In Appendix~\ref{app:wls_hajek}, we show that the bivariate weighted least squares estimator is a H\'{a}jek style estimator.

The standard working variance estimator for the weighted
difference-in-means is:

\begin{equation}
    \hat \Var_{bi}[\wdim] = \sum_{i} w_i^2 Z_i \left(Y_i -
    \overline{Y}_1^w\right)^2 + \sum_{i} w_i^2 (1 - Z_i) \left(Y_i - \overline{Y}_0^w\right)^2
    \label{eq:neymanwdim}
\end{equation}

where $\overline{Y}_z^w = \sum_{i:Z_i=z} w_i Y_i$ is the weighted mean outcome by treatment group.  We use the subscript $bi$ to indicate when this variance is estimated with the bivariate regression.  The Neyman-style correction bounds the covariance yielding a conservative estimator. This is numerically equivalent to the HC0 robust standard error from a weighted regression of $Y$ on $Z$ (with intercept), and serves as the baseline variance estimator against which our proposed residualized approach is
compared below.

\paragraph{Estimation and specification of weights.}
We proceed under the assumption that the weights lead to consistent effect estimation via Equation~\eqref{eqn:wtd_dim}. It can fail in two distinct ways. First, it is violated by failing to include in $X$ some variable(s) required to achieve conditional ignorability. Second, even with all the required variables in hand, it can be compromised by the functional form choices embedded in the weighting procedure. For propensity score methods that choice of functional form is explicit: the investigator specifies a model for $e(X)$. For balancing methods it is made implicitly, through the choice of $\phi(\cdot)$ and of the dispersion penalty.\footnote{\label{fn:dual} The balancing problem has a dual in which the weights are characterized as a generalized linear model for the inverse propensity score, with the dispersion penalty determining the link function \citep[\S9]{benmichael2021balancingactcausalinference}. Entropy balancing, for instance, yields inverse propensity weights that are exponential in a linear index, $\exp(\phi_i^\top \lambda)$; in the ATT case, where the control weights estimate the odds $e(X)/(1-e(X))$, this is exactly the form implied by a logistic model linear in $\phi$.}  A specification can be adequate in either of two ways. The first is available to any weighting approach: the explicit or implied model for $e(X)$ may be
correctly specified. The second is specific to balancing methods: if the $\phi$ chosen is rich enough that $\E[Y(z) \mid X]$ is linear in $\phi$, balancing on $\phi$ suffices regardless of the form of the treatment assignment model \citep{hazlett2020kernel, zhao2017entropy}, because these methods enforce balance on $\phi$ by construction rather than relying on the assignment model.

While we proceed under this assumed consistency of the estimator for the causal target estimand of interest, in practice investigators often use these approaches where they cannot rule out unobserved confounding, nor be certain that they have correctly specified the treatment assignment model or balanced on all functions of $X$ required for $\E[Y(z) \mid X]$ to be linear in them. In that case, the results we describe below nevertheless provide suitable standard errors for the resulting point estimates, whatever their biases relative to the causal target. What our results characterize is the uncertainty in an estimate under a given resampling process (here, design-based or model-based resampling, targeting either finite-sample or superpopulation estimands).  

Beyond this consistency assumption, the nature of the weights required for the results below differs slightly by resampling framework. In the design-based framework (Section~\ref{sec:design_based}), the derivations in Appendix~\ref{app:db_var_est} require weights asymptotically equivalent to inverse assignment probabilities, as in model-assisted estimation. Many weights satisfy this, including inverse propensity weights, balancing weights (through their dual representation; see footnote~\ref{fn:dual}), the covariate balancing propensity score \citep{imai2014covariate}, and regularized variants of each. Under the model-based framework (Section~\ref{sec:model_based}), the weights are fixed along with $X$ and $Z$, and the results require only an invertible weighted design matrix. 
Negative weights are also permitted in finite samples: in the model-based case they pose no conceptual problem, and in the design-based case the asymptotic equivalence to inverse probabilities can hold even when some weights are negative in a given sample.\footnote{With negative weights the weighted squared error is unbounded below, so the weighted least squares estimator we use below (Equation~\eqref{eq:wResModel}) is defined not as its minimizer but as the solution to the corresponding weighted normal equations.  Solutions for the coefficients and HC0 variance exist and are unique so long as the weighted design matrix remains nonsingular. Some software, including \texttt{lm} in R, implements weighted least squares by scaling each observation by $\sqrt{w_i}$ and so refuses negative weights; the estimator is nonetheless computable from the normal equations.}
For investigators wishing to study how estimates would appear given postulated degrees of effective residual confounding, sensitivity analysis can be combined with the inferential procedures here. Specifically, \citet{wainstein2025sensitivity} develop omitted variable bias sensitivity analysis for estimators of the kind we recommend here (weighted least squares regression of the outcome on the treatment and interacted covariates), characterizing how severe a departure from the assumption above would have to be in order to change one's conclusions. Because that approach asks how the outcome model is affected by what has been left out, rather than modeling how the weights themselves have been biased, it applies with any non-negative weights and speaks to both of the failure modes described above: bias from unobserved confounding and bias from misspecification.

\singlespacing
\section{Resampling frameworks: Where does uncertainty come from?}\label{sec:frameworks}
\doublespacing

Uncertainty estimates report how much variation we would expect under some notion of repeated sampling, but different traditions disagree on what is random in this process. We consider both of the dominant paradigms: ``design-based'' and ``model-based'' processes. Both of these approaches fix $\X$, and hence relate to a notion of a fixed sample, and sample-specific estimands.  We do not argue here for the supremacy of either approach, rather we discuss both and their implications for the variance estimation problem. In either framework we can further consider variation in $\X$ and $Z$ together to arrive at a  ``superpopulation'' framework.  Below we detail each of these approaches, taking care to clarify the target estimand of interest, the random variables that induce the sampling distribution, and the implications for the appropriate interpretation of the source of uncertainty and how to estimate it.

\subsection{Design-Based}

In the design-based framework, potential outcomes for each unit are considered fixed, and sampling uncertainty comes from hypothetical reassignment of treatment.  This framework is commonly used for experimental design, but is less often invoked in observational studies, which usually appeal to the superpopulation approach described below. Consider the fixed, finite set of $n$ units $\mathcal{S} = \{ \left(Y_i(1), Y_i(0), X_i \right)\}$ drawn iid from some infinite superpopulation with $p(Y(1), Y(0), \X)$.  Units are assigned treatment in a Bernoulli design with $\Pr(Z_i = 1) = e(X_i)$.  We then let $Z = (Z_1, \dots, Z_n) \in \{ 0, 1\}^n$ that takes on the realized value $(z_1, \dots, z_n)$ with probability $\prod_{i=1}^n e(X_i)^{z_i}(1- e(X_i))^{(1-z_i)}$.  Importantly, sampling uncertainty is defined with reference to $p(Z \mid \mathcal{S})$, the randomization distribution, which captures hypothetical perturbations to treatment assignment. This setting closely mirrors the nonresponse setting in survey sampling \citep{deville1992calibration, sarndal2005estimation, d2010linearization}, a feature we will leverage below.

\paragraph{Estimands.}  We begin by considering estimation of the Sample Average Treatment Effect (SATE).  In the design-based framework, we can define this as:

\[ \tau_{SATE} = \E \left[ \frac{1}{n} \sum_i (Y_i(1) - Y_i(0)) \mid \mathcal{S}\right] \]

where the expectation is taken over $p(Z \mid \mathcal{S})$. Note that $\frac{1}{n} \sum_i (Y_i(1) - Y_i(0))$  is a fixed quantity across all randomizations, and $\tau_{SATE} = \overline{Y}(1) - \overline{Y}(0)$, as it is more commonly defined.

We propose a SATT that marginalizes the average treatment effect on the treated across the randomization distribution:

\[ \tau_{SATT} = \E \left[ \frac{1}{\sum_i{Z_i}} \sum_i Z_i(Y_i(1) - Y_i(0)) \mid \mathcal{S}\right] \]

where the expectation is, again, taken over $p(Z \mid \mathcal{S})$.  This can be thought of as a ``design-weighted treatment effect on the treated'', or the expected SATT one would obtain in a fixed sample, given that different units would be treated with a probability governed by the design.  This differs slightly from the more common interpretation as ``the effect of treatment on those who \textit{received} treatment'', because it fixes the design, not the realized treatment assignment. We consider variance estimators and their coverage for sampling uncertainty targeting this quantity, rather than the oracle treatment effect among realized treated units in the observed sample  realization.\footnote{Empirically, in the simulations, we find that coverage of the oracle average treatment effect for treated units in a given realization of treatment also maintains nominal rates.}

\subsection{Model-Based}
The design-based view is natural when viewing causal inference through the lens of experiments or analytical approaches  analogized to experiments, as it is rooted in the act of randomization. On the other hand, in observational research, on which we focus here, $Z$ is not determined by intervention, but is  akin to another variable in $\X$. Our uncertainty is not taken over imagined reassignments of $Z$ given $\X$, but over perturbations in the value of the outcome, $Y$ given $Z = z$ and $\X = \mathbf{x}$. 

Concretely, consider a data generating process of the general form $Y  = g(\X, Z) + \epsilon$. As above, we assume throughout that the weights yield consistent estimation of the target (Section~\ref{subsec:weighting_methods}). The resampling exercise under the model-based approach is then to ask how an estimated parameter, $\hat{\tau}$, would vary under fixed values of $\bm{\phi}$ and $Z$, but where resampling would result in a different noise vector, $\epsilon$. The relevant source of variance is the \textit{conditional} variance, $\Var_{MB} [\hat{\tau}|\X,Z]$.

\paragraph{Estimands.} The notion of fixing $\X$ and $Z$ could be understood as ``fixing the sample'' (i.e. the identity of the units designated by $i$), with the resulting average $\tau$ representing a SATE, but differing from above because we do not have fixed $Y_i(z)$.  Alternatively, one may regard the integration over $p(\epsilon)$ as suggesting we are obtaining a population average treatment effect, but for a population defined to have exactly the $\X$ and $Z$ observed.  For these reasons, we do not argue for a meaningful SATE-ATE distinction in the model-based setting, and we use only the term ATE. There will be a separate superpopulation notion, however, below.  In this framework, the ATT is straightforward, since $Z$ remains fixed.  We denote our model-based estimands as:

\begin{align} 
\tau_{ATE} &= \E[Y(1) - Y(0) \mid \X, Z] \\
\tau_{ATT} &= \E[Y(1) - Y(0) \mid \X, Z = 1] 
\end{align}

\noindent where the expectation is taken over $p(\epsilon)$.

\subsection{Inference to the superpopulation or data generating process}

Both the design-based and model-based frameworks condition on aspects of the realized sample. The researcher may instead target an estimand defined over the superpopulation or data-generating process from which the sample is drawn \citep[e.g.][]{rosenbaum_central_1983, abadie_sampling-based_2020, ding2023coursecausalinference}. The population estimands are

\begin{align*}
\tau_{\mathrm{PATE}} &= \E[Y(1) - Y(0)], \\
\tau_{\mathrm{PATT}} &= \E[Y(1) - Y(0) \mid Z = 1].
\end{align*}

Each finite-sample estimand connects to its population counterpart by the law of iterated expectations. For the design-based SATE, which conditions on the realized sample,

\[
  \E_{\mathcal{S}}\bigl[\tau_{\mathrm{SATE}}\bigr]
  = \E_{\mathcal{S}}\bigl[\E[Y(1) - Y(0) \mid \mathcal{S}]\bigr]
  = \E[Y(1) - Y(0)]
  = \tau_{\mathrm{PATE}}.
\]

Similarly, for the model-based ATE, that conditions on $(\X, Z)$,

\[
  \E_{\X,Z}\bigl[\tau_{\mathrm{ATE}}\bigr]
  = \E_{\X,Z}\bigl[\E[Y(1) - Y(0) \mid \X, Z]\bigr]
  = \tau_{\mathrm{PATE}}.
\]

The same logic applies to the ATT/SATT and PATT. In each case, moving from the finite-sample to the population estimand introduces additional variance from sampling variability in the quantities that the finite-sample framework held fixed. Analogously, our approach to variance estimation below begins with the finite-sample variance, then amends it with a ``correction''  needed to account for this additional source of uncertainty in the population quantity.

\section{Quantifying sampling uncertainty}

\subsection{Overview of proposal}\label{subsec:proposal}
The central concept motivating our approach is the use of \textit{residualization} to isolate the variation in $Y$ \textit{conditionally} on $X$, as only this variation influences the point estimate under design-based and model-based frameworks.  Weighting will typically reduce the variation in effect estimates one would see across hypothetical resamples, yet the common variance estimator from a weighted bivariate regression fails to ``take credit'' for the reduced variation from the weighting procedure. The mechanics of weighted linear regression offer a specific way of achieving residualization to construct a variance estimate that appropriately accounts for the stabilizing impact of the weights. This leads to correct or conservative coverage, often with smaller standard errors.

With this motivation in hand, we propose the following. We employ the weights in a weighted ``Lin''-style regression of $Y$ on treatment, the centered covariates balanced upon ($\tilde{\phi}_i$), and their interaction \citep{lin_agnostic_2013}, then use a heteroskedasticity-robust (e.g.\ HC0) standard error for the coefficient on treatment. That is, we consider the point estimate and estimated (HC0) standard error for $\hat{\tau}$
in

\begin{equation}
(\hat{\tau}, \hat{\beta}_0, \hat{\beta}, \hat{\gamma})
= \operatorname*{argmin}_{\tau, \beta_0, \beta, \gamma}
\sum_{i=1}^n w_i \bigl(Y_i - \beta_0 - \tau Z_i
- \tilde{\phi}_i^{\top}\beta
- Z_i\, \tilde{\phi}_i^{\top}\gamma\bigr)^2
\label{eq:wResModel}
\end{equation}

where the weights $w_i$ may be designed to target the ATE or ATT. The demeaning of $\tilde{\phi}_i$ is done using the pooled means when seeking the ATE and the mean of the treated units when seeking the ATT. The standard error estimates are constructed using the values of the fitted residual, $\hat{\epsilon}_i = Y_i - \hat \beta_0 - \hat{\tau} Z_i - \tilde{\phi}_i^{\top} \hat{\beta} - Z_i \tilde{\phi}_i^\top \hat{\gamma}$, as detailed below.  We will refer to the regression in Equation~\eqref{eq:wResModel} as the ``residualized regression''.

We suggest this interacted regression per \citet{lin_agnostic_2013} principally because when treatment effects are heterogeneous in $X$ and the probability of treatment varies depending on $X$, a single regression without the interaction reports a weighted average of heterogeneous effect estimates not generally equal to the ATE or ATT. The interacted regression effectively weakens the specification assumption to allow $Y(1)$ and $Y(0)$ to be ``separately'' (and differently) linear in $\phi$, which resolves the problem (see e.g. \citealp{chattopadhyay2023implied,shinkrehazlettdemystifying}). The centering of $\tilde{\phi}$ is used so that the coefficient $\hat{\tau}$ can be interpreted as the estimate when the transformed covariates are at their means corresponding to the ATE (or ATT if centered only over the treated), rather than its value when the (untransformed) covariates take a value of 0.

In addition, the use of weighting has important implications for the variance of this regression. Ordinarily, adding covariates to a weighted regression can affect the precision of the treatment coefficient in two opposite ways:  improving it by shrinking the residual variance in proportion to the prognostic value of the covariates, or worsening it by introducing covariates that predict treatment, inflating variance in proportion to $1/(1-R^2_{Z \sim X, w})$ (the  ``variance inflation factor'').  However, all the weighting  approaches here seek to make $\phi(X)$ similar in the treated and control groups, thus driving $R^2_{Z \sim X, w}$ to be nearly (or in some cases exactly) zero under weighting. Accordingly, the addition of covariates in the weighted outcome model stands to reduce the standard error with virtually no risk of inflating it. 

We recommend this approach regardless of the origins of the weights, including exact balancing, approximate balancing, or IPW, and across a variety of estimands, e.g.\ ATT vs.\ ATE, and whether operating under a design- or model-based framework, though inference to the (super)population will require additional elements compared to inference for the sample estimands, described below.  Note that $\E[Y \mid X, Z]$ need not be linear in $\phi$ and $Z$ (or $\phi$, $Z$ and their interaction). That is, the regression in Equation~\eqref{eq:wResModel} need not be correctly specified for it to be the appropriate choice for purposes of learning the standard error of $\hat\tau$. The reason differs by framework, and we give it in each: under the design-based view in Section~\ref{sec:design_based}, and under the model-based view in Section~\ref{sec:model_based}.

\subsection{Design-based uncertainty}\label{sec:design_based}

Recall that under the design-based resampling framework, the sampling uncertainty arises from resampling the treatment assignment vector while holding the potential outcomes and covariates fixed.  In Appendix~\ref{app:db_var_est} we first show that the bivariate weighted least squares estimator is a type of H\'{a}jek estimator, and then, following \citet{aronow2013conservative} and using standard results from model-assisted estimation \citep{breidt_opsomer_2017}, we derive the asymptotic design-based variance for this bivariate weighted least squares estimator.  We do this under a Bernoulli design and show that with weights normalized to sum to one, the variance is:

\begin{equation}
     \Var_{DB}[\wdim] \le \frac{1}{n^2} \left( \sum_i \frac{1}{\pi_i} Y_i(1)^2 + \sum_i \frac{1}{1 - \pi_i} Y_i(0)^2 \right)  \label{eqn:db_var}
\end{equation}

We also show that a conservative estimator for the asymptotic design based variance, with normalized weights, is:

\begin{align}
    \hat \Var[\wdim] = \sum_i w^2_i Z_i (Y_i - \overline{Y}_1^w)^2 + \sum_i w^2_i (1 - Z_i) (Y_i - \overline{Y}_0^w)^2 \label{eqn:db_var_y}
\end{align}

where $\Var_{DB}$ refers to the design-based variance.  This matches the form of Equation~\eqref{eq:neymanwdim}, and can be estimated using HC0 standard errors in the weighted bivariate regression.

Our question concerns the design variance of the wDIM across treatment assignments, where each time we construct some sort of balancing weight.  Briefly consider the thought experiment in which we conduct exact mean balancing, targeting the SATE and balancing the mean of $\X$, across all repeated randomizations.  Recalling that the $\X$ are fixed in the sample, then exact balancing will fix the mean value of $\X$ in each and every randomization.  Intuitively, then, the variance of the wDIM estimator will be driven by the residual variance in $Y$, because weighting will remove the exact same mean of $\X$, and thus the variation in distributions of those $\X$ across randomizations, from the outcomes in the treatment and control group in each and every randomization.

We show in Appendix~\ref{app:db_var_est} that, following existing model-assisted estimation results, the asymptotic\footnote{Following the finite population notion of asymptotics, we need to consider sequences of finite populations, where the product of treatment assignment and selection is bounded away from zero and the potential outcomes and covariates have bounded higher moments \citep{fuller_2009, breidt_opsomer_2017, d2010linearization, abadie_sampling-based_2020}.} design based variance of our residualized model, using normalized weights that sum to one in each treatment arm, is 

\begin{equation}
    \hat{\Var}_{DB}[\hat \tau] =  \sum_i w_i^2 Z_i \hat{\epsilon}_i^2 + \sum_i w_i^2 (1 - Z_i) \hat{\epsilon}_i^2
    \label{eqn:db_var_res}
\end{equation}

where $\hat{\epsilon}_i$ are the residuals from our proposed interacted model in Equation~\eqref{eq:wResModel}.\footnote{\citet{d2010linearization} note that, in the case of nonrandom selection, it is important that we use the $w$-weighted residuals in our residualized variance estimator to reduce bias in the variance estimator, as opposed to uniform weights or design-weights.  In this literature, one can use either wLS or OLS coefficients for residualization, but we suggest using wLS coefficients of the residualized regression of $Y$ on $\bm{\phi}$.
}  This comes from the fact that our residualized model is a form of model-assisted estimator.   Appendix~\ref{sec:what_hc0_estimates_pate} shows that under exact balance the HC0 variance from the residualized regression equals Equation~\eqref{eqn:db_var_res} exactly, for any weights and any finite size; Appendix~\ref{sec:what_hc0_estimates_approx} then shows that under approximate balance (e.g. due to IPW), the ratio of the two converges in probability to one.  The variance in Equation~\eqref{eqn:db_var_res} will generally be less than Equation~\eqref{eqn:db_var_y} when the residuals from the population best linear predictor have less variation than the raw $y$ values \citep{breidt_opsomer_2017}. As foreshadowed in Section~\ref{subsec:proposal}, and as with the broader model-assisted estimation literature, the justification for linearized variance estimators does not rely on a linearity assumption for the outcome \citep{fuller_2009, breidt_opsomer_2017}.  For the SATT, the process and justification is similar, except the weights are constructed to target this estimand.

\subsection{Model-based uncertainty}\label{sec:model_based}

The model-based framework treats $\{X_i\}_{i=1}^n$ and $\{Z_i\}_{i=1}^n$ as fixed, with randomness arising solely from the error draws $\{\epsilon_i\}_{i=1}^n$. The inferential target is therefore $\Var_{MB}[\hat{\tau} \mid \X,Z]$ rather than the unconditional $\Var_{MB}[\hat{\tau}]$ (Equation~\eqref{eq:neymanwdim}).

Let $\mathbf{M}$ denote the $n \times p$ design matrix of the regression in Equation~\eqref{eq:wResModel} whose columns are the intercept, $Z_i$, $\tilde{\phi}_i$, and $Z_i\tilde{\phi}_i$. Let $\mathbf{W} = \mathrm{diag}(w_1, \ldots, w_n)$. The weighted least-squares coefficient vector is $\hat{\theta} = (\mathbf{M}^\top \mathbf{W} \mathbf{M})^{-1} \mathbf{M}^\top \mathbf{W} \mathbf{Y}$, and the variance of $\hat{\tau}$ is the corresponding diagonal element of,

\begin{equation}
    \Var_{MB}[\hat{\theta} \mid \X, Z]
    = (\mathbf{M}^\top \mathbf{W} \mathbf{M})^{-1}
      \mathbf{M}^\top \mathbf{W}\,
      \mathrm{diag}(\sigma_1^2, \ldots, \sigma_n^2)\,
      \mathbf{W} \mathbf{M}\,
      (\mathbf{M}^\top \mathbf{W} \mathbf{M})^{-1}
    \label{eq:mb.var}
\end{equation}

where $\sigma_i^2 = \Var_{MB}[Y_i \mid X_i, Z_i]$ is the conditional variance of each unit's outcome. Because the weights are fixed in this framework, $\mathbf{W}$ factors out of the stochastic component, yielding the familiar sandwich form. The resulting variance estimate will typically be  smaller than that in Equation~\eqref{eq:neymanwdim}, as it is driven by the estimated \textit{conditional} variance ($\Var[Y_i \mid X_i, Z_i]$), which respects resampling under fixed $X_i$ and $Z_i$. By contrast, the simple variance in Equation~\eqref{eq:neymanwdim} employs the unconditional variance of $Y$, that fixing $X$ and $Z$ reduces the conditional variance in proportion to the prognostic value of $X$ and $Z$.

The HC0 sandwich estimator from the residualized regression in Equation~\eqref{eq:wResModel} consistently estimates Equation~\eqref{eq:mb.var} by substituting the squared residuals $\hat{\epsilon}_i^2$ for the unknown $\sigma_i^2$. As previewed in Section~\ref{subsec:proposal}, this substitution does not require correct specification of $\E[Y_i \mid X_i, Z_i]$: \citet{white_using_1980} showed that the sandwich estimator produces asymptotically valid standard errors for linear projection coefficients even under misspecification (see \citealp{buja2019models} for further development).

\subsection{Extension to the superpopulation}\label{sec:superpop}

The design-based and model-based arguments above condition on aspects of the realized sample and target sample-level estimands. When the researcher instead targets a population estimand ($\tau_{\text{PATE}}$ or $\tau_{\text{PATT}}$), additional variance arises from sampling variability in the quantities those frameworks held fixed. As discussed in Section~\ref{sec:frameworks}, our approach begins with the finite-sample variance and adds a correction for this additional source of uncertainty.

\subsubsection{Analytical correction}

The HC0 robust standard error estimates are known to be anti-conservative for the superpopulation quantities as they do not account for variation that would arise when $\X$ is drawn from a population beyond the sample. We derive a correction to the robust standard error, following a proposal by \citet{berk_etal_2013}, \citet{negi_woolridge_2021}, and \citet{ding2023coursecausalinference}, as:

\begin{align}
    \hat\Var_{SP}[\hat \tau_{PATE}] &= \sum_i{ w_i^2 Z_i \hat \epsilon_i^2} + \sum_i{w_i^2 (1 - Z_i) \hat \epsilon_i^2 } + \underbrace{\hat\gamma^\top \hat S^{2}_{\phi} \hat \gamma / n}_{\text{PATE correction}} \label{eqn:sp_correction} \\
    \hat\Var_{SP}[\hat \tau_{PATT}] &= \sum_i w_i^2 Z_i \hat \epsilon_i^2 + \sum_i w_i^2 (1 - Z_i)  \hat \epsilon_i^2 + \underbrace{\hat \gamma^\top \hat S^{2}_{\phi_{Z = 1}} \hat \gamma / n_t}_{\text{PATT correction}}\label{eqn:sp_correction_ATT}
\end{align}

where $\hat \gamma$ are the interaction coefficients estimated using  residualized regression in Equation~\eqref{eq:wResModel}.  $\hat S^{2}_{\phi} = \frac{1}{n - 1}\sum_i (\phi_i - \overline{\phi})(\phi_i - \overline{\phi})^\top$ is the covariance matrix of the covariates in the realized sample and $\hat S^{2}_{\phi_{Z=1}} = \frac{1}{n_t - 1}\sum_i Z_i (\phi_i - \overline{\phi}_1)(\phi_i - \overline{\phi}_1)^\top$, with $\overline{\phi}_1 = n_t^{-1}\sum_i Z_i \phi_i$, is the covariance matrix among the treated units.  Note that this corresponds to the difference in coefficients from wLS regressions run separately by treatment group. While the correction for the PATE extends the derivation in \citet{negi_woolridge_2021} to the observational setting, the correction for the treatment effect among the treated has not been considered in the existing literature, which has focused on randomized trials.  A derivation for both corrections in Equations~\eqref{eqn:sp_correction} and~\eqref{eqn:sp_correction_ATT} can be found in Appendix~\ref{app:superpop_corr}.

This correction does not require a correctly specified outcome model. Appendix~\ref{app:superpop_corr} defines each arm's model as a linear projection \citep{negi_woolridge_2021}, so $\gamma$ is a difference in projection slopes, well-defined and consistently estimated whether or not $\E[Y(z) \mid X]$ is linear in $\phi$. The correction accounts for sampling variability in the heterogeneity those projections represent, $\phi^\top\gamma$; heterogeneity they miss is contained in the residuals and hence appears in the HC0 terms.

\subsubsection{Bootstrap}

Researchers can also use the bootstrap as a computational alternative to the superpopulation variance estimator. A simple and valid approach generates bootstrap resamples and for each resample, constructs a new set of weights \citep{austin2016variance, austin2022bootstrap, kostouraki2024variance}. We expect, and find in simulation, that these bootstrapped standard errors would recover the uncertainty in estimating these superpopulation quantities \citep{hernan2020causal, ding2023coursecausalinference}.  See Appendix~\ref{app:bootstrap} for further discussion and simulations.

\subsection{M-estimation}\label{sec:mestimation}

An alternative approach to estimation that also accounts for sampling uncertainty in the estimation of the weights due to the random nature of the sample is through M-estimation \citep{stefanski2002calculus}, which estimates parameters that achieve a set of moment conditions. When estimating a difference-in-means (and, with additional stacked score conditions, augmented outcome models) using IPW weights, M-estimation provides a consistent variance estimator for the PATE estimator \citep[see e.g.][]{lunceford_davidian_2004, reifeis_hudgens_2022, gabriel2024inverse}.

In general, the M-estimator for the parameters \(\hat \theta\) is the solution for
the estimating equations \(\sum_{i = 1}^n \psi(O_i, \theta) = 0\) with
\(O_i = \{Z_i, X_i, Y_i\}\).  What will vary for each of our estimators is the $\psi(O_i, \theta)$, defined by the score conditions for our weighting and outcome models, which we will outline below.  For notational simplicity, in this section we let $\phi_i = X_i$, and consider regressions on the untransformed covariates.  One virtue of M-estimation is that under standard regularity conditions, it provides a valid closed-form estimate for the variance of $\hat{\theta}$ with the plug-in form

\begin{equation*}
    \begin{array}{ccc}
        \widehat{\Var}_M[\hat{\theta}] = \frac{1}{n} \hat{A}^{-1} \hat{B} \hat{A}^{-\top} & \text{where} & 
        \hat{A} = \frac{1}{n}\sum_{i=1}^n \frac{\partial}{\partial \theta^\top} \psi(O_i, \hat{\theta}),
\quad
\hat{B} = \frac{1}{n}\sum_{i=1}^n \psi(O_i, \hat{\theta})\psi(O_i, \hat{\theta})^\top.
    \end{array}
\end{equation*}

This is the usual \emph{sandwich} variance estimator.\footnote{The matrices $A$ and $B$ play roles analogous to the expected Hessian and the variance of the score in likelihood theory. When the estimating equations correspond to the score of a correctly specified likelihood, the information identity implies $A = B$, collapsing the sandwich variance to the familiar inverse information matrix.} We use the \texttt{geex} package to implement our stacked estimating equations and to directly compute the plug-in variance estimator.

We consider two weighting models here, inverse propensity weights (using logistic regression) and entropy balancing weights, for which the score conditions are well known.  Alternative weighting estimators can be analogously defined assuming you can derive the associated estimating equations.  We use weights

\begin{align}
W_i = \begin{cases}
			\frac{Z_i}{n \cdot \operatorname{expit}(G_i^\top \alpha)} + \frac{1 - Z_i}{n (1 - \operatorname{expit}(G_i^\top \alpha))}, & \text{if IPW}\\
            \frac{Z_i}{n_t}\exp(G_i^\top \eta_t) + \frac{1 - Z_i}{n_c}\exp(G_i^\top \eta_c), & \text{if Entropy Balancing}
		 \end{cases} \label{eq:mest_weights}
\end{align}

where $G_i = [1, X_i]^\top$, $\alpha$ are the parameters weighting model, and $\eta_t$ and $\eta_c$ are the parameters in the dual form of the entropy balancing optimization problem, for the treated and control weights, respectively \citep{hainmueller_entropy_2012}.

We focus on our proposed residualized model and provide additional stacked estimating equations for the bivariate regression model in Appendix~\ref{app:mest}.  For the residualized regression, let $J_i = [1, Z_i, X_i - \mu_x, Z_i(X_i - \mu_x)]^\top$, where the target causal quantity of interest, the ATE, will be the coefficient on $Z$.  The stacked estimating equations for our M-estimators are then

\begin{equation}
    \begin{array}{ccc}
        \text{IPW} & & \text{Entropy Balancing} \\
        \left[ \begin{array}{c} \psi_{\mu_{x_c}} \\  \psi_{\mu_{zg}} \\ \psi_\alpha \\ \psi_\lambda \end{array} \right] = \left[ \begin{array}{c} (1 - Z_i) (X_i - \mu_{x_c})\\ (Z_iG_i - \mu_{zg}) \\ \operatorname{expit}(G_i^{\top}\alpha)G_i - \mu_{zg} \\ \{ W_i(Y_i - {J_i}^\top \lambda)\} J_i  \end{array} \right]
        & ; &   \left[ \begin{array}{c}  \psi_{\mu_x} \\ \psi_{\eta_t} \\ \psi_{\eta_c} \\ \psi_\lambda \end{array} \right] = \left[ \begin{array}{c}  X_i - \mu_x \\ \frac{Z_i}{n_t}\left(\exp(G_i^{\top}\eta_t)\, G_i - [1\;\; \mu_x^\top]^\top\right) \\ \frac{1 - Z_i}{n_c}\left(\exp(G_i^{\top}\eta_c)\, G_i - [1\;\; \mu_x^\top]^\top\right) \\ \{ W_i(Y_i - {J_i}^\top \lambda)\} J_i  \end{array} \right].
\end{array} \label{eq:mest_stacks_main}
\end{equation}

For IPW, $\psi_{\mu_{x_c}}$ are the score condition for estimating $\mu_{x_c}$, the mean $X$ among the control units ; $\psi_{\mu_{zg}}$ and $\psi_\alpha$ are the score conditions for the logistic regression that produces the IPW weights (the $\mu_x$ used to center $J_i$ is a function of $\mu_{zg}$ and $\mu_{x_c}$; see Appendix~\ref{app:mest}). For entropy balancing, $\psi_{\mu_x}$ corresponds to the score conditions for estimating $\mu_x$, the population mean of the $X$; $\psi_{\eta_t}$ and $\psi_{\eta_c}$ are the score conditions for estimating the treatment and control entropy balancing weights, respectively.  Finally, $\psi_\lambda$ gives the score conditions for estimating the residualized regression using weights $W_i$ defined above.

Note that the IPW score conditions in Equation~\eqref{eq:mest_stacks_main} for the logit model are in a form different than the existing literature. In Appendix~\ref{app:mest}, we rewrite the existing IPW estimating equations to show they implicitly estimate (a transformation of) the target population means (i.e. those for the treated), and use this to motivate our stacked estimating equations for entropy balancing.  We refer to this reframing as making clear that ``uncertainty in weights'' in IPW estimation is really related to the ``uncertainty in population means'' problem. To the best of our knowledge, the appropriate score conditions for exact balancing weights have not previously been worked out in the existing literature.\footnote{Though the covariate balancing propensity score similarly employs the generalized method of moments \citep{imai2014covariate}.}  This ``uncertainty in population means'' is also essential in our residualized regression, which must re-center using the superpopulation parameter covariate means for superpopulation estimands.

In addition to providing the appropriate M-estimation stacked estimating equations for our superpopulation quantities, we provide a fixed sample M-estimator appropriate for our design-based and model-based resampling frameworks.  In a fixed sample, the target means for balancing constraints are known; we show how to update the estimating equations in Equation~\eqref{eq:mest_stacks_main} to incorporate this fact in Appendix~\ref{app:mest}.  In short, $J$ is updated to recenter on the fixed sample covariate means  $\overline{X}$; we update $\psi_\alpha$ to use $\overline{ZG}$ in place of $\mu_{zg}$ and $\psi_{\eta_t}$ and $\psi_{\eta_c}$ to use $\overline{X}$ in place of $\mu_X$; and we can remove the estimating equations related to estimating these covariate means ($\psi_{\mu_{x_c}}$ for IPW, $\psi_{\mu_{x}}$ for entropy balancing).

In Appendix~\ref{app:mest} we detail the M-estimation approach for both superpopulation and fixed sample resampling methods; provide the appropriate stacked estimating equations for both bivariate regression and residualized regressions; and adapt these for either IPW or balancing weights.  We do this for both ATE and ATT estimands.  We conduct simulations and show that the M-estimation approach performs nearly identically to our proposed residualized regression approach.  This shows that while M-estimation can be used to appropriately estimate standard errors, both in scenarios where weights are estimated and where they are considered ``fixed'', it is not the only way, nor a necessary way, to account for uncertainty in weighted causal estimators.  Our simple proposed residualized regression, with superpopulation correction when necessary, can also appropriately account for this uncertainty.

\subsection{Effect of weighted regression on point estimates and efficiency}\label{sec:exact_v_approx}

\paragraph{Imbalance and the weighted regression point estimate.}
To concretely see how the point estimate of $\hat\tau$ compares with the wDIM note that (Appendix~\ref{app:AIPW}),

\begin{equation}
\hat\tau = \wdim - \hat\delta_1^\top (\hat\beta + \hat\gamma) + \hat\delta_0^\top \hat\beta,
\qquad \hat\delta_z := \overline\phi^w_z - \phibstar,
\label{eq:decomp}
\end{equation}

\noindent where $\bar\phi^w_z = \sum_{i:Z_i=z} w_i \phi_i$ is the weighted feature mean in arm $z$, so that $\hat\delta_z$ describes the residual imbalance of covariates in arm $z$ relative to the common balancing target $\bar\phi^\star$. Equation \eqref{eq:decomp} is exact in any sample, for any weights and clarifies that the point estimate moves away from the wDIM estimate to the extent that (i) each arm is not fully balanced to the common target, and (ii) those imbalances predict the outcome in each arm. 

This immediately produces the result that under exact mean balancing (so that $\hat\delta_1 = \hat\delta_0 = 0$), the weighted regression does not change the point estimate, $\hat\tau = \wdim$. 
This has been noted previously, e.g.\ \citet{hainmueller_entropy_2012} and \citet{chattopadhyay2023implied}, and is easy to show directly via Frisch-Waugh-Lovell (FWL) theorem \citep{frisch1933partial, lovell1963seasonal}, see Appendix~\ref{app:exactBalwLSequiv}.
This result also formalizes the logic we discuss in Appendix \ref{app:application-disc}: large changes in the point estimate require covariates that are both residually imbalanced and prognostic.

\paragraph{IPW: Equivalence to AIPW} 

Since IPW does not produce exact balancing weights, $\hat\tau \neq \wdim$ in general. Fortunately, as we describe here, this change to the point estimate is not merely benign, but results in an efficiency-enhancement because the weighted outcome regression is exactly an augmented inverse propensity weighting (AIPW) estimator. 
We emphasize that under the assumption above that the wDIM already consistently estimates the target estimand, this augmentation plays no role in identification of the point estimate or in obtaining ``double-robustness'' \citep[e.g.][]{robins1994estimation}; the benefit is instead efficiency. 

Recall that because the regression in Equation~\eqref{eq:wResModel} is fully interacted with $Z$, it is numerically identical to fitting separate weighted regressions 

\begin{equation}
\hat{m}_0(\phi) = \hat\beta_0 + (\phi - \phibstar)^\top \hat\beta,
\qquad
\hat{m}_1(\phi) = (\hat\beta_0 + \hat\tau) + (\phi - \phibstar)^\top (\hat\beta + \hat\gamma),
\label{eq:armfits}
\end{equation}

\noindent The $\hat{\tau}$ estimated in Equation~\eqref{eq:wResModel} is then equal to performing g-computation/regression imputation with these two fitted models, i.e.  $\hat{\tau} = \frac{1}{n}\sum_{i=1}^n \big[\hat{m}_1(\phi_i) - \hat{m}_0(\phi_i)\big]$. 

The AIPW estimator for the ATE, employing outcome models $\hat{m}_1, \hat{m}_0$ and with weights normalized (summing to 1 per arm) is given by

\begin{equation}
\hat\tau_{AIPW} =\frac{1}{n}\sum_{i=1}^n \big[\hat{m}_1(\phi_i) - \hat{m}_0(\phi_i)\big]
+ \left[\sum_i w_i Z_i \big(Y_i - \hat{m}_1(\phi_i)\big) - \sum_i w_i (1-Z_i)\big(Y_i - \hat{m}_0(\phi_i)\big)\right]
\label{eq:aipw}
\end{equation}

The first term in \eqref{eq:aipw} is simply our $\hat{\tau}$ while the second is the ``augmentation'' term, contrasting the weighted mean residuals by arm. Both of these weighted mean residuals, however, are made identically zero by the first-order conditions of the weighted regressions. That is, because each arm's weighted regression freely estimates an intercept, they solve the first-order conditions $\sum_{i:Z_i=z} w_i \big(Y_i - \hat{m}_z(\phi_i)\big) = 0$ exactly, thus zeroing out the entire augmentation term in \eqref{eq:aipw}. Accordingly, $\hat\tau = \hat\tau_{AIPW}$ exactly in every sample. The analogous statement holds for the ATT, with the predicted effects averaged over the treated units (Appendix~\ref{app:AIPW}).

This equivalence allows us to leverage well-established theory regarding efficiency and variance estimation for AIPW. Recall that we assume only consistency, which holds by either of two routes (Section~\ref{subsec:weighting_methods}): correct specification of the explicit or implied model for $e(X)$, or a choice of $\phi$ rich enough that $\E[Y(z) \mid X]$ is linear in $\phi$. Semiparametric efficiency demands more: that \textit{both} hold. The first ensures that the weights converge to the true inverse propensity weights; the second is exactly the condition that the arm-specific outcome models be correctly specified. Under both, the resulting AIPW estimator (and thus the identical $\hat\tau$) attains the semiparametric efficiency bound, so that no regular estimator has smaller asymptotic variance \citep{hahn1998role, robins1994estimation}. Accordingly, while the proposed estimate will differ from the wDIM point estimate under IPW, this difference moves the estimate in the direction of the efficient estimator, attaining it when both conditions hold. 

Regarding variance estimation, writing $\epsilon(z)$ for the projection errors of each arm's model, this standard variance is \citep[e.g.][]{robins1994estimation, lunceford2004stratification, glynn2010introduction}

\begin{equation}
\Var_{SP}[\hat\tau] \approx \frac{1}{n} \E\left[\frac{\epsilon^2(1)}{e(X)}\right]
+ \frac{1}{n}\E\left[\frac{\epsilon^2(0)}{1-e(X)}\right]
+ \frac{1}{n}\Var_{SP}\big[\phi^\top \gamma \big],
\label{eq:aipwvar}
\end{equation}

\noindent which has sample analog

\begin{equation}
\widehat{\Var}_{SP}[\hat\tau] \;=\; \sum_i w_i^2 Z_i\, \hat\epsilon_i^2
\;+\; \sum_i w_i^2 (1-Z_i)\, \hat\epsilon_i^2
\;+\; \frac{\hat\gamma^\top \hat{S}^2_{\phi}\, \hat\gamma}{n},
\label{eq:aipwvarhat}
\end{equation}

where $\hat{S}^2_{\phi}$ is the sample covariance of the features, so that $\hat\gamma^\top \hat{S}^2_{\phi} \hat\gamma$ estimates $\Var_{SP}[\phi^\top \gamma]$.\footnote{Because the weights as defined in Equation~\eqref{eqn:wtd_dim} are normalized to sum to one within each arm, $w_i \approx 1/(n\,\hat{e}(X_i))$ for treated units. Then $\E[\epsilon^2(1)/e(X)] = \E[Z\,\epsilon^2(1)/e(X)^2]$ by conditional ignorability, whose sample analog scaled by $1/n$ is $n^{-2}\sum_i Z_i \hat\epsilon_i^2 / \hat{e}(X_i)^2 = \sum_i w_i^2 Z_i \hat\epsilon_i^2$; the control term follows identically. See Appendix~\ref{app:superpop_corr} for the full argument.} Fortunately, this exactly matches Equation~\eqref{eqn:sp_correction}, the HC0 standard error of the weighted Lin regression (the first two terms) together with the proposed superpopulation correction (the third term).

\section{Simulation evidence}\label{sec:sims}

\subsection{Data Generating Process}
Our simulations build on those conducted in~\citet{hainmueller_entropy_2012}.  There are six covariates $X_j$ with $j \in (1, \dots, 6)$. $X_1$, $X_2$, and $X_3$ are multivariate normal with means zero, variances of $(2, 1, 1)$ with $\Cov[X_1, X_2] = 1$, $\Cov[X_1, X_3] = -1$, and $\Cov[X_2, X_3] = -0.5$. $X_4$ is uniformly distributed on $[-3, 3]$; $X_5$ is $\chi^2$ distributed with one degree of freedom, and $X_6$ is Bern($0.5$).  

The treatment status of a unit is: 

\vspace{-1em}
\[ Z_i = \mathbf{1}\{X_{1i} + 2X_{2i} - 2X_{3i} - X_{4i} - 0.5X_{5i} + X_{6i} + \epsilon_i > 0\} \]

where the proportion of treated units is, in expectation, 1/2 and $\epsilon \sim N(0, 200)$. We consider a moderately non-linear outcome: $Y_i(0) = X_{1i} + X_{2i} + 0.2 X_{3i} X_{4i} - \sqrt{X_{5i}} + \nu_i$, where $\nu \sim N(0, 17.1)$ chosen so as to calibrate the overall linear signal to noise ratio to around 0.25 for the potential outcomes under control. Note that $X_6$, while relevant to treatment probability, is not prognostic.

Finally, we consider two scenarios, one in which the treatment effect is homogeneous, with $\tau_i = 0 \ \forall i$, or heterogeneous, with $\tau_i = X_{3i} + X_{5i} + 2 X_{3i}X_{5i}$. This definition ensures that, under heterogeneity, covariates are correlated with individual-level treatment effects, although treatment effect is not linear in those covariates.  We then construct the potential outcome under treatment as $Y_i(1) = Y_i(0) + \tau_i$ for all outcome models described above.

We consider the design-based, model-based, and superpopulation frameworks as described in Section~\ref{sec:frameworks}.  Under the design-based process, $\X$ is fixed in the sample, and all variation across simulations arises due to imagined reassignment of $Z$, which is driven by redrawing $\epsilon$ in each realization.  In the model-based process, both $\X$ and $Z$ are fixed across simulations, but $\nu$ is redrawn for each iteration, thus generating stochastic potential outcomes.  For superpopulation resampling, we redraw $\X$, $\nu$, and $\epsilon$ (and thus $Z$).  We fix the sample size at $n = 1000$ and conduct 4,000 simulations for each resampling paradigm.

\subsection{Estimators}
We construct estimators for both the ATE and ATT, and employ three weighting approaches: entropy balancing, inverse propensity weighting, and matching.  For entropy balancing, we conduct exact mean balancing. For IPW, we construct weights by estimating a probit model of treatment on marginal $\X$, stabilizing as appropriate per target estimand.  For matching, we conduct generalized full matching on all covariates $\X$ using the quick match algorithm of \citet{savje2021generalized} as implemented in the \texttt{MatchIt} package in R \citep{hoetal_matchit_2011}.  The other approaches are implemented using the \texttt{WeightIt} 
\citep{greifer2025} package in R.

Using the weights described above, we estimate the standard errors in two ways. The ``\textcolor{neyman}{bivariate}'' standard errors are estimated using HC0 standard errors from the weighted regression of outcome on treatment.   The ``\textcolor{separate}{residualized}'' standard errors are HC0 standard errors from the residualized regression in Equation~\eqref{eq:wResModel} with appropriate centering.  For the superpopulation  resampling framework, we also include the corrections described in Equation~\eqref{eqn:sp_correction}, also referred to as \textcolor{separate}{Residualized}.\footnote{Some researchers may consider a pooled regression, where the centered covariates are not interacted with treatment.  We do not consider this as it is, in general, less efficient than the residualized regression under heterogeneity and nearly equivalent under homogeneity.  However, it may be a prudent choice for researchers balancing a large number of covariates with relatively small sample sizes, where degrees of freedom in the regression may be a concern.  The superpopulation correction would need to be adjusted appropriately, see \citet{negi_woolridge_2021}.}

\subsection{Results}

\begin{figure}[ht]
\centering
\includegraphics[width=\linewidth]{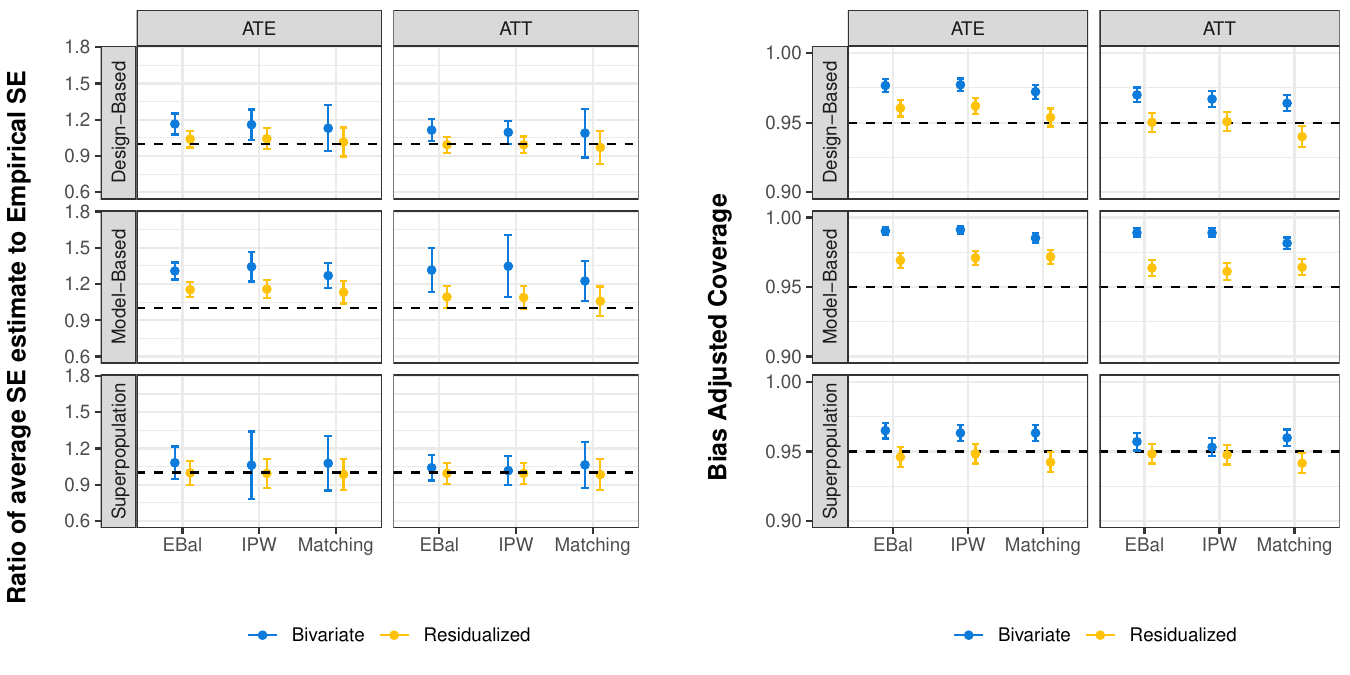}
\begin{subfigure}[b]{0.48\textwidth}
  \centering
  \caption{Ratio of the average estimated standard error to the empirical standard error. }
  \label{fig:ratio_main}
\end{subfigure}
\hspace{0.01\textwidth}
\begin{subfigure}[b]{0.015\textwidth}
    
\end{subfigure}
\begin{subfigure}[b]{0.48\textwidth}
  \centering
  \caption{Bias-adjusted coverage results for Normal approximation 95\% confidence intervals.}
  \label{fig:ba_cov_main}
\end{subfigure}
\caption{Simulation results under heterogeneous treatment effects.}\label{fig:main_results}
\end{figure}

Across simulations we evaluate average performance of the estimated standard error relative to the simulated empirical standard error as well as coverage for the target estimand.  We present the heterogeneous treatment effect results in this section (Figure~\ref{fig:main_results}).

We begin with a discussion of the precision gains possible from residualization.  We present the ratio of the average estimated standard error to the empirical standard error (i.e. the standard deviation of estimates across simulation draws) in Figure~\ref{fig:ratio_main}. An ideal standard error estimator would be close to one, or slightly greater than one for a conservative estimator.  Our proposed residualized estimator is generally closer to one than the bivariate variance estimator, although as it is a conservative estimator in the model-based and design-based settings, it is greater than one in some settings.  The residualized regression leads to reductions of up to 30\% in the estimated standard error as compared to the bivariate regression estimator, depending on the data-generating process.  Importantly, we see that gains persist across different weighting approaches and for both of the target estimands.

Next we turn to (bias-adjusted) coverage of the target estimands in Figure~\ref{fig:ba_cov_main}. We should expect our suggested residualized standard error to have nominal or conservative coverage, with nominal coverage in the superpopulation setting. We note that: 1) the common practice using the robust standard errors from the bivariate regression exhibits significant overcoverage, even under homogeneity, and 2) all of the estimated standard errors under residualization are nominal or conservative.  While both the bivariate and residualized regression may be conservative under a Neyman-style bounding exercise on the covariance term, the bivariate regression estimator's conservativeness is greatly exacerbated by a failure to account for the precision gains from weighting.  Results under homogeneity and for alternative outcomes (including a linear and highly non-linear outcome), as well as numerical results, are presented in Appendix~\ref{app:sim_figures}.

\section{Empirical Applications}\label{sec:app}
We demonstrate the precision gains from using our suggested residualized standard error on replication data from three studies. Appendix Table \ref{tab:replication_summary} includes original results from the studies.

\subsection{Ladd and Lenz (2009)}
The first study from \citet{ladd2009exploiting} examines the persuasive power of the news media in the 1997 British election. At the beginning of the election campaign, several newspapers switched their support from the ruling Conservative party to endorse the Labour candidate. The authors compare respondents that reported reading one of these newspapers to those who report reading newspapers whose partisan endorsements did not change. Since selection into these groups is nonrandom, the original study uses genetic matching with 39 control variables and exact matching with 9 categorical variables. These 39 variables include measures of prior ideology, prior support for the Labour party, socioeconomic status, gender, age, region, and occupation.  A reanalysis by \citet{hainmueller_entropy_2012} uses entropy balancing. 
 
\subsection{Chan et al. (2016)}
We additionally reexamine a childhood nutrition study from \citet{chan2016globally}. The authors study whether participation in school breakfast and lunch meal programs leads to an increase in body mass index. They use data from the 2007--2008 National Health and Nutrition Examination Survey which includes data on 2,330 youths aged 4 to 17 years old of which about 55\% received meals at school. Since participation in school meal programs depends on the income of the student's family, treatment assignment is nonrandom. Following \citet{chan2016globally}, we adjust for 11 covariates including the child's age, gender, race, family income being above 200\% of federal poverty line, participation in food stamp or `Special Supplemental Nutrition Program for Women, Infants, and Children,' a food security index, health insurance coverage, and the survey responder's age and gender. 

\subsection{Black and Owens  (2016)}
\citet{black2016courting} study the effect of promotion prospects to the Supreme Court on the behavior of circuit court judges. The authors define a binary treatment as vacancy vs. non-vacancy periods.   They investigate subgroup effects for judges with promotion prospects, referred to as contender judges, and non-contender judges, which we replicate here. The authors use coarsened exact matching on seven covariates (six for non-contenders), including the judge's Judicial Common Space score, ideological alignment with the president, and whether the circuit court reversed the case decision.  This data is also reanalyzed by \citet{xu2023hierarchically} using hierarchically regularized entropy balancing. The authors find that contender judges  are more likely to vote in line with the president's preferences during vacancy periods. The same is not true among non-contender judges.

\subsection{Results}

Table \ref{table:application_tab} includes the results from our reanalyses. For each study, we conduct exact mean balancing with an entropy loss, inverse propensity score weighting, and fast generalized full matching. We target the sample ATE for \citet{chan2016globally} and the sample ATT for \citet{ladd2009exploiting} and \citet{black2016courting}.  We present point estimates and standard errors for three sets of estimators: unweighted OLS, bivariate wLS, and our proposed residualized regression.  For the residualized estimator, we include all covariates used in the weighting or matching. Unweighted estimates are from a simple regression of outcome on treatment. Appendix \ref{app:application-superpop-correction} contains superpopulation corrected standard errors that target the PATE and PATT using Equations~\eqref{eqn:sp_correction} and~\eqref{eqn:sp_correction_ATT}. To show potential precision gains, we present the ratio of the standard errors from the bivariate wLS and our residualized regression to the unweighted robust standard errors. We additionally report $R^2_{Y\sim X, w}$ from a weighted regression of outcome on covariates, which summarizes how prognostic the covariates are collectively and so governs the precision gain from residualization. We also include $R^2_{Z\sim X, w}$ from a weighted regression of treatment on covariates, which summarizes residual imbalance.

These studies represent different settings faced by applied researchers in terms of how prognostic covariates are and how well balance is achieved. Based on $R^2_{Y\sim X, w}$, we broadly categorize how prognostic covariates are in each study. Covariates in \citet{chan2016globally} are quite prognostic ($R^2_{Y\sim X, w} \approx 0.3$), and those in \citet{ladd2009exploiting} are even more so  ($R^2_{Y\sim X, w} \approx 0.5$). \citet{black2016courting} covariates are not particularly prognostic for either subgroup analysis ($R^2_{Y\sim X, w} \approx 0.01 \text{ to }0.15$). $R^2_{Z\sim X, w}$ indicates how well balance was achieved by weighting. It is exactly zero under entropy balancing, and never exceeds $0.009$ under IPW or matching weights. As discussed in Section~\ref{subsec:proposal}, this allows residualization to reduce standard errors without the risk of inflation due to collinearity that additional covariates may otherwise generate.  Appendix \ref{app:application-diagnostics} includes the remaining imbalance and partial-$R^2$ for each covariate.

\begin{table}[ht]
\centering
\resizebox{\columnwidth}{!}{
\begin{tabular}{ll cc ccc ccc cc}
\toprule

& & \multicolumn{2}{c}{\textbf{Unweighted}} & \multicolumn{3}{c}{\textcolor{neyman}{\textbf{Bivariate}}} & \multicolumn{3}{c}{\textcolor{separate}{\textbf{Residualized}}} & \multicolumn{2}{c}{} \\
\cmidrule(lr){3-4} \cmidrule(lr){5-7} \cmidrule(lr){8-10}

\textbf{Method} & \textbf{Study} & 
$\bm{\hat\tau}$ & $\mathbf{\widehat{se}_{unw}}$ & 
$\bm{\hat\tau}$ & $\mathbf{\widehat{se}}$ & $\mathbf{\widehat{se}/\widehat{se}_{unw}}$ & 
$\bm{\hat\tau}$ & $\mathbf{\widehat{se}}$ & $\mathbf{\widehat{se}/\widehat{se}_{unw}}$ & 
$\mathbf{R^2_{Y\sim\mathbf{X}, w}}$ & $\mathbf{R^2_{Z\sim\mathbf{X},w}}$ \\
\midrule

\multirow{5}{*}{\rotatebox{90}{\textbf{Entropy}}} 
& Ladd and Lenz (2009) & 0.15 & 0.04 & 0.12 & 0.04 & 1.11 & 0.12 & 0.03 & 0.84 & 0.49 & 0.000 \\
& Chan et al. (2016) & 0.53 & 0.23 & -0.05 & 0.28 & 1.24 & -0.05 & 0.22 & 0.99 & 0.32 & 0.000 \\
& Black and Owens (2016) & & & & & & & & & & \\
& $\quad$ Contender & 0.10 & 0.01 & 0.03 & 0.01 & 1.36 & 0.03 & 0.01 & 1.18 & 0.14 & 0.000 \\
& $\quad$ Non-contender & 0.03 & 0.01 & 0.04 & 0.01 & 1.01 & 0.04 & 0.01 & 1.00 & 0.01 & 0.000 \\
\midrule
\addlinespace
\multirow{5}{*}{\rotatebox{90}{\textbf{IPW}}} 
& Ladd and Lenz (2009) & 0.15 & 0.04 & 0.10 & 0.04 & 1.10 & 0.12 & 0.03 & 0.84 & 0.47 & 0.005 \\
& Chan et al. (2016) & 0.53 & 0.23 & -0.17 & 0.30 & 1.34 & -0.05 & 0.23 & 1.01 & 0.33 & 0.002 \\
& Black and Owens (2016) & & & & & & & & & & \\
& $\quad$ Contender & 0.10 & 0.01 & 0.02 & 0.01 & 1.36 & 0.03 & 0.01 & 1.18 & 0.14 & 0.003 \\
& $\quad$ Non-contender & 0.03 & 0.01 & 0.04 & 0.01 & 1.01 & 0.04 & 0.01 & 1.00 & 0.01 & 0.001 \\
\midrule
\addlinespace
\multirow{5}{*}{\rotatebox{90}{\textbf{Matching}}} 
& Ladd and Lenz (2009) & 0.15 & 0.04 & 0.08 & 0.05 & 1.28 & 0.12 & 0.03 & 0.90 & 0.54 & 0.009 \\
& Chan et al. (2016) & 0.53 & 0.23 & -0.30 & 0.32 & 1.41 & -0.20 & 0.26 & 1.14 & 0.31 & 0.005 \\
& Black and Owens (2016) & & & & & & & & & & \\
& $\quad$ Contender & 0.10 & 0.01 & 0.02 & 0.03 & 3.23 & 0.03 & 0.02 & 2.49 & 0.15 & 0.009 \\
& $\quad$ Non-contender & 0.03 & 0.01 & 0.05 & 0.01 & 1.36 & 0.04 & 0.01 & 1.21 & 0.02 & 0.005 \\
\addlinespace
\bottomrule
\end{tabular}
}
\caption{Unweighted, weighted bivariate, and weighted residualized estimators for three studies and three weighting methods. Bivariate and residualized estimators include a column $\widehat{se}/\widehat{se}_{unw}$ which is the ratio of each estimator's standard errors, relative to the unweighted standard error estimate. }
\label{table:application_tab}
\end{table}

Most notably, our residualized estimator decreases standard errors across all studies and weighting methods relative to the weighted bivariate regression. Although weighting increases the common bivariate variances,\footnote{Standard errors after matching for the contender subgroups in \citet{black2016courting} are particularly large  due to a large reduction in the effective sample size. There are 5,681 control units and 4,490 treated units, but after matching, the effective sample size for the control group drops to 245.} our residualized estimator offsets this variance inflation. In some cases, the ratio of the residualized standard error to the unweighted is even below one, indicating the residualized estimator can even improve precision relative to the unweighted estimator. The gains track the prognostic value of the covariates, as the residualized variance in Equation~\eqref{eqn:db_var_res} replaces outcome deviations with residuals: they are largest in \citet{ladd2009exploiting} ($R^2_{Y\sim X, w} \approx 0.5$), smaller in \citet{chan2016globally} ($\approx 0.3$), and negligible for the \citet{black2016courting} non-contender analysis ($\approx 0.01$).

Adding covariates in the residualized regression does not change the point estimates when using entropy balancing.  However, IPW and matching do not exactly balance all covariates, so the point estimates from the two estimators may be different.  This is particularly noticeable in the \citet{chan2016globally} study, where the IPW and matching point estimates are substantially different from the wDIM in the residualized regression, yet the difference is not statistically significant.  We discuss this further in Appendix \ref{app:application-disc}. This demonstrates an important consideration for residualization when using approximate balancing weights, namely that the point estimate may change when controlling for covariates that are highly prognostic and residually imbalanced.

\section{Discussion}\label{sec:discussion}

Residualization or linearization approaches are not novel, and are even standard practice in some areas. For example, in the seminal entropy balancing paper, \citet{hainmueller_entropy_2012} notes, ``[as] indicated above, the entropy balancing weights can be easily combined with almost any standard estimator that the researcher may want to use to model the outcome in the preprocessed data. In particular, the entropy balancing weights are easily passed to regression models that may further address the correlation between the outcome and covariates in the reweighted data and also provide variance estimates for the treatment effects (which treat the weights as fixed).''

In addition, the well-known and widely used \texttt{WeightIt} package in \texttt{R} allows various options and employs different defaults in different settings but broadly calls for residualization, noting for example,  ``[when] doing g-computation after weighting... the outcome model should be fit incorporating the estimated weights'', and ``we use weighted g-computation when possible for all effect estimates, even if there are simpler methods that would yield the same estimates''.\footnote{See documentation at \href{https://ngreifer.github.io/WeightIt/articles/estimating-effects.html}{https://ngreifer.github.io/WeightIt/articles/estimating-effects.html}.}  Similar arguments are made for adjustment to matched datasets \citep{greifer2021matching}. Likewise, the \texttt{WeightIt} documentation advises that ``estimation (i.e., of SEs, confidence intervals, and p-values) may consider the variety of sources of uncertainty present in the analysis, including (but not limited to!) estimation of the propensity score (if used) and estimation of the treatment effect (i.e., because of sampling error)''. 

However, while our prescriptions overlap with previously offered guidance, there remains at best incomplete consensus, a lack of complete discussion, and most notably, differences in theoretical motivations. Across the accounts cited above and others, three types of problems are often mentioned to motivate estimation choices:
(i) the need to account for residual imbalances, especially for approximate balancing and IPW approaches; (ii) the need to account for sampling uncertainty in the weights, in the sample, and (iii) the need to account for uncertainty in $p(X)$ or in $p(X,Z)$ not captured in the sample.

Our analysis has produced important lessons and answers to these open questions. First, the problem of ``uncertainty in the weights'' can be addressed in different ways, but first has to be refined to distinguish two possible sources of uncertainty. Within a given sample there is sampling uncertainty in the weights in the design-based approach, as reassigning treatment would produce different weights; however, our analysis proposes to side-step this by appealing to asymptotics.  In the model-based approach, since $\X$ and $Z$ are fixed, the weights are fixed, and no notion of uncertainty should be entertained for purposes of variance estimation.  However, ``sampling uncertainty in the weights'' can also be an appropriate term to describe the added issues with variance estimation for the PATE or PATT: here, the $\X$ and $Z$ available in our sample cannot stand in for the $\X$ and $Z$ in the superpopulation of presumed interest, highlighting that it is really an ``uncertainty in population means'' issue.  We do not assume the investigator is wedded to a superpopulation view and instead separate the inferential goals related to the sample and those related to a postulated superpopulation. If one wishes to make inferences about a superpopulation from which the obtained sample has been drawn (as for the PATE or PATT), then additional adjustments are required to achieve this. Fortunately, our simple proposed superpopulation correction is available for doing so. One notable consequence of the last two points above is that, by our current analysis, bootstrapping, existing M-estimation methods and similar approaches intended to incorporate sampling uncertainty in the weights are not necessary for the (S)ATE and (S)ATT, and we show how to adapt M-estimation to the fixed sample setting.

Second, and perhaps most remarkably, despite the differences between model- and design-based notions of uncertainty, both cases offer straightforward justifications for the ``residualized'' standard error obtained by using a weighted outcome model, with the details of the weight construction and outcome model depending on the estimand. Further, in both model- and design-based settings, these results can be adjusted to account for uncertainty over the $p(\X,Z)$ in a superpopulation, using a simple approach described in Equations~\eqref{eqn:sp_correction} and~\eqref{eqn:sp_correction_ATT}. 

This convenience is made still more fortunate as our proposed residualized estimators (i) are familiar, having been proposed and used in various causal inference settings, (ii) are consistent with a well-established tradition in the survey literature, and finally, (iii) are easy to implement with standard software. 
The theoretical expectations regarding our proposed residualized estimator are borne out in simulations, and demonstrated by application. In short, we hope that this work helps investigators consider not just whether residualized standard errors should be used, but also why.

\clearpage
\singlespacing
\printbibliography
\doublespacing
\end{refsection}

\clearpage

\appendix 
\pagenumbering{arabic}
\renewcommand*{\thepage}{A\arabic{page}}
\setcounter{section}{0}
\renewcommand*{\thesection}{A\arabic{section}}
\setcounter{figure}{0}
\renewcommand*{\thefigure}{A\arabic{figure}}
\setcounter{table}{0}
\renewcommand*{\thetable}{A\arabic{table}}
\setcounter{equation}{0} 
\renewcommand{\theequation}{A\arabic{equation}}

\singlespacing
\begin{center}
\Large
\textbf{Appendix for: Inference with weights: Residualization produces short, valid intervals for varying estimands and varying resampling processes}
\end{center}

\label{sec:appendix}
\begin{refsection}

\section{Equivalency of proposed regression and wDIM point estimate with exact balancing weights} \label{app:exactBalwLSequiv}

Consider for simplicity the regression of $Y$ on $Z$ and $X$, without transforming $X$ into $\phi$ or interacting it with $Z$. Using the Frisch--Waugh--Lovell (FWL) theorem \citep{frisch1933partial, lovell1963seasonal}, the coefficient on $Z$ from the weighted regression of $Y$ on $Z$ and $X$ is

\[ \widehat{\tau_w} = \frac{\Cov_w(Z^{\perp_w X},Y)}{\Var_w(Z^{\perp_w X})}\] 

\noindent where $Z^{\perp_w X}$ is the residual from the weighted regression of $Z$ on $\X$. However, when $w$ is exact mean balancing, $\X$ is unpredictive of $Z$ in the weighted regression, so $Z^{\perp_w X}$ is just $Z-\overline{Z}_w$. The coefficient is thus equal to $\frac{\Cov_w(Z,Y)}{\Var_w(Z)}$, which is precisely the weighted difference-in-means.

\section{Design-based variance estimators}\label{app:db_var_est}

\subsection{Weighted least squares as H\'{a}jek estimators}\label{app:wls_hajek}

We first show that least squares is a H\'{a}jek-style estimator.  To be consistent with standard design-based inference notation, let $\pi_i = e(X_i)$ be the probability unit $i$ is in treatment and $(1 - \pi_i) = 1 - e(X_i)$ be the probability unit $i$ is assigned to control.  For now, we let $w_i = Z_i / \pi_i + (1 - Z_i) / (1 - \pi_i)$, that is we do not normalize the weights.  We will return to normalized weights below.

Assume a population model where

\[Y_i = \beta_0 + \tau_{wDIM} Z_i + \epsilon_i\]

where we assume $\E[\epsilon_i \mid Z_i] = 0$.  Another way of stating this is that we can mean deviate $Y$, with mean-zero residuals, in each group.

The weighted least squares estimator solves

\begin{equation}
(\hat{\tau}_{wDIM}, \hat{\beta}_0)
= \operatorname*{argmin}_{\tau_{wDIM}, \beta_0}
\sum_{i=1}^n w_i \bigl(Y_i - \beta_0 - \tau_{wDIM} Z_i\bigr)^2
\label{eq:wBiModel}
\end{equation}

and the associated score conditions are

\begin{align}
-2 \sum_i w_i (Y_i - \hat \beta_0 - \hat \tau_{wDIM} Z_i) &= 0 \label{eq:score_control} \\
-2 \sum_i w_i Z_i (Y_i - \hat \beta_0 - \hat \tau_{wDIM} Z_i) &= 0. 
\label{eq:score_treated}
\end{align}

Rearranging Equation~\eqref{eq:score_treated}, we get:

\begin{equation}
\sum_i w_i Z_i Y_i = (\hat\beta_0 + \hat \tau_{wDIM}) \sum_i w_i Z_i \implies (\hat\beta_0 + \hat \tau_{wDIM}) = \frac{\sum_i w_i Z_i Y_i} {\sum_i w_i Z_i}
\label{eq:treated_wls_mean}
\end{equation}

which is the H\'{a}jek estimator for the treated group mean.  Turning to Equation~\eqref{eq:score_control}, 

\begin{align}
    \sum_i w_i Y_i &= \hat \beta_0 \sum_i w_i + \hat \tau_{wDIM} \sum_i w_i Z_i \nonumber\\
    \implies \sum_i w_i Z_i Y_i + \sum_i w_i (1 - Z_i) Y_i &= \hat \beta_0 \sum_i w_i (1 - Z_i) + \hat \beta_0 \sum_i w_i Z_i  + \hat \tau_{wDIM} \sum_i w_i Z_i \nonumber\\
    \intertext{Using the results from Equation~\eqref{eq:treated_wls_mean}:}\nonumber\\
    \implies \sum_i w_i Z_i Y_i + \sum_i w_i (1 - Z_i) Y_i &= \hat \beta_0 \sum_i w_i (1 - Z_i) + \sum_i w_i Z_i Y_i \nonumber\\
    \implies \sum_i w_i (1 - Z_i) Y_i &=  \hat \beta_0 \sum_i w_i (1 - Z_i) \nonumber\\
    \implies \hat \beta_0 &= \frac{\sum_i w_i (1 - Z_i) Y_i}{\sum_i w_i (1 - Z_i)}
\end{align}

which is the H\'{a}jek estimator for the control group mean.

Finally, we can rewrite 

\begin{equation}
    \hat \tau_{wDIM} = (\hat\beta_0 + \hat \tau_{wDIM}) - \hat\beta_0 = \frac{\sum_i w_i Z_i Y_i} {\sum_i w_i Z_i} -  \frac{\sum_i w_i (1 - Z_i) Y_i}{\sum_i w_i (1 - Z_i)}
\end{equation}

This shows that the bivariate weighted least squares treatment effect estimator can equivalently be written as the difference in the H\'{a}jek estimators for treated and control groups.

\subsection{Horvitz-Thompson variance estimators}

To derive the design-based variance for the H\'{a}jek-style weighted least squares estimator, we first need to derive the variance for the Horvitz-Thompson style wDIM estimator, which is a design-linear estimator.  The Horvitz-Thompson estimator is 

\[\hat \tau_{HT} = \frac{1}{n} \left( \sum_i w_i Z_i Y_i - \sum_i w_i (1 - Z_i)Y_i\right).\]

The difference from the H\'{a}jek is that we do not divide by the sum of the weights.  From standard design-based inference theory, the variance of this estimator is \citep{aronow2013conservative, fuller_2009}:

\begin{align}
    \Var_{DB}[\hat \tau_{HT}] &= \frac{1}{n^2} \left( \Var\left[\sum_i \frac{Z_i Y_i}{\pi_i}\right] + \Var\left[\sum_i \frac{(1 - Z_i) Y_i}{1 - \pi_i}\right] - 2 \Cov\left[\sum_i \frac{Z_i Y_i}{\pi_i}, \sum_i \frac{(1 - Z_i) Y_i}{1 - \pi_i}\right] \right) \nonumber\\
    & = \frac{1}{n^2} \left( \sum_i \frac{1 - \pi_i}{\pi_i} Y_i(1)^2 + \mathop{\sum\sum}_{i\neq j} (\pi_{ij} - \pi_i \pi_j)\frac{Y_i(1)}{\pi_i}\frac{Y_j(1)}{\pi_j} \right) \label{eq:y1_var} \\
    &\quad + \frac{1}{n^2} \left( \sum_i \frac{\pi_i}{1 - \pi_i} Y_i(0)^2 + \mathop{\sum\sum}_{i\neq j} (\pi_{ij} - \pi_i \pi_j)\frac{Y_i(0)}{1 - \pi_i}\frac{Y_j(0)}{1 - \pi_j} \right) \label{eq:y0_var} \\
    &\quad - \frac{2}{n^2}\left( \sum_i \sum_j \left(\pi_{i!j} - \pi_i (1 - \pi_j) \right)\frac{Y_i(1)}{\pi_i}\frac{Y_j(0)}{1 - \pi_j} \right) \label{eqn:db_var_cov_true}
\end{align}

where $\pi_i$ is the probability unit $i$ is in treatment, $\pi_{ij}$ is the probability that unit $i$ and unit $j$ are both assigned to treatment, and $\pi_{i!j} \text{ for } i \ne j \text{ and } \pi_{i!i} = 0$ is the probability that unit $i$ will be in the treatment group and unit $j$ will be in the control group.

This cannot generally be estimated due to the fundamental problem of causal inference, in which we cannot jointly observe the $Y_i(1) Y_i(0)$ term that appears in the covariance in \eqref{eqn:db_var_cov_true}.  However, focusing on the covariance term, \citet{aronow2013conservative} show that:

\begin{align}
\sum_i \sum_j \left(\pi_{i!j} - \pi_i (1 - \pi_j) \right)\frac{Y_i(1)}{\pi_i}\frac{Y_j(0)}{1 - \pi_j} &= \sum_i (\pi_{i!i} - \pi_i (1 - \pi_i)) \frac{Y_i(1)}{\pi_i}\frac{Y_i(0)}{1 - \pi_i} \nonumber\\
& \qquad + \mathop{\sum\sum}_{i\neq j} \left(\pi_{i!j} - \pi_i (1 - \pi_j) \right)\frac{Y_i(1)}{\pi_i}\frac{Y_j(0)}{1 - \pi_j} \nonumber\\
& = \sum_i (0 - \pi_i (1 - \pi_i)) \frac{Y_i(1)}{\pi_i}\frac{Y_i(0)}{1 - \pi_i} \nonumber \\
& \qquad +  \mathop{\sum\sum}_{i\neq j} \left(\pi_{i!j} - \pi_i (1 - \pi_j) \right)\frac{Y_i(1)}{\pi_i}\frac{Y_j(0)}{1 - \pi_j} \nonumber\\
& = \mathop{\sum\sum}_{i\neq j} \left(\pi_{i!j} - \pi_i (1 - \pi_j) \right)\frac{Y_i(1)}{\pi_i}\frac{Y_j(0)}{1 - \pi_j} - \sum_i Y_i(1) Y_i(0) \nonumber \\
& \ge \mathop{\sum\sum}_{i\neq j} \left(\pi_{i!j} - \pi_i (1 - \pi_j) \right)\frac{Y_i(1)}{\pi_i}\frac{Y_j(0)}{1 - \pi_j} \nonumber \\
&\qquad - \sum_i \frac{Y_i^2(1)}{2} - \sum_i \frac{Y_i^2(0)}{2} \label{eq:cov}
\end{align}

where the inequality comes from Young's Inequality.  The lines after the inequality are estimable with observed data.

We assume that each joint inclusion probability $\pi_{ij} > 0$ and $\pi_{i!j} > 0$ under a Bernoulli design.  Under a Bernoulli design $\pi_{ij} = \pi_i \pi_j$.  This means that in Equations~\eqref{eq:y1_var} and ~\eqref{eq:y0_var} the cross unit terms cancel out due to independence.  Similarly, in Equation~\eqref{eq:cov}, $\pi_{i!j} = \pi_i (1 - \pi_j)$.

\begin{align}
    \Var_{DB}[\hat \tau_{HT}] &\le  \frac{1}{n^2} \left( \sum_i \frac{1 - \pi_i}{\pi_i} Y_i(1)^2 + \sum_i \frac{\pi_i}{1 - \pi_i} Y_i(0)^2 - 2 \left( - \sum_i \frac{Y_i^2(1)}{2} - \sum_i \frac{Y_i^2(0)}{2} \right) \right) \nonumber \\
     & =\frac{1}{n^2} \left( \sum_i \frac{1}{\pi_i} Y_i(1)^2 - \sum_i  Y_i(1)^2 + \sum_i Y_i(1)^2 + \sum_i \frac{\pi_i}{1 - \pi_i} Y_i(0)^2 + \sum_i Y_i(0)^2 \right) \nonumber  \\
     & =\frac{1}{n^2} \left( \sum_i \frac{1}{\pi_i} Y_i(1)^2 + \sum_i \frac{\pi_i + 1 - \pi_i}{1 - \pi_i} Y_i(0)^2 \right) \nonumber  \\
     & =\frac{1}{n^2} \left( \sum_i \frac{1}{\pi_i} Y_i(1)^2 + \sum_i \frac{1}{1 - \pi_i} Y_i(0)^2 \right)
\end{align}

Noting these are design-linear estimators, an unbiased estimator for this conservative variance is \citep{aronow2013conservative, fuller_2009}:

\begin{align}
   \hat{\Var}_{DB}[\hat \tau_{HT}] &= \frac{1}{n^2} \sum_i \left (\frac{ Z_i Y_i^2}{\pi^2_i} + \frac{(1 - Z_i) Y_i^2}{(1 - \pi_i)^2} \right) \nonumber\\
   &= \frac{1}{n^2} \left( \sum_i w_i^2 Z_i Y_i^2 + \sum_i w_i^2 (1 - Z_i) Y_i^2 \right) \label{app:var_db_estimator_HT}
\end{align} 

where the second line uses our weights $w_i = Z_i / e(X_i) + (1 - Z_i)/(1 - e(X_i))$, so that $w_i = 1/\pi_i$ when $Z_i = 1$ and $w_i = 1/(1 - \pi_i)$ when $Z_i = 0$.

When weights are normalized to sum to one, then the $\frac{1}{n^2}$ in Equation~\eqref{app:var_db_estimator_HT} disappears.

\subsection{Design-based variance for H\'{a}jek estimator}

The variance estimator in Equation~\eqref{app:var_db_estimator_HT} is for the Horvitz-Thompson estimator, not our H\'{a}jek estimator which normalizes by the sum of the weights.  In general, it is difficult to analytically derive the design-based variance for the H\'{a}jek estimator since it is a ratio estimator.  However, it is well known that in model-assisted estimation, the asymptotic variance of the model-assisted estimator inherits the form of the Horvitz-Thompson estimator as applied to the residuals from the model.

In this case, recall our population model is $Y_i = \beta_0 + \tau_{wDIM} Z_i + \epsilon_i$, which we estimate with the wLS coefficients derived above.  First, let $\overline{Y}_1^w = \frac{\sum_i w_i Z_i Y_i} {\sum_i w_i Z_i}$  and let $\overline{Y}_0^w = \frac{\sum_i w_i (1 - Z_i) Y_i}{\sum_i w_i (1 - Z_i)}$ represent the treated and control group H\'{a}jek estimators, respectively.  This implies that, using Equation~\eqref{app:var_db_estimator_HT}, our estimator for the asymptotic\footnote{We work under standard regularity conditions for finite sample asymptotics.  See \citet{fuller_2009} for more details.} variance of the bivariate wLS regression is:

\begin{align}
    \hat \Var[\hat \tau_{wDIM}] = \frac{1}{n^2}\left( \sum_i w^2_i Z_i (Y_i - \overline{Y}_1^w)^2 + \sum_i w^2_i (1 - Z_i) (Y_i - \overline{Y}_0^w)^2\right)
\end{align}

The last thing to note is that when we normalize the weights to sum to one in each treatment arm, $\frac{1}{n}\sum_i Z_i w_i \overset{p}{\rightarrow} 1$, so $\frac{Z_i w_i}{\sum_j Z_j w_j} = \frac{Z_i}{n\pi_i}(1 + o_p(1))$ and a similar argument holds that $\frac{Z_i w_i}{\sum_j Z_j w_j} = \frac{1 - Z_i}{n(1 - \pi_i)}(1 + o_p(1))$.  If one uses these weights in the derivation above, $1/n^2$ in the variance equations cancels out, and we will be left with a variance estimator of the form in Equation~\eqref{eq:neymanwdim},

\begin{align}
    \hat \Var[\hat \tau_{wDIM}] = \sum_i w^2_i Z_i (Y_i - \overline{Y}_1^w)^2 + \sum_i w^2_i (1 - Z_i) (Y_i - \overline{Y}_0^w)^2
\end{align}

\subsection{Variance estimator for residualized regression}

Finally, noting that our residualized regression is also a model-assisted estimator, and with weights normalized to sum to one in each treatment arm, we get that

\begin{equation}
    \hat{\Var}_{DB}[\hat \tau] =  \sum_i w_i^2 Z_i \hat{\epsilon}_i^2 + \sum_i w_i^2 (1 - Z_i) \hat{\epsilon}_i^2
\end{equation}

where the $\hat \epsilon$ come from the regression described in Equation~\eqref{eq:wResModel}.

Confirming the above intuition, \citet[][p.\ 282]{abadie_sampling-based_2020} show that the design-based variance for a multiple regression estimator, in the unweighted case, is driven by the residual variation, under finite sample asymptotics in the design-based setting.

\section{Superpopulation Corrections}\label{app:superpop_corr}

In this section we derive the superpopulation corrections for the variance estimators for the Lin-style weighted least squares estimator of causal effects. In particular, we consider the variance of the fully-interacted weighted regression of $Y$ on treatment and de-meaned covariates, where covariates are de-meaned using the weighted mean of either the full sample or treated units depending on if the researcher is interested in estimating the ATE or ATT, respectively.  This proof closely follows the variance derivation for estimating the PATE using a classic Lin-estimator in a randomized controlled trial in \citet{negi_woolridge_2021}.  We build on this derivation in two key ways.  First, the original derivation considers an equiprobable Bernoulli design.  We update this to allow for unit-level differential probability of assignment, meaning our results would hold for general Bernoulli randomized trials.  Second, we adapt the derivation to consider the variance of estimators for the PATT, an estimand of more common interest in observational studies and not considered in the \citet{negi_woolridge_2021}.

\subsection{Definitions, Assumptions, and Preliminaries}

Following Section \ref{sec:notation}, we denote the potential outcomes as $\{Y(1), Y(0)\}$, we let $Z$ denote treatment, and we let $Y = Z Y(1) + (1 - Z)Y(0)$ under the common SUTVA assumption \citep{rubin1980randomization}.  We define covariates $\X = (X_1, X_2, \dots, X_k)$.  We will let $X$ be a vector that represents a row of $\X$.  While in the main manuscript we allow for transformations of $X$ in our regression, $\phi$, here we will focus on a design matrix that only accounts for $\X$ (i.e. we let $\bm{\phi} = \X$).  The results hold if $\X$ contains transformations so long as it retains full rank.  We will also not normalize the weights to sum to one here, but rather consider IPW-style weights.  Additionally, with a slight abuse of notation, we will redefine $\alpha$ and $\beta$ in Sections~\ref{app:sp_pate} and \ref{app:sp_patt} relative to the main manuscript.  These definitions only apply in this appendix.

Researchers are then typically interested in one of two target parameters:

\begin{align*}
\tau = \E[Y(1) - Y(0)] = \mu_1 - \mu_0 & & \text{(PATE)} \\
\tau_{PATT} = \E[Y(1) - Y(0) \mid Z = 1] = \mu_{1, 1} - \mu_{0, 1} & & \text{(PATT)}
\end{align*}

where, for the PATT, we denote the population means as $\mu_{*, 1}$ to indicate it is the mean potential outcome among the $Z = 1$ treated group.  We assume that our data come from a Bernoulli design, or more formally the following:

\begin{assumption}[\emph{Bernoulli Design.}]\label{assump:bern_design}
    The binary assignment indicator, $Z$, is a Bernoulli random variable.  We let $e(X) = \Pr[Z = 1 \mid X]$ be the conditional probability that a unit with covariates $X$ is assigned to treatment, and assume that $0 < e(X) < 1$.
\end{assumption}

Note that Assumption~\ref{assump:bern_design} includes the common ``strong overlap'' assumption, which bounds the propensity score, $e(X)$ away from 0 and 1.  We acknowledge that we use a slight abuse of notation in defining $e(X)$, as it is a fixed quantity and not a random variable. With this definition of the propensity score, we can also define $\rho = \E[\E[e(X) \mid X]] = \Pr[Z = 1]$.  We assume that $e(X)$ and $\rho$ are unknown, but that under random sampling we have consistent and asymptotically normal estimators for them.  Finally, we consider inverse propensity score weights in our estimator defined as $w_i = Z_i/ e(X_i) + (1 - Z_i)/(1 - e(X_i))$.

We maintain two additional key assumptions in our analysis.

\begin{assumption}[\emph{Conditional Ignorability.}]\label{assump:cond_ignorability}
    The binary assignment indicator, $Z$, is independent of $\{Y(1), Y(0)\}$ conditional on observable covariates $\X$.  That is
    
    \[ \{Y(1), Y(0)\} \indep Z \mid \X\]
    
\end{assumption}

Unlike in our design- and model-based settings, the superpopulation estimands are defined over an infinite superpopulation rather than conditional on a set of units.  To identify the target superpopulation parameters, and estimate the associated uncertainty, we additionally require a sampling process for our observed sample.  We let $n$ be the total sample size, with $n_t$ denoting the number of realized treated units and $n_c$ denoting the number of realized control units.

\begin{assumption}[\emph{Random Sampling.}]\label{assump:random_sampling}
For a nonrandom integer $n$, \newline $\{Y_i(1), Y_i(0), Z_i, X_i, e(X_i) \}_{i = 1}^n$ are independently and identically distributed draws from a superpopulation.
\end{assumption}

Finally, we require a set of regularity conditions, namely that at least the second \emph{weighted} moments of the potential outcomes and covariates ($\E[w (X - \E[wX])^2] \le c$) are finite, required for the law of large numbers and central limit theorem to our weighted least squares estimator. The fourth-order moments, $\E[w^2 \epsilon^2(z)X^2]$, must also be finite for consistency of the variance estimator. Let $\mathbf{\Omega}_X$ be the variance-covariance matrix of $\X$ and $\mathbf{\Omega}_{X, 1}$ be the variance-covariance matrix of $\X \mid Z = 1$.  We assume that these matrices are nonsingular, meaning that no covariates are perfectly collinear in the superpopulation, or among treated units for the PATT.

We begin by deriving the asymptotic variance and estimator for the superpopulation correction for the PATE followed by the derivation for the PATT.

\subsubsection{PATE Weights}

For the PATE, we define the inverse propensity weights as:

    \[ w_i = \frac{1}{Z_i e(X_i) + (1-Z_i)(1 - e(X_i))} \]

Before we begin, we note one useful fact that will recur, namely that with this definition of $w_i$:

\begin{align}
    \E[w_iZ_i \mid X_i] &= \E\left[ \frac{Z_i}{Z_i e(X_i) + (1 - Z_i)(1 - e(X_i))} \mid X_i, Z_i = 1\right]\Pr[Z_i = 1 \mid X_i] + \nonumber \\
    &\qquad \E\left[ \frac{Z_i}{Z_i e(X_i) + (1 - Z_i)(1 - e(X_i))} \mid X_i, Z_i = 0\right]\Pr[Z_i = 0 \mid X_i]  & \text{(LIE)} \nonumber\\
    &= \E[1/e(X_i) \mid X_i] \cdot e(X_i) + 0 \cdot (1 - e(X_i)) & \text{(Asm.~\eqref{assump:bern_design})} \nonumber \\
    &= 1/e(X_i) \cdot e(X_i) = 1 \text{ and by a similar logic }\E[w_i(1 - Z_i) \mid X_i] = 1\label{note_e_wZ}
\end{align}

Useful consequences of this are that:

\begin{align}
n^{-1}\sum_i w_iZ_i &\overset{p}{\to} \E[w_iZ_i] & \text{(WLLN)} \nonumber\\
&= \E\big[\E[w_iZ_i \mid X_i]\big] = 1 & \text{(LIE, Eqn.~\eqref{note_e_wZ})} \nonumber\\
\implies \sum_i w_iZ_i &= n\,(1 + o_p(1)) \label{note_sum_wZ_ate}
\end{align}

and

\begin{align}
n^{-1}\sum_i w_i = n^{-1}\sum_i w_i Z_i + n^{-1}\sum_i w_i (1 - Z_i) &\overset{p}{\to} 2 \nonumber \\
\quad \implies \quad \sum_i w_i &= 2n\,(1 + o_p(1)).\label{note_sum_w_ate}
\end{align}

\subsubsection{PATT Weights}

For the PATT, we define the inverse propensity weights as:

    \[ w_i = \frac{e(X_i)}{Z_i e(X_i) + (1-Z_i)(1 - e(X_i))} \]

Before we begin, we note one useful fact  will recur, namely that with this definition of $w_i$:

\begin{align}
    \E[w_iZ_i \mid X_i] &= \E\left[ \frac{Z_i e(X_i)}{Z_i e(X_i) + (1 - Z_i)(1 - e(X_i))} \mid X_i, Z_i = 1\right]\Pr[Z_i = 1 \mid X_i] + \nonumber \\
    &\qquad \E\left[ \frac{Z_i e(X_i)}{Z_i e(X_i) + (1 - Z_i)(1 - e(X_i))} \mid X_i, Z_i = 0\right]\Pr[Z_i = 0 \mid X_i]  & \text{(LIE)} \nonumber\\
    &= \E[1 \mid X_i] \cdot e(X_i) + 0 \cdot (1 - e(X_i)) & \text{(Asm.~\eqref{assump:bern_design})} \nonumber \\
    &= e(X_i) = \E[Z_i \mid X_i] \label{note_e_wZ_att}
\end{align}

And by similar logic

\begin{equation}
    \E[w_i(1 - Z_i) \mid X_i] = e(X_i) = \E[Z_i \mid X_i] \label{note_e_w1_Z_att}
\end{equation}

A useful consequence of this is:

\begin{align}
n^{-1}\sum_i w_i(1 - Z_i) &\overset{p}{\to} \E[w_i(1 - Z_i)] & \text{(WLLN)} \nonumber\\
&= \E\big[\E[w_i(1 - Z_i) \mid X_i]\big] = \E\big[\E[Z_i \mid X_i]\big] = \rho & \text{(LIE, Eqn.~\eqref{note_e_w1_Z_att})} \nonumber\\
\implies \sum_i w_i(1 - Z_i) &= n\rho\,(1 + o_p(1)) \label{note_sum_wZ_att}
\end{align}

and

\begin{align}
n^{-1}\sum_i w_i = n^{-1}\sum_i w_i Z_i + n^{-1}\sum_i w_i (1 - Z_i) &\overset{p}{\to} 2\rho \nonumber \\
\quad \implies \quad \sum_i w_i &= 2n\rho\,(1 + o_p(1)).\label{note_sum_w_att}
\end{align}

\subsection{Weighted regression HC0 standard errors}
Before we derive the variance for the PATE and PATT estimators, we begin by deriving the HC0 standard errors for our proposed interacted, demeaned weighted least squares estimator.  While \citet{lin_agnostic_2013} shows that the HC0 variance for the coefficient on $Z$ in the interactive, centered regression is $\frac{\hat \sigma_1^2}{n_t} + \frac{\hat \sigma_0^2}{n_c}$, that result applies (i) without weights, and (ii) under randomization.  Intuitively, we might expect that with weights chosen to effectively mimic randomization, the variance in the so-weighted regression would have this form as well. Here we show this and the assumptions on which it relies. The result allows us to complete the connection between the variances derived below and what regression produces. 

For notational simplicity, assume we have sorted observations such that the first $1:n_t$ rows correspond to treated units (i.e. those for which $Z_i = 1$) and the remaining $(n_t + 1):n$ rows correspond to control units (i.e. those for which $Z_i = 0$).   We let $\W$ be the matrix of weights, with $\W = \operatorname{diag}(w)$ where $w$ is a vector in which the first $1:n_t$ observations correspond to weights for treated units and the remaining $(n_t + 1):n$ observations correspond to the control weights.  Finally, let $\hat \epsilon$ correspond to the estimated vector of residuals, where the first $1:n_t$ correspond to the residuals for treated units ($\hat \epsilon_i(1)$) and the remaining $(n_t + 1):n$ correspond to the residuals for the control units ($\hat \epsilon_i(0)$).

Recall that we are regressing $Y$ on $Z$, $X$, and $ZX$ (without weights). Throughout, the covariates $X$ entering the regression are exactly those balanced upon, as in Equation~\eqref{eq:wResModel}. Including additional covariates not balanced upon is possible but effectively places the result in the approximate-balance category (Appendix~\ref{sec:what_hc0_estimates_approx}) even when the weights exactly balance on the $X$ they employ. We want the HC0 variance on the coefficient for $Z$, which will be the second component of  

\begin{align} 
\hat \Var_{HC}[\hat \theta_{unwtd}] &= (\M^{\top}\M)^{-1}\M^{\top}  \hat{\BSigma}_{HC} \M (\M^{\top}\M)^{-1} \label{eq:lshc0}
\end{align}

\noindent where $\hat{\BSigma}_{HC} = \hat{\BSigma}_{HC} = \operatorname{diag}(\hat\epsilon_1^2, \ldots, \hat\epsilon_n^2)$, where $\hat \epsilon$ is the vector of residuals from the unweighted regression, and $\M=[1 \ Z \ X \ ZX]$. Presuming $X$ to be mean zero,  $\M^{\top}\M$ will be

\[
\M^{\top}\M =
\begin{pmatrix}
n & \sum_i Z_i & 0 & \sum_i(Z_iX_i) \\
\sum_i Z_i & \sum_i Z_i^2 & \sum_i (Z_iX_i) & \sum_i (Z_i^2 X_i) \\
0 & \sum_i (Z_iX_i) & \sum_i X_i^2 & \sum_i(Z_i X_i^2) \\ 
\sum_i(Z_iX_i) & \sum_i(Z_i^2 X_i) & \sum_i(Z_i X_i^2) & \sum_i(Z_i^2 X_i^2)
\end{pmatrix}
\]

\noindent Scaling by $1/n$, noting $Z_i^2=Z_i$ in the binary case, and applying the WLLN,

\begin{equation}
\M^{\top}\M/n \overset{p}{\to}
\begin{pmatrix}
1 & \rho & 0 & \Cov[Z,X] \\
\rho & \rho & \Cov[Z,X] & \Cov[Z,X] \\
0 & \Cov[Z,X] & \Var[X] & \E[ZX^2] \\
\Cov[Z,X] & \Cov[Z,X] & \E[ZX^2] & \E[ZX^2]
\end{pmatrix}
\end{equation}

This recovers the results from \citet{lin_agnostic_2013}, but we employ weights in our regression. For weights $w$ with $0 < \E[w] < \infty$, define the weighted mean $\mu^w_A = \E[wA]/\E[w]$, the weighted covariance $\Cov_w[A,B] = \E\big[w\,(A - \mu^w_A)(B - \mu^w_B)\big]$, and $\Var_w[A] = \Cov_w[A,A]$. Their sample analogs, e.g.\ $n^{-1}\sum_i w_i (A_i - \overline{A}^w)(B_i - \overline{B}^w)$ with $\overline{A}^w = \sum_i w_iA_i / \sum_i w_i$, converge in probability to these quantities by the WLLN. Then, with $X$ centered at its weighted mean so that $\E[wX] = 0$,

\begin{equation}
\M^{\top}\W\M/n \overset{p}{\to}
\begin{pmatrix}
\E[w] & \E[wZ] & 0 & \Cov_w[Z,X] \\
\E[wZ] & \E[wZ] & \Cov_w[Z,X] & \Cov_w[Z,X] \\
0 & \Cov_w[Z,X] & \Var_w[X] & \E[wZX^2] \\
\Cov_w[Z,X] & \Cov_w[Z,X] & \E[wZX^2] & \E[wZX^2]
\end{pmatrix}\label{eq:mvmovern}
\end{equation}

\subsubsection{HC0 standard errors under exact balance}\label{sec:what_hc0_estimates_pate}

Recall that in our proposed estimator we mean center $X$ with weights, so $\sum_i w_i X_i = 0$. For weights that achieve exact finite-sample balance, as in exact mean balancing weights or subclassification weights with categorical, low-dimensional $X$, the weighted mean of $X$ in each arm equals the common target $\phibstar$, so after centering

\begin{equation}
\sum_i w_i Z_i X_i = 0 \qquad \text{and} \qquad \sum_i w_i (1 - Z_i) X_i = 0.
\label{eq:exact_balance_sample}
\end{equation}

Let $\overline{w} = n^{-1}\sum_i w_i$ and $\overline{wz} = n^{-1}\sum_i w_i Z_i$. The finite sample evaluation of $\frac{\M^{\top}\W\M}{n}$ is thus

\begin{equation}
\frac{\M^{\top}\W\M}{n} =
\begin{pmatrix}
\overline{w} & \overline{wz} & 0 & 0 \\
\overline{wz} & \overline{wz}  & 0  & 0 \\
0 & 0 & \widehat{\Var}_w[X] & n^{-1}\sum_i w_i Z_i X_i^2 \\
0 & 0 & n^{-1}\sum_i w_i Z_i X_i^2 &  n^{-1}\sum_i w_i Z_i X_i^2
\end{pmatrix} \label{var_w_mWm_exact}
\end{equation}

This matrix is exact in any given sample. It is also block diagonal, so its inverse is  block diagonal with each block being the inverse of the original block.  In the expression

\begin{align}
\hat \Var_{HC}[\hat \theta] &= (\M^{\top}\W \M)^{-1}\M^{\top}\W  \hat{\BSigma}_{HC}\W \M (\M^{\top}\W\M)^{-1} \nonumber \\
&= n^{-1}\left(\frac{\M^{\top}\W\M}{n}\right)^{-1}\M^{\top}\W  \hat{\BSigma}_{HC} \W \M \left(\frac{\M^{\top}\W \M}{n}\right)^{-1} n^{-1}  \label{eq:hc0wls}
\end{align}

\noindent we are interested in the variance on the $Z$ coefficient, element $([2,2])$. Due to the block structure this relies only on the local block inverse ([1:2],[1:2]),

\begin{equation}
\begin{pmatrix}
\overline{w} & \overline{wz} \\
\overline{wz} & \overline{wz}
\end{pmatrix}^{-1}
= \frac{1}{\overline{wz}(\overline{w} - \overline{wz})}
\begin{pmatrix}
\overline{wz} & -\overline{wz} \\
-\overline{wz} & \overline{w}
\end{pmatrix}.
\end{equation}

Since we want only the $[2,2]$ element, we need only the second row of this inverse, applied to $\begin{pmatrix} \mathbbm{1} & Z \end{pmatrix}$ on each side. The variance estimator in Expression~\eqref{eq:hc0wls} simplifies to 

\begin{equation}
\hat \Var_{HC}[\hat \tau] = n^{-2}\, \ell^\top \W \hat{\BSigma}_{HC} \W \ell = n^{-2} \sum_{i=1}^n \ell_i^2 w_i^2 \hat\epsilon_i^2.
\end{equation}

\noindent where $\ell$ is a length $n$ vector with entries $\ell_i = (Z_i\overline{w} - \overline{wz}) / \overline{wz}(\overline{w} - \overline{wz})$. Among treated units this is simply $1/ \overline{wz}$, and among untreated units it is $-1/(\overline{w}-\overline{wz})$. Writing out the contribution from treated units, 

\begin{equation}
\frac{\sum_{i:Z_i=1} w_i^2 \hat\epsilon_i^2}{n^2\, \overline{wz}^{\,2}}
= \frac{\sum_{i=1}^n Z_i w_i^2 \hat\epsilon_i^2}{\left(\sum_{i=1}^n w_i Z_i \right)^2}
\end{equation}

\noindent Likewise over control units the contribution is

\begin{equation}
\frac{\sum_{i:Z_i=0} w_i^2 \hat\epsilon_i^2}{n^2 (\overline{w} - \overline{wz})^2}
= \frac{\sum_{i=1}^n (1-Z_i) w_i^2 \hat\epsilon_i^2}{\left(\sum_{i=1}^n (1-Z_i) w_i\right)^2}
\end{equation}

\noindent Putting these together over all units, we have

\begin{equation}
\hat \Var_{HC}[\hat \tau] = \frac{\sum_{i=1}^{n} Z_i w_i^2 \hat{\epsilon}_i^2}{\left(\sum_{i=1}^n Z_i w_i\right)^2}  + \frac{\sum_{i=1}^{n}(1-Z_i) w_i^2 \hat{\epsilon}_i^2 }{\left(\sum_{i=1}^n (1-Z_i) w_i\right)^2}
\label{eq:hc0_armwise}
\end{equation}

That is, under exact balance the HC0 standard error from the residualized regression is the standard error of $\wdim$ computed with the residuals $\hat\epsilon_i$ in place of the outcomes.
Equation~\eqref{eq:hc0_armwise} is an identity: it requires no limiting argument, no particular origin for the weights, and no condition on their scale, since rescaling the weights within either arm leaves it unchanged. Most simply, when weights are scaled to sum to one by arm, as used in the paper, $\sum_i Z_i w_i = \sum_i (1-Z_i) w_i = 1$ and Equation~\eqref{eq:hc0_armwise} is

\begin{equation}
\hat \Var_{HC}[\hat \tau] = \sum_{i=1}^{n} w_i^2 Z_i \hat{\epsilon}_i^2  +  \sum_{i=1}^{n} w_i^2 (1-Z_i)\hat{\epsilon}_i^2 = \sum_{i=1}^n w_i^2 \hat{\epsilon}_i^2,
\label{reg_var_consistent_ate}
\end{equation}

\noindent which is Equation~\eqref{eqn:db_var_res} of the main text, i.e.\ Equation~\eqref{eq:neymanwdim} with the residuals in place of the outcomes. For the ATT, the controls are instead balanced to the treated mean, and the treated units have uniform weights, so the first term of Equation~\eqref{eq:hc0_armwise} reduces to $\sum_i Z_i \hat\epsilon_i^2/n_t^2$.

\subsubsection{HC0 standard errors under approximate balance}\label{sec:what_hc0_estimates_approx}

Inverse propensity weights, matching weights, and balancing weights with tolerances $\xi_k > 0$ do not satisfy Equation~\eqref{eq:exact_balance_sample} exactly. For these weights $\M^{\top}\W\M/n$ is not precisely block diagonal in the sample, and Equation~\eqref{eq:hc0_armwise} is not an identity. It is, however, asymptotically equivalent to the HC0 variance. Recall that our weights have the ``balancing property'' where $\widehat \Cov_w [Z,X] \xrightarrow{p} 0$, as with IPW, for which $\Cov_w[Z,X] = 0$ in the population.

Write $\hat{\mathbf{Q}} = \M^{\top}\W\M/n$, and let $\hat{\mathbf{Q}}_{\mathrm{bd}}$ denote $\hat{\mathbf{Q}}$ with its two off-diagonal blocks set to zero. The entries of those blocks are $n^{-1}\sum_i w_i X_i = 0$ and $n^{-1}\sum_i w_i Z_i X_i = \widehat \Cov_w[Z,X]$, so by the balancing property $\hat{\mathbf{Q}} - \hat{\mathbf{Q}}_{\mathrm{bd}}$ converges in probability to zero, and both matrices converge in probability to the matrix in Equation~\eqref{eq:mvmovern} with $\Cov_w[Z,X] = 0$, which is invertible under our regularity conditions. Under the same conditions $n^{-1}\M^{\top}\W\hat{\BSigma}_{HC}\W\M$ converges in probability to a finite matrix. Letting $e_2$ select the coefficient on $Z$, the continuous mapping theorem then implies that

\begin{align*}
\operatorname{plim} (n\, \hat \Var_{HC}[\hat \tau]) &=  \operatorname{plim} \;e_2^\top \hat{\mathbf{Q}}^{-1} \left(n^{-1}\M^{\top}\W\hat{\BSigma}_{HC}\W\M\right) \hat{\mathbf{Q}}^{-1} e_2 \\
&= \operatorname{plim} \;  e_2^\top \hat{\mathbf{Q}}_{\mathrm{bd}}^{-1} \left(n^{-1}\M^{\top}\W\hat{\BSigma}_{HC}\W\M\right) \hat{\mathbf{Q}}_{\mathrm{bd}}^{-1} e_2
\end{align*}

The second quantity is computed from a block diagonal matrix whose top-left block is $\begin{pmatrix} \overline{w} & \overline{wz} \\ \overline{wz} & \overline{wz} \end{pmatrix}$, exactly as in Section~\ref{sec:what_hc0_estimates_pate}, and is therefore $n$ times the right-hand side of Equation~\eqref{eq:hc0_armwise}. Hence for approximately balancing weights the two sides of Equation~\eqref{eq:hc0_armwise} are no longer equal, but their ratio converges in probability to one:

\begin{equation}
\hat \Var_{HC}[\hat \tau] \Big/ \left[\frac{\sum_{i=1}^{n} Z_i w_i^2 \hat{\epsilon}_i^2}{\left(\sum_{i=1}^n Z_i w_i\right)^2}  + \frac{\sum_{i=1}^{n}(1-Z_i) w_i^2 \hat{\epsilon}_i^2 }{\left(\sum_{i=1}^n (1-Z_i) w_i\right)^2}\right] \xrightarrow{p} 1.
\label{eq:hc0_armwise_approx}
\end{equation}

\noindent The finite-sample discrepancy is driven entirely by the residual imbalance $\widehat \Cov_w[Z,X]$, the same quantity that governs the gap between $\hat\tau$ and $\wdim$ in Equation~\eqref{eq:decomp} of the main text.

Unnormalized IPW weights add one further step, because $\sum_i w_i$ and $\sum_i w_i Z_i$ are then random rather than fixed by construction. One option is simply to normalize the weights to sum to one within each arm, after which the denominators in Equation~\eqref{eq:hc0_armwise} equal one and Equation~\eqref{eq:hc0_armwise_approx} reads $\hat \Var_{HC}[\hat \tau] / \sum_i w_i^2 \hat\epsilon_i^2 \xrightarrow{p} 1$. Alternatively, with the raw weights for the ATE, $\sum_i w_i Z_i / n$ and $\sum_i w_i (1 - Z_i)/n$ each converge in probability to one (Equations~\eqref{note_sum_wZ_ate} and~\eqref{note_sum_w_ate}), so replacing both denominators by $n^2$ changes the bracketed expression only by a factor converging to one, and

\begin{equation}
\hat \Var_{HC}[\hat \tau] \Big/ \left[\sum_{i=1}^n \frac{w_i^2 \hat{\epsilon}_i^2}{n^2}\right] \xrightarrow{p} 1.
\label{reg_var_consistent_ate_ipw}
\end{equation}

\noindent With raw IPW weights for the ATT, the treated weights equal one, so $\sum_i Z_i w_i = n_t$ exactly, while $\sum_i w_i (1 - Z_i)/n_t$ converges in probability to one (Equation~\eqref{note_sum_wZ_att}), giving

\begin{equation}
\hat \Var_{HC}[\hat \tau_{PATT}] \Big/ \left[\sum_{i=1}^{n} \frac{Z_i \hat{\epsilon}_i^2}{n_t^2}  + \sum_{i=1}^{n} \frac{w_i^2 (1-Z_i)\hat{\epsilon}_i^2}{n_t^2}\right] \xrightarrow{p} 1.
\label{reg_var_consistent_att}
\end{equation}

\subsection{Superpopulation correction for PATE}\label{app:sp_pate}

We begin by defining linear projections for the potential outcomes, following \citet{negi_woolridge_2021} and \citet[Sec. 2.3]{wooldridge2010econometric}, again assuming our regularity conditions above, although we only require at least the unweighted second moments are finite for the existence of the linear projections.

\begin{align*}
        m_1(X) &= \alpha_{1} + X \beta_{1}  \\
        m_0(X) &= \alpha_{0} + X \beta_{0}
    \end{align*}

And therefore we can write the PATE as

\[ \tau = \mu_{1} - \mu_{0} = (\alpha_{1} - \alpha_{0}) + \mu_{X}(\beta_{1} - \beta_{0})\]

Where $\mu_{X}$ is the vector of covariate means in the population.  It is also helpful to rewrite our linear projections in terms of demeaned covariates, such as we use in our Lin-style estimator.  First, note that

\begin{align*}
        Y(0) = \mu_0 + \nu(0)  \\
        Y(1) = \mu_1 + \nu(1) 
\end{align*}

where $\nu(0)$ and $\nu(1)$ have mean zero.  Define demeaned covariates $\dotXt = X - \mu_{X}$ to be the $X$, demeaned using true population means of $X$.   We can linearly project the error terms onto our demeaned covariates.

\begin{align*}
    \nu(0) = \dotXt \beta_0 + \epsilon(0) \\
    \nu(1) = \dotXt \beta_1 + \epsilon(1)
\end{align*}

where the intercepts are zero due to the de-meaning.  We can then rewrite our potential outcomes as:

\begin{align}
    Y(0) = \mu_0 + \dotXt \beta_0 + \epsilon(0) \label{eq:lp_y0}\\
    Y(1) = \mu_1 + \dotXt \beta_1 + \epsilon(1) \label{eq:lp_y1}
\end{align}

This formulation has the following useful properties due to the definition of linear projections.

\begin{align}
    \E[\epsilon(0)] = \E[\epsilon(1)] = 0 \label{prop:LP_mean0} \\
    \E[\dotXt^\top \epsilon(0)] = \E[\dotXt^\top \epsilon(1)] = 0 \label{prop:LP_orthog}
\end{align}

Finally, let $\Var_{SP}[\epsilon(0)] = \sigma_0^2$ and $\Var_{SP}[\epsilon(1)] = \sigma_1^2$ be the unconditional variances of the linear projection errors, which are mean zero by construction.

\subsubsection{Variance of wLS for PATE}

We begin by estimating $\mu_{1}$ using the weighted regression of $Y$ on 1, $X$, for the $Z = 1$ sample:

\[\hat \mu_{1} = \hat\alpha_{1} + \overline{X} \hat\beta_{1}\]

By algebraic equivalence, we can also estimate $\hat\mu_{1}$ from the weighted regression on demeaned covariates. Let $\ddotXt = X - \overline{X}$, then more specifically, we can also estimate $\hat\mu_{1}$ from the weighted regression of $Y$ on 1, $\ddotXt$ for $Z = 1$.

Define $\ddotRt = ( 1, \ddotXt{}_i )$, then, by regression mechanics, our de-meaned regression estimates

\vspace{-1em}
\begin{align}
    \hat\theta_{1} = \genfrac{(}{)}{0pt}{}{\hat\mu_{1}}{ \hat \bbeta_{1}} &= \left( \sum_{i = 1}^{n} w_i Z_i \ddotRt^\top \ddotRt  \right)^{-1} \left( \sum_{i = 1}^{n} w_i  Z_i \ddotRt^\top Y_i  \right) \nonumber \\
    &= \left( n^{-1}\sum_{i = 1}^{n} w_i Z_i \ddotRt^\top \ddotRt  \right)^{-1} \left( n^{-1} \sum_{i = 1}^{n} w_i Z_i \ddotRt^\top Y_i(1)  \right) \label{eq:lp_reg_1}
\end{align} 

Focusing on $Y(1)$, we can rewrite it as:

\begin{align*}
    Y_i(1) &= \mu_{1} + \dotXt{}_i \bbeta_{1} + \epsilon_i(1) & \text{from \eqref{eq:lp_y1}} \\
    &= \mu_{1} + \ddotXt{}_i\bbeta_{1} - \ddotXt{}_i\bbeta_{1} + \dotXt{}_i \bbeta_{1} + \epsilon_i(1) & \text{add 0}\\
    &= \ddotRt \theta_{1} + (\overline{X} - \mu_{X})\bbeta_{1} + \epsilon_i(1) & \text{simplifying}
\end{align*}

Plugging this into the numerator of equation \eqref{eq:lp_reg_1}, we get:

\begin{align*} n^{-1} \sum_{i = 1}^{n} w_i Z_i \ddotRt^\top Y_i(1) = \left( n^{-1} \sum_{i = 1}^{n} w_i Z_i \ddotRt^\top \ddotRt \right) \theta_{1}  + \left(  n^{-1} \sum_{i = 1}^{n} w_i Z_i \ddotRt^\top\right)(\overline{X} - \mu_{X})\beta_{1} \\+ n^{-1} \sum_{i = 1}^{n} w_i Z_i \ddotRt^\top \epsilon_i(1)
\end{align*}

and therefore we can simplify equation \eqref{eq:lp_reg_1} as:

\[ \hat\theta_{1} = \theta_{1} + \left( n^{-1}\sum_{i = 1}^{n} w_i Z_i \ddotRt^\top \ddotRt  \right)^{-1} \left[ \left(  n^{-1} \sum_{i = 1}^{n} w_i Z_i \ddotRt^\top\right)(\overline{X} - \mu_{X})\beta_{1} + n^{-1} \sum_{i = 1}^{n} w_i Z_i \ddotRt^\top \epsilon_i(1) \right]\]

Which gives us:

\begin{align} \sqrt{n}(\hat\theta_{1} - \theta_{1}) = \left( n^{-1}\sum_{i = 1}^{n} w_i Z_i \ddotRt^\top \ddotRt  \right)^{-1} \left[ \left(  n^{-1} \sum_{i = 1}^{n} w_i Z_i \ddotRt^\top\right)\sqrt{n}(\overline{X} - \mu_{X})\beta_{1} + \right.\nonumber \\ \left.n^{-1/2} \sum_{i = 1}^{n} w_iZ_i \ddotRt^\top \epsilon_i(1) \right] \label{eq:full_y1}\end{align}

Now note that $\overline{X} \overset p \to \mu_{X}$ by Assumption~\ref{assump:random_sampling} and the law of large numbers, combined with Slutsky's theorem, we get:

\[ n^{-1} \sum_{i = 1}^n w_i Z_i \ddotRt^\top \ddotRt = n^{-1} \sum_{i = 1}^n w_i Z_i \dotRt^\top \dotRt + o_p(1)\]

where

$\dotRt = ( 1, \dotXt{}_i ) = ( 1, X_i - \mu_{X} )$. 

And, further,

\begin{align*}
    n^{-1} \sum_{i = 1}^n w_i Z_i \dotRt^\top \dotRt \overset p \to&\ \E[w_i Z_i \dotRt^\top \dotRt] & \text{(WLLN)}\\
    &= \E[\E[w_i Z_i \dotRt^\top \dotRt \mid X_i]] & \text{(LIE)} \\
    &= \E[\E[w_i Z_i \mid X_i] \E[\dotRt^\top \dotRt \mid X_i]] & \text{(Asm.~\eqref{assump:cond_ignorability})} \\
    &= \E[\E[\dotRt^\top \dotRt \mid X_i]] & \text{(Eqn~\eqref{note_e_wZ})}\\
    &= \E[\dotRt^\top \dotRt] & \text{(LIE)}  
\end{align*} 

Letting

\[ \bm{A} \equiv \begin{pmatrix}
    1 & 0 \\
    0 & \E[\dotXt{}_i^\top \dotXt{}_i]
\end{pmatrix}
= \begin{pmatrix}
    1 & 0 \\
    0 & \bm{\Omega}_{X}
\end{pmatrix} \]

which comes from rewriting the covariance of $\E[\dotRt^\top \dotRt]$ and using the fact that the first column is 1s.

Noting that the terms $\sqrt{n}(\overline{X} - \mu_{X})\beta_{1}$ and $n^{-1/2}\sum_{i = 1}^n w_i Z_i \ddotRt^\top \epsilon_i(1)$ are $O_p(1)$ under our regularity conditions,

\[ \sqrt{n}(\hat\theta_{1} - \theta_{1}) = \bm{A}^{-1}\left[ \left(  n^{-1} \sum_{i = 1}^{n} w_i Z_i \ddotRt^\top\right)\sqrt{n}(\overline{X} - \mu_{X})\beta_{1} + n^{-1/2} \sum_{i = 1}^{n} w_i Z_i \ddotRt^\top \epsilon_i(1) \right] + o_p(1)\] 

Now, considering the first element of $n^{-1} \sum_{i = 1}^{n} w_i Z_i \ddotRt^\top$:

\[n^{-1} \sum_{i = 1}^{n} w_i Z_i \ddotRt^\top = n^{-1} \sum_{i = 1}^n w_i Z_i \begin{pmatrix} 1 \\ \ddotXt{}_i\end{pmatrix}\]

where the first element is then 

\begin{equation}  n^{-1} \sum_{i = 1}^n w_i Z_i \overset p \to \E[w_iZ_i] = \E[\E[w_iZ_i \mid X_i]] = \E[1/e(X_i) \E[Z_i \mid X_i]] = \E[e(X_i)/e(X_i)] = 1 \label{eq:norm_const_y1}\end{equation}

Also, 

\[ n^{-1/2} \sum_{i = 1}^{n} w_i Z_i \ddotRt^\top \epsilon_i(1) = n^{-1/2} \sum_{i = 1}^{n} w_i Z_i \begin{pmatrix} 1 \\ \ddotXt{}_i\end{pmatrix} \epsilon_i(1)  \]

where the first element is then:

\[ n^{-1/2} \sum_{i = 1}^n w_i Z_i \epsilon_i(1)\]

And since $\bm{A}$ is block diagonal, the first element of $\sqrt{n}(\hat\theta_{1} - \theta_{1})$, $\sqrt{n}(\hat\mu_{1} - \mu_{1})$ satisfies

\begin{align}
\sqrt{n}(\hat\mu_{1} - \mu_{1}) &= \sqrt{n}(\overline{X} - \mu_{X})\beta_{1} + n^{-1/2}\sum_{i = 1}^n w_i Z_i \epsilon_i(1) + o_p(1) \nonumber \\
&= n^{-1/2}\sum_{i = 1}^n \left[(X_i - \mu_{X})\beta_{1} + w_i Z_i \epsilon_i(1) \right] + o_p(1) \label{eq:final_y1}
\end{align}

An analogous argument comes for estimating the $\hat\mu_0$ from the regression of $Y_i$ on 1, $\ddotXt$ among the control units, yielding

\begin{align}
\sqrt{n}(\hat\mu_{0} - \mu_{0}) &=  \sqrt{n}(\overline{X} - \mu_{X})\beta_{0} + n^{-1/2}\sum_{i = 1}^n w_i(1 - Z_i)\epsilon_i(0) + o_p(1) \nonumber \\
& =n^{-1/2}\sum_{i = 1}^n \left[(X_i - \mu_{X})\beta_{0} + w_i(1 - Z_i)\epsilon_i(0)\right] + o_p(1)  \label{eq:final_y0}
\end{align}

Combining equation \eqref{eq:final_y1} and \eqref{eq:final_y0}, we thus get:

\begin{align*}
\sqrt{n}(\hat\tau - \tau) = n^{-1/2} \sum_{i = 1}^n \left[ \dotXt{}_i(\beta_{1} - \beta_{0}) + w_i Z_i \epsilon_i(1) - w_i(1 - Z_i)\epsilon_i(0) \right] + o_p(1)
\end{align*}

By CLT and asymptotic equivalence lemma, we get:

\begin{align*}
\sqrt{n}\,(\hat\tau - \tau) &\overset{d}{\to} \mathcal{N}(0, \omega^2) \\
\hat\tau &\overset{a}{\sim} \mathcal{N}\left(\tau, \frac{\omega^2}{n}\right)
\end{align*}

where $\omega^2 = \Var_{SP}[\dotXt{}_i(\beta_{1} - \beta_{0}) + w_i Z_i \epsilon_i(1) - w_i(1 - Z_i)\epsilon_i(0)]$.

\vspace{1em}
To simplify the expression for $\omega^2$, we begin by noting that the terms are pairwise uncorrelated.  First note that, in the expansion, the last two terms interacted drop out because $Z_i (1 - Z_i) = 0$.  We next consider:

\vspace{-1em}
\begin{align*}
\E[(\beta_{1} -\beta_{0})^\top \dotXt{}_i^\top w_i Z_i \epsilon_i(1)] 
&= (\beta_{1} -\beta_{0})^\top  \E[\E[\dotXt{}_i^\top w_i Z_i \epsilon_i(1) \mid X_i]] & \text{(LIE)}\\
&= (\beta_{1} -\beta_{0})^\top  \E[\dotXt{}_i^\top \E[ w_i Z_i \epsilon_i(1) \mid X_i]] \\
&= (\beta_{1} -\beta_{0})^\top  \E[\dotXt{}_i^\top \E[ w_i Z_i \mid X_i ] \E[\epsilon_i(1) \mid X_i]] & \text{(Asm.~\eqref{assump:cond_ignorability})} \\
&= (\beta_{1} -\beta_{0})^\top  \E[\dotXt{}_i^\top \E[\epsilon_i(1) \mid X_i]] & \text{(Eqn.~\eqref{note_e_wZ})}\\
&= (\beta_{1} -\beta_{0})^\top  \E[ \E[\dotXt{}_i^\top \epsilon_i(1) \mid X_i]] \\
&= (\beta_{1} -\beta_{0})^\top  \E[\dotXt{}_i^\top \epsilon_i(1)] & \text{(LIE)}\\
&=0 & \text{(Eqn~\eqref{prop:LP_orthog})}
\end{align*}

A similar argument holds for $\E[(\beta_{1} -\beta_{0})^\top \dotXt{}_i^\top w_i(1-Z_i)\epsilon_i(0)] = 0$.

Finally, note that:

\vspace{-1em}
\begin{align*}
\Var_{SP}[w_i Z_i \epsilon_i(1)]
&= \E\big[(w_i Z_i \epsilon_i(1))^2\big] - \big(\E[w_i Z_i \epsilon_i(1)]\big)^2 \\
&= \E\big[(w_i Z_i \epsilon_i(1))^2\big] - \big(\E\big[\E[w_i Z_i \epsilon_i(1) \mid X_i]\big]\big)^2 & \text{(LIE)}\\
&= \E\big[(w_i Z_i \epsilon_i(1))^2\big] - \big(\E\big[\E[w_i Z_i \mid X_i]\, \E[\epsilon_i(1) \mid X_i]\big]\big)^2 & \text{(Asm.~\ref{assump:cond_ignorability})}\\
&= \E\big[(w_i Z_i \epsilon_i(1))^2\big] - \big(\E\big[\E[\epsilon_i(1) \mid X_i]\big]\big)^2 & \text{(Eqn.~\eqref{note_e_wZ})}\\
&= \E\big[(w_i Z_i \epsilon_i(1))^2\big] - \big(\E[\epsilon_i(1)]\big)^2 & \text{(LIE)}\\
&= \E\big[(w_i Z_i \epsilon_i(1))^2\big] - 0 & \text{(Eqn.~\eqref{prop:LP_mean0})}\\
&= \E[w_i^2 Z_i \epsilon_i(1)^2]
\end{align*}

where a similar argument holds for $\Var_{SP}[w_i (1 - Z_i) \epsilon_i(0)] = \E[w_i^2 (1 - Z_i) \epsilon_i(0)^2]$. This leaves us, then, with the following expression for the variance:

\begin{align}
\frac{\omega^2}{n} &= \frac{(\beta_{1} -\beta_{0})^\top \bm{\Omega}_{X} (\beta_{1} -\beta_{0})}{n} + \frac{\E[w_i^2 Z_i \epsilon_i^2(1)]}{n} + \frac{\E[w_i^2 (1 - Z_i) \epsilon_i^2(0)]}n \nonumber\\
&=\frac{(\beta_{1} -\beta_{0})^\top \bm{\Omega}_{X} (\beta_{1} -\beta_{0})}{n} + \frac{\rho\ \E[w_i^2 \epsilon_i^2(1) \mid Z_i = 1]}{n} + \frac{(1 - \rho)\ \E[w_i^2 (1 - Z_i) \epsilon_i^2(0) \mid Z_i = 0]}{n} \label{var_pate_simple}
\end{align}

\subsubsection{Variance Estimators for PATE}
We now turn to estimating the variance.  In addition to our regularity conditions above, we assume the researcher has a consistent estimator for $e(X_i)$.  While this can be estimated nonparametrically with low dimensional $X$, in general it requires imposing additional functional form assumptions for most common estimation strategies in finite samples, such as logit-based IPW or entropy balancing \citep[e.g.][]{zubizarreta2015, benmichael2021balancingactcausalinference, hartman2024kpop}.  Using Equation~\eqref{var_pate_simple}, a plug-in estimator for the standard error of $\hat\tau$ for the PATE is:

\begin{align} \hat se(\hat \tau)
&= \sqrt{\frac{(\hat\beta_{1} -\hat\beta_{0})^\top \bm{\hat\Omega}_{X} (\hat\beta_{1} -\hat\beta_{0})}{n} + \frac{\hat\rho\ \hat \E[w_i^2 \epsilon_i^2(1) \mid Z_i = 1]}{n} + \frac{(1 - \hat \rho)\ \hat \E[ w_i^2 (1 - Z_i) \epsilon_i^2(0) \mid Z_i = 0]}{n} } \nonumber \\
&= \sqrt{\frac{(\hat\beta_{1} -\hat\beta_{0})^\top \bm{\hat\Omega}_{X} (\hat\beta_{1} -\hat\beta_{0})}{n} + \frac{n_t/n\ \hat \E[ w_i^2  \epsilon_i^2 \mid Z_i = 1]}{n} + \frac{n_c/n\ \hat \E[ w_i^2 (1 -  Z_i)  \epsilon_i^2 \mid Z_i = 0]}{n} } \nonumber \\
&= \sqrt{\frac{(\hat\beta_{1} -\hat\beta_{0})^\top \bm{\hat\Omega}_{X} (\hat\beta_{1} -\hat\beta_{0})}{n} + \frac{n_t/n\ \sum{ w_i^2 Z_i \hat \epsilon_i^2 }/n_t}{n} + \frac{n_c/n\ \sum{ w_i^2 (1 - Z_i) \hat \epsilon_i^2 }/n_c}{n} } \nonumber \\
&= \sqrt{\underbrace{\frac{(\hat\beta_{1} -\hat\beta_{0})^\top \bm{\hat\Omega}_{X} (\hat\beta_{1} -\hat\beta_{0})}{n}}_{\text{correction}} + \frac{ \sum{w_i^2 Z_i} \hat \epsilon_i^2 }{n^2} + \frac{ \sum{w_i^2 (1 - Z_i) \hat \epsilon_i^2 }}{n^2} 
}\label{eq:plugin_var_ate}
\end{align}

We show in Sections~\ref{sec:what_hc0_estimates_pate} and~\ref{sec:what_hc0_estimates_approx} that the HC0 standard error from our proposed residualized regression equals the last two terms in Equation~\eqref{eq:plugin_var_ate} exactly under exact balance, and is consistent for it under approximate balance. Therefore, the only additional component a researcher must estimate is the ``correction".  In our residualized regression in Equation~\eqref{eq:wResModel}, $(\hat\beta_{1} -\hat\beta_{0})$ is estimated by $\hat \gamma$, i.e. the coefficients on the interaction terms.  $\bm{\hat\Omega}_{X}$ is estimated as $\hat S^{2}_{\phi}$, or the covariance matrix of the included covariates.

Note that when we have a completely randomized trial or equiprobable Bernoulli design with $e(X_i) = n_t/n\ \forall i$, then we have weights $w_i = n/n_t\ \forall i$, and our estimator is:

\begin{align*}
    \widehat{se}(\hat \tau) &= \sqrt{\frac{\hat \omega^2}{n}}  \\
&= \sqrt{\frac{(\hat\beta_{1} -\hat\beta_{0})^\top \bm{\hat\Omega}_{X} (\hat\beta_{1} -\hat\beta_{0})}{n} + \frac{\sum_i w_i^2 Z_i \hat\epsilon_i^2}{n^2} + \frac{ \sum{w_i^2 (1 - Z_i) \hat \epsilon_i^2 }}{n^2} }\\
    &= \sqrt{\frac{(\hat\beta_{1} -\hat\beta_{0})^\top \bm{\hat\Omega}_{X} (\hat\beta_{1} -\hat\beta_{0})}{n} + \frac{\ \sum{n^2/n_t^2 Z_i} \hat \epsilon_i^2 }{n^2} + \frac{ \sum{ n^2/n_c^2 (1 - Z_i) \hat \epsilon_i^2 }}{n^2} }\\
    &= \sqrt{\frac{(\hat\beta_{1} -\hat\beta_{0})^\top \bm{\hat\Omega}_{X} (\hat\beta_{1} -\hat\beta_{0})}{n} + \frac{\sum{Z_i \hat \epsilon_i^2}}{n_t^2} + \frac{\sum{(1 - Z_i) \hat \epsilon_i^2}}{n_c^2} } \\
    &= \sqrt{\frac{(\hat\beta_{1} -\hat\beta_{0})^\top \bm{\hat\Omega}_{X} (\hat\beta_{1} -\hat\beta_{0})}{n} + \frac{\hat \sigma_1^2}{n_t} + \frac{\hat \sigma_0^2}{n_c}}
\end{align*}

which matches the results in \citet{negi_woolridge_2021}.  \citet{lin_agnostic_2013} shows that the HC0 variance returned by (the unweighted) regression of $Y$ on treatment, demeaned covariates, and their interaction is $\frac{\hat \sigma_1^2}{n_t} + \frac{\hat \sigma_0^2}{n_c}$.

\renewcommand{\dotXt}{\dot{X}_{1}}
\renewcommand{\ddotXt}{\ddot{X}_{1}}
\renewcommand{\ddotRt}{\ddot{R}_{1,i}}

\renewcommand{\dotRt}{\dot{R}_{1,i}}

\subsection{Superpopulation correction for PATT}\label{app:sp_patt}

We begin by defining linear projections for the potential outcomes for treated units, again assuming our regularity conditions above, although we only require at least the unweighted second moments are finite for the existence of the linear projections.

\begin{align*}
       m_{1,1}(X) &= \alpha_{1, 1} + X \beta_{1, 1}  \\
        m_{0,1}(X) &= \alpha_{0, 1} + X \beta_{0, 1}
    \end{align*}

And therefore we can write the PATT as

\[ \tau_{PATT} = \mu_{1, 1} - \mu_{0, 1} = (\alpha_{1, 1} - \alpha_{0, 1}) + \mu_{X, 1}(\beta_{1, 1} - \beta_{0, 1})\]

Where $\mu_{X, 1}$ is the vector of covariate means in the population among treated units, i.e. those with $Z_i = 1$.  It is also helpful to rewrite our linear projections in terms of de-meaned covariates, such as we use in our Lin-style estimator.  First, note that, \emph{among our treated units}

\begin{align*}
        Y(0) = \mu_{0, 1} + \nu_1(0) \text{ when } Z = 1 \\
        Y(1) = \mu_{1, 1} + \nu_1(1) \text{ when } Z = 1
\end{align*}

where $\nu_1(0)$ and $\nu_1(1)$ have mean zero when $Z = 1$.  Define $\dotXt = X - \mu_{X, 1}$ to be the demeaned $X$ using true population means of $X$ among the treated units (i.e. $\E[X \mid Z = 1]$).   We can linearly project these demeaned covariates onto our error terms for those units with $Z = 1$.

\begin{align*}
    \nu_1(0) = \dotXt \beta_{0, 1} + \epsilon_1(0) \text{ when } Z = 1 \\
    \nu_1(1) = \dotXt \beta_{1, 1} + \epsilon_1(1) \text{ when } Z = 1
\end{align*}

where the intercepts are zero due to the de-meaning.  We can then rewrite our potential outcomes for our treated units as:

\begin{align}
    Y(0) = \mu_{0,1} + \dotXt \beta_{0, 1} + \epsilon_1(0) \text{ when } Z = 1 \label{eq:lp_y01}\\
    Y(1) = \mu_{1,1} + \dotXt \beta_{1, 1} + \epsilon_1(1) \text{ when } Z = 1 \label{eq:lp_y11}
\end{align}

This formulation has the following useful properties due to the definition of linear projections.

\begin{align}
    \E[\epsilon_1(0) \mid Z = 1] = \E[\epsilon_1(1) \mid Z = 1] = 0 \label{prop:LP_mean0_att} \\
    \E[\dotXt^\top \epsilon_1(0) \mid Z = 1] = \E[\dotXt^\top \epsilon_1(1) \mid Z = 1] = 0 \label{prop:LP_orthog_att}
\end{align}

Finally, let $\Var_{SP}[\epsilon_1(0) \mid Z = 1] = \sigma_{0,1}^2$ and $\Var_{SP}[\epsilon_1(1) \mid Z = 1] = \sigma_{1, 1}^2$ be variances among the treated units of the linear projection errors, which are mean zero by construction.

We begin by estimating $\mu_{1, 1}$ using the weighted regression of $Y$ on 1, $X$, for the $Z = 1$ sample.

\[\hat \mu_{1, 1} = \hat\alpha_{1, 1} + \overline{X}_1 \hat\beta_{1, 1}\]

By algebraic equivalence, we can also estimate $\hat\mu_{1, 1}$ from the weighted regression on demeaned covariates. Let $\ddotXt = X - \overline{X}_1$ where $\overline X_1$ is the mean among the treated units.  Then more specifically, we can also estimate $\hat\mu_{1, 1}$ from the weighted regression of $Y$ on 1, $\ddotXt$ for $Z = 1$.

Define $\ddotRt = ( 1, \ddotXt{}_i )$, then, by regression mechanics, our demeaned regression estimates

\begin{align}
    \hat\theta_{1, 1} = \genfrac{(}{)}{0pt}{}{\hat\mu_{1, 1}}{ \hat \bbeta_{1, 1}} &= \left( \sum_{i = 1}^{n} w_i Z_i \ddotRt^\top \ddotRt  \right)^{-1} \left( \sum_{i = 1}^{n} w_i  Z_i \ddotRt^\top Y_i  \right) \nonumber \\
    &= \left( n^{-1}\sum_{i = 1}^{n} w_i Z_i \ddotRt^\top \ddotRt  \right)^{-1} \left( n^{-1} \sum_{i = 1}^{n} w_i Z_i \ddotRt^\top Y_i(1)  \right) \label{eq:lp_reg_11}
\end{align}

Note we can write

\begin{equation*}
    Y_i(1) = \mu_{1, 1} + \dotXt{}_i \bbeta_{1, 1} + \epsilon_{1,i}(1) = \ddotRt \theta_{1, 1} + (\overline{X}_1 - \mu_{X, 1})\bbeta_{1, 1} + \epsilon_{1,i}(1)
\end{equation*}

Plugging this into the numerator of equation \eqref{eq:lp_reg_11}, we get:

\begin{align*}
n^{-1} \sum_{i = 1}^{n} w_i Z_i \ddotRt^\top Y_i(1) &= \left( n^{-1} \sum_{i = 1}^{n} w_i Z_i \ddotRt^\top \ddotRt \right) \theta_{1, 1}  + \left(  n^{-1} \sum_{i = 1}^{n} w_i Z_i \ddotRt^\top\right)(\overline{X}_1 - \mu_{X, 1})\beta_{1, 1} +\\
&\qquad n^{-1} \sum_{i = 1}^{n} w_i Z_i \ddotRt^\top \epsilon_{1,i}(1)
\end{align*}

and therefore we can simplify equation \eqref{eq:lp_reg_11} as:

\begin{align*} \hat\theta_{1, 1} = \theta_{1, 1} + \left( n^{-1}\sum_{i = 1}^{n} w_i Z_i \ddotRt^\top \ddotRt  \right)^{-1} \left[ \left(  n^{-1} \sum_{i = 1}^{n} w_i Z_i \ddotRt^\top\right)(\overline{X}_1 - \mu_{X, 1})\beta_{1, 1} \right.\\
\left.+ n^{-1} \sum_{i = 1}^{n} w_i Z_i \ddotRt^\top \epsilon_{1,i}(1) \right]
\end{align*}

Which gives us:

\begin{align} \sqrt{n}(\hat\theta_{1, 1} - \theta_{1, 1}) = \left( n^{-1}\sum_{i = 1}^{n} w_i Z_i \ddotRt^\top \ddotRt  \right)^{-1} \left[ \left(  n^{-1} \sum_{i = 1}^{n} w_i Z_i \ddotRt^\top\right)\sqrt{n}(\overline{X}_1 - \mu_{X,1})\beta_{1, 1} + \right. \nonumber \\
\left. n^{-1/2} \sum_{i = 1}^{n} w_iZ_i \ddotRt^\top \epsilon_{1,i}(1) \right] \label{eq:full_y11}
\end{align}

Now note that $\overline{X}_1 \overset p \to \mu_{X, 1}$ by Assumption~\ref{assump:random_sampling} and the law of large numbers, combined with Slutsky's theorem, we get:

\[ n^{-1} \sum_{i = 1}^n w_i Z_i \ddotRt^\top \ddotRt = n^{-1} \sum_{i = 1}^n w_i Z_i \dotRt^\top \dotRt + o_p(1)\]

where

$\dotRt = ( 1, \dotXt{}_i ) = ( 1, X_i - \mu_{X, 1} )$. 

And, further,

\begin{align*}
    n^{-1} \sum_{i = 1}^n w_i Z_i \dotRt^\top \dotRt \overset p \to& \E[w_i Z_i \dotRt^\top \dotRt] & \text{(WLLN)}\\
    &= \E\big[\E[w_i Z_i \mid X_i]\, \dotRt^\top \dotRt\big] & \text{(LIE; $\dotRt$ is a function of $X_i$)} \\
    &= \E\big[\E[Z_i \mid X_i]\, \dotRt^\top \dotRt\big] & \text{(Eqn.~\eqref{note_e_wZ_att})} \\
    &= \E[Z_i\, \dotRt^\top \dotRt] & \text{(LIE)} \\
    &= \rho\, \E[\dotRt^\top \dotRt \mid Z_i = 1] & \text{($Z_i$ binary)}
\end{align*}

Let

\[ \bm{A} \equiv \begin{pmatrix}
    1 & 0 \\
    0 & \E[\dotXt{}_i^\top \dotXt{}_i \mid Z_i = 1]
\end{pmatrix}
= \begin{pmatrix}
    1 & 0 \\
    0 & \bm{\Omega}_{X,1}
\end{pmatrix} \]

which comes from rewriting the covariance of $\E[\dotRt^\top \dotRt]$ and using the fact that the first column is 1s.

Noting that the terms $\sqrt{n}(\overline{X}_1 - \mu_{X, 1})\beta_{1, 1}$ and $n^{-1/2}\sum_{i = 1}^n w_i Z_i \ddotRt^\top \epsilon_{1i}(1)$ are $O_p(1)$ under our regularity conditions,

\begin{align*} \sqrt{n}(\hat\theta_{1, 1} - \theta_{1, 1}) = \frac{1}{\rho}\bm{A}^{-1}\left[ \left(  n^{-1} \sum_{i = 1}^{n} w_i Z_i \ddotRt^\top\right)\sqrt{n}(\overline{X}_1 - \mu_{X, 1})\beta_{1, 1} + \right.\\
\left. n^{-1/2} \sum_{i = 1}^{n} w_i Z_i \ddotRt^\top \epsilon_{1i}(1) \right] + o_p(1)
\end{align*}

Now, considering the first element of $n^{-1} \sum_{i = 1}^{n} w_i Z_i \ddotRt^\top$:

\[n^{-1} \sum_{i = 1}^{n} w_i Z_i \ddotRt^\top = n^{-1} \sum_{i = 1}^n w_i Z_i \begin{pmatrix} 1 \\ \ddotXt{}_i\end{pmatrix}\]

where the first element is then 

\[  n^{-1} \sum_{i = 1}^n w_i Z_i \overset p \to \E[w_iZ_i] = \E[\E[w_iZ_i \mid X_i]] = \E[\E[Z_i \mid X_i]] = \E[Z_i] = \rho\]

Also, 

\[ n^{-1/2} \sum_{i = 1}^{n} w_i Z_i \ddotRt^\top \epsilon_{1i}(1) = n^{-1/2} \sum_{i = 1}^{n} w_i Z_i \begin{pmatrix} 1 \\ \ddotXt{}_i\end{pmatrix} \epsilon_{1i}(1)  \]

where the first element is then:

\[ n^{-1/2} \sum_{i = 1}^n w_i Z_i \epsilon_{1i}(1)\]

And since $\bm{A}$ is block diagonal, the first element of $\sqrt{n}(\hat\theta_{1, 1} - \theta_{1, 1})$, $\sqrt{n}(\hat\mu_{1, 1} - \mu_{1, 1})$ satisfies

\begin{align}
\sqrt{n}(\hat\mu_{1, 1} - \mu_{1, 1}) &= \sqrt{n}(\overline{X}_1 - \mu_{X, 1})\beta_{1, 1} + \frac{1}{\rho}\, n^{-1/2}\sum_{i = 1}^n w_i Z_i \epsilon_{1i}(1) + o_p(1) \nonumber \\
&= n^{-1/2}\sum_{i = 1}^n \left[\frac{Z_i}{\rho}(X_i - \mu_{X, 1})\beta_{1, 1} + \frac{w_i Z_i \epsilon_{1i}(1)}{\rho} \right] + o_p(1) \label{eq:final_y11}
\end{align}

This comes from noting that $\overline{X}_1 - \mu_{X,1} = \frac{1}{n_t}\sum_i Z_i (X_i - \mu_{X,1})$, so

\begin{equation}
\sqrt{n}(\overline{X}_1 - \mu_{X,1})
= \frac{n}{n_t}\, n^{-1/2}\sum_{i=1}^n Z_i (X_i - \mu_{X,1})
= \frac{1}{\rho}\, n^{-1/2}\sum_{i=1}^n Z_i (X_i - \mu_{X,1}) + o_p(1),
\end{equation}
since $n_t/n \overset{p}{\to} \rho$ and $n^{-1/2}\sum_i Z_i(X_i - \mu_{X,1}) = O_p(1)$, as it has mean $\rho\,\E[X_i - \mu_{X,1} \mid Z_i = 1] = 0$.

This comes from noting that $\overline X_1 = \frac{\sum_{i:Z_i=1} X_i}{n_t} = \frac{\sum_i Z_i X_i}{n}\frac{n}{n_t} =\frac{1}{n} \sum_i \frac{Z_i X_i}{\rho}$.

An analogous argument comes for estimating the $\hat\mu_{0, 1}$ from the regression of $Y_i$ on 1, $\ddotXt$ among the control units, yielding

\begin{align}
\sqrt{n}(\hat\mu_{0, 1} - \mu_{0, 1}) &= \sqrt{n}(\overline{X}_1 - \mu_{X, 1})\beta_{0, 1} + \frac{1}{\rho}\, n^{-1/2}\sum_{i = 1}^n w_i(1 - Z_i)\epsilon_{1i}(0) + o_p(1) \nonumber \\
&= n^{-1/2}\sum_{i = 1}^n \left[\frac{Z_i}{\rho}(X_i - \mu_{X, 1})\beta_{0, 1} + \frac{w_i(1 - Z_i)\epsilon_{1i}(0)}{\rho}\right] + o_p(1) \label{eq:final_y01}
\end{align}

Combining equation \eqref{eq:final_y11} and \eqref{eq:final_y01}, we thus get:

\begin{align*}
\sqrt{n}(\hat\tau_{PATT} - \tau_{PATT}) = n^{-1/2} \sum_{i = 1}^n \left[ \frac{Z_i}{\rho}\dotXt{}_i(\beta_{1, 1} - \beta_{0, 1}) + w_i Z_i \epsilon_{1i}(1)/\rho - \right. \\  w_i(1 - Z_i)\epsilon_{1i}(0)/\rho \Big] + o_p(1)
\end{align*}

By CLT and asymptotic equivalence lemma, we get:

\begin{align*}
\sqrt{n} (\hat\tau_{PATT} - \tau_{PATT}) &\overset d \to \mathcal{N}(0, \omega^2) \\
\hat\tau_{PATT} &\overset{a}{\sim} \mathcal{N}\left(\tau_{PATT}, \frac{\omega^2}{n}\right)
\end{align*}

where $\omega^2 = \Var_{SP}[\frac{Z_i}{\rho}\dotXt{}_i(\beta_{1, 1} - \beta_{0, 1}) + w_i Z_i \epsilon_{1i}(1)/\rho - w_i(1 - Z_i)\epsilon_{1i}(0)/\rho]$.

To simplify the expression for $\omega^2$, we begin by noting that the terms are pairwise uncorrelated.  First note that, in the expansion, the last two terms interacted drop out because $Z_i (1 - Z_i) = 0$.  We next consider:

\vspace{-1em}
\begin{align*}
&\E\left[\frac{Z_i}{\rho}(\beta_{1, 1} -\beta_{0, 1})^\top \dotXt{}_i^\top w_i Z_i \epsilon_{1i}(1)/\rho\right] \\
&= (\beta_{1, 1} -\beta_{0, 1})^\top  \E[\E[\dotXt{}_i^\top w_i Z_i^2 \epsilon_{1i}(1) \mid X_i]]/\rho^2 & \text{(LIE)}\\
&= (\beta_{1, 1} -\beta_{0, 1})^\top  \E[\dotXt{}_i^\top \E[ w_i Z_i \epsilon_{1i}(1) \mid X_i]]/\rho^2 \\
&= (\beta_{1, 1} - \beta_{0, 1})^\top  \E[\dotXt{}_i^\top \E[ w_i Z_i \mid X_i ] \E[\epsilon_{1i}(1) \mid X_i]]/\rho^2 & \text{(Asm.~\eqref{assump:cond_ignorability})} \\
&= (\beta_{1, 1} -\beta_{0, 1})^\top  \E[\dotXt{}_i^\top \E[Z_i \mid X_i] \E[\epsilon_{1i}(1) \mid X_i]]/\rho^2 & \text{(Eqn.~\eqref{note_e_wZ_att})}\\
&= (\beta_{1, 1} - \beta_{0, 1})^\top  \E[\E[Z_i \dotXt{}_i^\top \epsilon_{1i}(1) \mid X_i]]/\rho^2\\
&= (\beta_{1, 1} - \beta_{0, 1})^\top  \E[Z_i \dotXt{}_i^\top \epsilon_{1i}(1)]/\rho^2 & \text{(LIE)}\\
&= (\beta_{1, 1} -\beta_{0, 1})^\top  \rho\ \E[\dotXt{}_i^\top \epsilon_{1i}(1) \mid Z_i = 1]/\rho^2 + (1 - \rho) \cdot 0 & \text{(LIE)}\\
&= (\beta_{1, 1} -\beta_{0, 1})^\top \E[\dotXt{}_i^\top \epsilon_{1i}(1) \mid Z_i = 1]/\rho &  \\
&=0 & \text{(Eqn.~\eqref{prop:LP_orthog_att})}
\end{align*}

where a similar argument holds for $\E\left[\frac{Z_i}{\rho}(\beta_{1, 1} -\beta_{0, 1})^\top \dotXt{}_i^\top w_i (1 - Z_i) \epsilon_{1i}(0)/\rho\right] = 0$.

Now we consider 

\vspace{-1em}

\begin{align*}
\Var_{SP}[w_i Z_i \epsilon_{1i}(1)/\rho]
&= \E\big[(w_i Z_i \epsilon_{1i}(1)/\rho)^2\big] - \big(\E[w_i Z_i \epsilon_{1i}(1)/\rho]\big)^2 \\
&= \E\big[(w_i Z_i \epsilon_{1i}(1)/\rho)^2\big] - \big(\E\big[\E[w_i Z_i \epsilon_{1i}(1) \mid X_i]\big]/\rho\big)^2 & \text{(LIE)}\\
&= \E\big[(w_i Z_i \epsilon_{1i}(1)/\rho)^2\big] - \big(\E\big[\E[w_i Z_i \mid X_i]\, \E[\epsilon_{1i}(1) \mid X_i]\big]/\rho\big)^2 & \text{(Asm.~\ref{assump:cond_ignorability})}\\
&= \E\big[(w_i Z_i \epsilon_{1i}(1)/\rho)^2\big] - \big(\E\big[\E[Z_i \mid X_i]\, \E[\epsilon_{1i}(1) \mid X_i]\big]/\rho\big)^2 & \text{(Eqn.~\eqref{note_e_wZ_att})}\\
&= \E\big[(w_i Z_i \epsilon_{1i}(1)/\rho)^2\big] - \big(\E[Z_i\, \epsilon_{1i}(1)]/\rho\big)^2 & \text{(Asm.~\ref{assump:cond_ignorability}, LIE)}\\
&= \E\big[(w_i Z_i \epsilon_{1i}(1)/\rho)^2\big] - \Big(\big(\rho\, \E[\epsilon_{1i}(1) \mid Z_i = 1] + (1 - \rho)\cdot 0\big)/\rho\Big)^2 & \text{(LIE over $Z_i$)}\\
&= \E\big[(w_i Z_i \epsilon_{1i}(1)/\rho)^2\big] - \big(\E[\epsilon_{1i}(1) \mid Z_i = 1]\big)^2 \\
&= \E\big[(w_i Z_i \epsilon_{1i}(1)/\rho)^2\big] & \text{(Eqn.~\eqref{prop:LP_mean0_att})}\\
&= \E[w_i^2 Z_i \epsilon_{1i}^2(1)]/\rho^2
\end{align*}

where a similar argument holds for $\Var_{SP}[w_i (1 - Z_i) \epsilon_{1i}(0)/\rho] = \E[w_i^2 (1 - Z_i) \epsilon_{1i}^2(0)]/\rho^2$.

Finally, considering the first term in $\omega^2$,

\begin{align*}
\Var_{SP}\left[\frac{Z_i}{\rho}\dotXt{}_i(\beta_{1, 1} - \beta_{0, 1})\right] &= \E\left[\frac{Z_i^2}{\rho^2}(\beta_{1, 1} - \beta_{0, 1})^\top\dotXt{}_i^\top\dotXt{}_i(\beta_{1, 1} - \beta_{0, 1})\right] - \E\left[\frac{Z_i}{\rho}\dotXt{}_i(\beta_{1, 1} - \beta_{0, 1})\right]^2 \\
&= \frac{1}{\rho^2}\left[\E[Z_i(\beta_{1, 1} - \beta_{0, 1})^\top(X_i - \mu_{X, 1})^\top(X_i - \mu_{X, 1})(\beta_{1, 1} - \beta_{0, 1})] \right. \\
&\qquad \left. - \E[Z_i(X_i - \mu_{X, 1})(\beta_{1, 1} - \beta_{0, 1})]^2\right] \\
&= \frac{1}{\rho^2}\left[\rho\ \E[(\beta_{1, 1} - \beta_{0, 1})^\top(X_i - \mu_{X, 1})^\top(X_i - \mu_{X, 1})(\beta_{1, 1} - \beta_{0, 1}) \mid Z_i = 1] \right. \\
&\qquad + (1 - \rho) \cdot 0 \left. - \left(\rho\ \E[(X_i - \mu_{X, 1})(\beta_{1, 1} - \beta_{0, 1}) \mid Z_i = 1] + (1 - \rho) \cdot 0\right)^2\right] & \text{(LIE)}\\
&= \frac{1}{\rho}\left[\E[(\beta_{1, 1} - \beta_{0, 1})^\top(X_i - \mu_{X, 1})^\top(X_i - \mu_{X, 1})(\beta_{1, 1} - \beta_{0, 1}) \mid Z_i = 1] \right. \\
&\qquad \left. - \left(\underbrace{\E[(X_i - \mu_{X, 1}) \mid Z_i = 1] }_{=0}(\beta_{1, 1} - \beta_{0, 1})\right)^2\right] \\
&= \frac{1}{\rho}\left[\E[(\beta_{1, 1} - \beta_{0, 1})^\top(X_i - \mu_{X, 1})^\top(X_i - \mu_{X, 1})(\beta_{1, 1} - \beta_{0, 1}) \mid Z_i = 1]\right] \\
&= \frac{1}{\rho}(\beta_{1, 1} - \beta_{0, 1})^\top\E[(X_i - \mu_{X, 1})^\top(X_i - \mu_{X, 1}) \mid Z_i = 1](\beta_{1, 1} - \beta_{0, 1}) \\
&= \frac{1}{\rho}(\beta_{1, 1} - \beta_{0, 1})^\top\bm{\Omega}_{X, 1}(\beta_{1, 1} - \beta_{0, 1})
\end{align*}

Combining these results leaves us with:

\begin{align}
\frac{\omega^2}{n} &= \frac{(\beta_{1, 1} - \beta_{0, 1})^\top\bm{\Omega}_{X, 1}(\beta_{1, 1} - \beta_{0, 1})}{n\rho} + \frac{\E[w_i^2 Z_i \epsilon_{1i}^2(1)]}{n\rho^2} + \frac{\E[w_i^2 (1 - Z_i) \epsilon_{1i}^2(0)]}{n\rho^2} \nonumber\\
&= \frac{(\beta_{1, 1} - \beta_{0, 1})^\top\bm{\Omega}_{X, 1}(\beta_{1, 1} - \beta_{0, 1})}{n\rho} + \frac{\rho \E[w_i^2 \epsilon_{1i}^2(1)\mid Z_i = 1]}{n\rho^2} + \nonumber\\
&\qquad \frac{(1 - \rho)\ \E[w_i^2 \epsilon_{1i}^2(0) \mid Z_i = 0]}{n\rho^2} & \text{(LIE)}\nonumber\\
\nonumber\\
&= \frac{(\beta_{1, 1} - \beta_{0, 1})^\top\bm{\Omega}_{X, 1}(\beta_{1, 1} - \beta_{0, 1})}{n_t} + \frac{\E[w_i^2 \epsilon_{1i}^2(1)\mid Z_i = 1]}{n_t} +  \nonumber\\
&\qquad  \frac{n_c\ \E[w_i^2 \epsilon_{1i}^2(0) \mid Z_i = 0]}{n_t^2}\nonumber\\
&= \frac{(\beta_{1, 1} - \beta_{0, 1})^\top\bm{\Omega}_{X, 1}(\beta_{1, 1} - \beta_{0, 1})}{n_t} + \frac{\E[\epsilon_{1i}^2(1)\mid Z_i = 1]}{n_t} +  \nonumber\\
&\qquad  \frac{n_c\ \E[w_i^2 \epsilon_{1i}^2(0) \mid Z_i = 0]}{n_t^2} & (w_i = 1 \forall Z_i = 1) \nonumber\\
&= \frac{(\beta_{1, 1} - \beta_{0, 1})^\top\bm{\Omega}_{X, 1}(\beta_{1, 1} - \beta_{0, 1})}{n_t} + \frac{\sigma_{1, 1}^2}{n_t} + \frac{n_c\ \E[w_i^2 \epsilon_{1i}^2(0) \mid Z_i = 0]}{n_t^2} \label{var_patt_simple}
\end{align}

\subsubsection{Variance Estimators for PATT}

We now turn to estimating the variance.  Again, in addition to our regularity conditions above, we assume the researcher has a consistent estimator for $e(X_i)$.  Using the form in Equation~\eqref{var_patt_simple}, a plug-in estimator for the standard error of $\hat\tau_{PATT}$ for the PATT is:

\begin{align}
\hat{se}(\hat\tau_{PATT}) &= \sqrt{\frac{\hat\omega^2}{n}} \nonumber \\
&= \sqrt{\frac{(\hat \beta_{1, 1} - \hat \beta_{0, 1})^\top\bm{\hat \Omega}_{X, 1}(\hat \beta_{1, 1} - \hat \beta_{0, 1})}{n_t} + \frac{\hat \sigma_{1, 1}^2}{n_t} + \frac{n_c\ \hat \E[w_i^2 \epsilon_{1i}^2(0) \mid Z_i = 0]}{n_t^2}}\nonumber \\
&= \sqrt{\frac{(\hat \beta_{1, 1} - \hat \beta_{0, 1})^\top\bm{\hat \Omega}_{X, 1}(\hat \beta_{1, 1} - \hat \beta_{0, 1})}{n_t} + \frac{\sum_i Z_i \hat \epsilon_{1i}^2}{n_t^2} + \frac{n_c\ \sum_i w_i^2 (1 - Z_i)  \hat \epsilon_{1i}^2}{n_cn_t^2}} \nonumber \\
&= \sqrt{\underbrace{\frac{(\hat \beta_{1, 1} - \hat \beta_{0, 1})^\top\bm{\hat \Omega}_{X, 1}(\hat \beta_{1, 1} - \hat \beta_{0, 1})}{n_t}}_{\text{correction}} + \frac{\sum_i Z_i \hat \epsilon_{1i}^2}{n_t^2} + \frac{\ \sum_i w_i^2 (1 - Z_i)  \hat  \epsilon_{1i}^2}{n_t^2}} \label{eq:plugin_var_att}
\end{align}

As with the PATE weights, Equation~\eqref{reg_var_consistent_att} shows that the HC0 standard error from our residualized regression estimates the last two terms in Equation~\eqref{eq:plugin_var_att} when PATT weights are used.  Therefore, the only additional component a researcher must calculate is the ``correction".  $(\hat \beta_{1, 1} - \hat \beta_{0, 1})$ is estimated by $\hat \gamma$ in our proposed residualized regression in Equation~\eqref{eq:wResModel} when using ATT weights.  $\bm{\hat\Omega}_{X, 1}$ is estimated as $\hat S^{2}_{\phi_{Z = 1}}$, or the covariance matrix among the treated units of the included covariates.

\section{M-Estimation} \label{app:mest}

In this section we provide the estimating equations for various M-estimation problems.  We consider when the weights are estimated using logit to estimate IPW weights as well as entropy balancing weights.  We also consider two outcome models---a weighted bivariate regression and our proposed residualized model.  We also show variations relevant to finite/fixed sample vs. superpopulation queries.
In particular, 
regarding the sample as fixed, or the moment conditions as known, implies that $p(X)$ is fixed, and so sample estimates of quantities like the means can be used directly for forming moment conditions. By contrast, to replicate the uncertainty engaged by considering the ``superpopulation'' or data-generating process beyond the sample, the sample moments used to construct moment conditions must themselves be constructed as estimates within the M-estimation process.  To the best of our knowledge, the stacked estimating equation approach for balancing weights has not previously been described in the existing literature.  

Accounting for uncertainty in the estimation of the means of $X$ is necessary to recover standard errors under the superpopulation framework (e.g. for the PATE).  We show how existing M-estimation approaches using IPW implicitly estimate population means for balancing covariates (of the treated units) by reformulating the logit score conditions into two separate conditions, one for estimation of population means and one for estimating the logit coefficients.  This turns it from a problem of ``uncertainty in the weights'', as it is seen in the existing literature, to one of ``uncertainty in the population targets'', which is the same as the problem we tackle with the proposed superpopulation correction.  We then use this formulation as a starting point for the entropy balancing weights, where we show how the analogous score conditions should properly account for estimation of the population balancing covariate means.  We then provide the ``known means'' estimating equations as a point of comparison, where we treat the population means for the balancing constraints as known and remove score conditions associated with their estimation.  We show that M-estimation that accounts for the estimation of the population means for balancing covariates is appropriate for superpopulation estimands, whereas M-estimation that treats these quantities as fixed is appropriate for our design-based and model-based estimands which rely on fixed populations.  

We evaluate performance through simulation, with respect to our superpopulation estimands, and compare performance to our proposed residualized standard errors with our proposed superpopulation correction.

\subsection{Notation}

We follow the notation outlined in Section~\ref{sec:mestimation}.  For notational simplicity we consider features $\phi(\X) = \X$, but the results generalize to any transformed features as well.  However, in this appendix we consider two outcome models: the weighted bivariate regression and our proposed residualized model.  For the weighted bivariate regression, we let the design matrix be $H_i = [1, Z_i]^\top$.  For both, the target causal quantity of interest will be the coefficient on $Z_i$.  Note that, when we treat the sample means as known, then we will let $J_i = [1, Z_i, X_i - \overline{X}, Z_i(X_i - \overline{X})]^\top$.

\subsection{Estimating Equations}

\subsubsection{Superpopulation Estimands}\label{sec:m_est_unknown_means} We begin with the estimating equations appropriate for uncertainty associated with superpopulation parameters. These require that any sample moments involved in writing score conditions (typically covariate means, globally or by treatment group) are estimated through their own score conditions. 

\paragraph{IPW+Bivariate.} Superpopulation M-estimation for the (population) average treatment effect with inverse propensity weights is interesting as the procedure does not explicitly engage sample means or covariances in the score conditions, and thus no such quantities need to be estimated via additional score conditions. The appropriate estimating equations in the existing literature are:

\begin{equation} \label{eq:unk-ipw-bi-twostack}
    \psi(O_i; \theta) = \left[ \begin{array}{c} \psi_\alpha \\ \psi_\lambda \end{array} \right] = \left[ \begin{array}{c} \{Z_i - \operatorname{expit}(G_i^\top \alpha)\}G_i \\ \{ W_i(Y_i - H_i^\top \lambda)\} H_i  \end{array} \right]
\end{equation}

where the $\psi_\alpha$ corresponds to the score equations for the logit model for treatment. The resulting choice of $\hat{\alpha}$ defines $W_i$ through Equation~\eqref{eq:mest_weights}.  $\psi_\lambda$ corresponds to the score equations for the bivariate regression outcome model using those weights $W_i$ (see e.g. \citealp{stefanski2002calculus, reifeis_hudgens_2022, ross2024m}).

Implicitly, the score conditions for $\psi_\alpha$ include a set of balancing constraints.  Note that we can rewrite the first score condition, which estimates the logit model, as:

\begin{align*}
    \sum_i \{Z_i - \operatorname{expit}(G_i^\top \alpha)\}G_i &= 0 \\
    \sum_i \big(Z_i G_i - \mu_{zg}\big) + \sum_i \big(\mu_{zg} - \operatorname{expit}(G_i^\top \alpha)G_i\big) &= 0
    \intertext{which we can rewrite as two conditions:}\\
    \sum_i \big(Z_i G_i - \mu_{zg}\big) = 0 \quad \text{and} \quad \sum_i \big(\operatorname{expit}(G_i^\top \alpha)G_i - \mu_{zg}\big) &= 0
\end{align*}

\noindent where $\mu_{zg}$ is a vector where the first element is the average treatment probability and the remaining elements are the average treated means for each balancing covariate multiplied by the overall treatment probability. Thus, we can rewrite the estimating equations to explicitly account for these balancing constraints as:

\begin{equation} \label{eq:unk-ipw-bi-threestack}
\psi(O_i; \theta) = \left[ \begin{array}{c} 
\psi_{\mu_{zg}} \\ 
\psi_\alpha \\ \psi_\lambda \end{array} \right] = \left[ \begin{array}{c} 
(Z_iG_i - \mu_{zg}) \\
\operatorname{expit}(G_i^{\top}\alpha)G_i - \mu_{zg} \\ 
\{ W_i(Y_i - H_i^\top \lambda)\} H_i  \end{array} \right]
\end{equation}

This form parallels the estimating equations we will present below for entropy balancing, with the exception that the balancing constraints for IPW are not separated by treatment and control group, as they are for entropy balancing.

\paragraph{Entropy Balancing+Bivariate.} For entropy balancing weights used in a bivariate regression, our estimating equations are:

\begin{equation}\label{eq:unk-eb-bi}
\psi(O_i; \theta) = \left[ \begin{array}{c} \psi_{\mu_x} \\ \psi_{\eta_t} \\ \psi_{\eta_c} \\ \psi_\lambda \end{array} \right] = \left[ \begin{array}{c} X_i - \mu_x \\ \frac{Z_i}{n_t}\left(\exp(G_i^{\top}\eta_t)\, G_i - [1\;\; \mu_x^\top]^\top\right) \\ \frac{1 - Z_i}{n_c}\left(\exp(G_i^{\top}\eta_c)\, G_i - [1\;\; \mu_x^\top]^\top\right) \\ \{ W_i(Y_i - H_i^\top \lambda)\}H_i  \end{array} \right]
\end{equation}

where the $\psi_{\mu_x}$ corresponds to the score conditions for estimating $\mu_X$, $\psi_{\eta_t}$ and $\psi_{\eta_c}$ are the score conditions for estimating the treatment and control entropy balancing weights, respectively, accounting for the uncertainty in the estimation of the target means.  Finally, $\psi_\lambda$ gives the score conditions for estimating the weighted bivariate regression as the outcome model using weights $W_i$.

\paragraph{IPW+Residualized.} For our proposed residualized model using IPW weights, we can update the score conditions in Equation~\eqref{eq:unk-ipw-bi-threestack} to account for the need to estimate the population means of the $X$ used in centering the residualized estimator as:

\begin{equation} \label{eq:unk-ipw-res}
    \psi(O_i; \theta) = \left[ \begin{array}{c} \psi_{\mu_{zg}} \\  \psi_{\mu_{x_c}} \\ \psi_\alpha \\ \psi_\lambda \end{array} \right] = \left[ \begin{array}{c} (Z_iG_i - \mu_{zg}) \\ (1 - Z_i)(X_i - \mu_{x_c}) \\ \operatorname{expit}(G_i^{\top}\alpha)G_i - \mu_{zg} \\ \{ W_i(Y_i - {J_i}^\top \lambda)\} J_i  \end{array} \right]
\end{equation}

where $\psi_{\mu_{zg}}$ and $\psi_\alpha$ are the score conditions for the logistic regression that produces the IPW weights, $\psi{\mu_{x_c}}$ are score conditions for the mean $X$ among the control units, and $\psi_\lambda$ are the score conditions for the proposed residualized regression.  To account for the estimation of the population means for recentering around $\mu_X$ in the definition of $J$, it is important to note that

\begin{align*}
    \mu_x &= \Pr[Z = 1]\, \E[X \mid Z = 1] + \Pr[Z = 0]\, \E[X \mid Z = 0] \\
    &= \E[ZX] + (1 - \Pr[Z = 1])\, \mu_{x_c},
\end{align*}

where $\Pr[Z = 1]$ is the first element of $\mu_{zg}$ and $\E[ZX]$ is the vector of its remaining elements.

Another way of saying this is that once we have estimated $\mu_{zg}$, and noting that the first element is the probability of treatment, then the only additional parameters we need to know to estimate the population means of $X$ are the means among the control units.  One can directly estimate the means of $X$ instead of the control means and recover the same estimate, however we prefer this formulation because it simplifies the estimation of the ATT which does not require estimating the control means because the treated population means used for recentering are already estimated in $\mu_{zg}$ (see Section~\ref{app:mest_att}).

An alternative approach is taken by \citet{shook2025double}, who derive an M-estimator for a doubly robust weighted least squares estimator.  Their approach estimates the mean potential outcomes for treated and controls separately using a g-computation estimator rather than our proposed residualized regression to estimate the same quantity in one regression through the interaction (see e.g. \citealp{chattopadhyay2023implied, shinkrehazlettdemystifying} regarding equivalence of these approaches). This approach also accounts for the uncertainty in the sample of $X$ through the projection step.

\paragraph{Entropy Balancing+Residualized.} For our proposed residualized model using entropy balancing weights, the estimating equations are:

\begin{equation} \label{eq:unk-eb-res}
\psi(O_i; \theta) = \left[ \begin{array}{c} \psi_{\mu_x} \\ \psi_{\eta_t} \\ \psi_{\eta_c} \\ \psi_\lambda \end{array} \right] = \left[ \begin{array}{c} X_i - \mu_x \\ \frac{Z_i}{n_t}\left(\exp(G_i^{\top}\eta_t)\, G_i - [1\;\; \mu_x^\top]^\top\right) \\ \frac{1 - Z_i}{n_c}\left(\exp(G_i^{\top}\eta_c)\, G_i - [1\;\; \mu_x^\top]^\top\right) \\ \{ W_i(Y_i - J_i^\top \lambda)\}J_i  \end{array} \right]
\end{equation}

where the score conditions match those for the bivariate case, except we now have $\psi_\lambda$ for the score conditions for the residualized outcome model.

\subsubsection{Fixed Sample Means}\label{sec:m_est_known_means}

As discussed above, when targeting the ATE, such as in the model-based framework or the SATE in the design-based framework, it would be appropriate to treat the sample means as known when estimating IPW or entropy balancing weights or a residualized model, but it would generally be inappropriate when estimating uncertainty for superpopulation estimators.  In this section we consider a ``fixed-mean'' version of the estimating equations that correspond to estimators for the S/ATE.

To minimize changes in notation, for the residualized model, we will redefine $J_i = [1, Z_i, X_i-\overline{X}, Z_i(X_i - \overline{X})]^\top$.  That is, we will assume that $J$ is centered on the sample quantities and not the population quantities for this section.

\paragraph{IPW+Bivariate.} Using score conditions for IPW + bivariate in Equation~\eqref{eq:unk-ipw-bi-threestack}, for sample estimands $\mu_{zg}$ need not be estimated, but rather $\overline{ZG}$ (the sample analog) can be used directly in estimation of the weights.  When we do this, we will estimate the same coefficients in the weighting and outcome models as the superpopulation form, but we do not need to account for uncertainty in the estimation of these means.  This implies that the estimating equations take the form:

\begin{equation} \label{eq:known-ipw-bi}
\psi(O_i; \theta) = \left[ \begin{array}{c} 
\psi_\alpha \\ \psi_\lambda \end{array} \right] = \left[ \begin{array}{c} 
\operatorname{expit}(G_i^{\top}\alpha)G_i - \overline{ZG} \\ 
\{ W_i(Y_i - H_i^\top \lambda)\} H_i  \end{array} \right]    
\end{equation}

\noindent where $\overline{ZG}$ is the vector of means of $Z$ interacted with each column of $G$.  Note that we no longer have the equations corresponding to $\psi_{\mu_{zg}}$ from the superpopulation quantity, which leads to a different form of score conditions than the typical IPW approach in the existing literature.

\paragraph{Entropy Balancing+Bivariate.} For entropy balancing weights used in the bivariate regression and with known means our estimating equations are:

\begin{equation} \label{eq:known-eb-bi}
\psi(O_i; \theta) = \left[ \begin{array}{c} \psi_{\eta_t} \\ \psi_{\eta_c} \\ \psi_\lambda \end{array} \right] = \left[ \begin{array}{c} \frac{Z_i}{n_t}\left(\exp(G_i^{\top}\eta_t)\, G_i - [1\;\; \overline{X}^\top]^\top\right) \\ \frac{1 - Z_i}{n_c}\left(\exp(G_i^{\top}\eta_c)\, G_i - [1\;\; \overline{X}^\top]^\top\right) \\ \{ W_i(Y_i - H_i^\top \lambda)\}H_i  \end{array} \right]
\end{equation}

where the only change from Equation~\eqref{eq:unk-eb-bi} is that $\overline{X}$ appears in the score conditions for the entropy weights, as opposed to $\mu_X$. There are no score conditions corresponding to the estimation of $\mu_X$; it is obtained directly from the sample and regarded as fixed.  Again, this results in the same weights from the estimating equations for the superpopulation quantities, but updates the uncertainty to account for known sample means.

\paragraph{IPW+Residualized.} We update our residualized model in the fixed-means case with IPW using the following estimating equations:

\begin{equation} \label{eq:known-ipw-res}
\psi(O_i; \theta) = \left[ \begin{array}{c} 
\psi_\alpha \\ \psi_\lambda \end{array} \right] = \left[ \begin{array}{c} 
\operatorname{expit}(G_i^{\top}\alpha)G_i - \overline{ZG} \\ 
\{ W_i(Y_i - J_i^\top \lambda)\} J_i  \end{array} \right]
\end{equation}

where the only changes from Equation~\eqref{eq:unk-ipw-res} are that we no longer have the $\psi_{\mu_{zg}}$ or $\psi_{\mu_{x_c}}$ score conditions, we treat $\mu_{zg}$ as fixed at $\overline{ZG}$, and $J_i$ treats the sample mean as a fixed quantity in the recentering.

\paragraph{Entropy Balancing+Residualized.}  We update our residualized model in the fixed-means case with entropy balancing using the following estimating equations:

\begin{equation} \label{eq:known-eb-res}
\psi(O_i; \theta) = \left[ \begin{array}{c} \psi_{\eta_t} \\ \psi_{\eta_c} \\ \psi_\lambda \end{array} \right] = \left[ \begin{array}{c} \frac{Z_i}{n_t}\left(\exp(G_i^{\top}\eta_t)\, G_i - [1\;\; \overline{X}^\top]^\top\right) \\ \frac{1 - Z_i}{n_c}\left(\exp(G_i^{\top}\eta_c)\, G_i - [1\;\; \overline{X}^\top]^\top\right) \\ \{ W_i(Y_i - J_i^\top \lambda)\}J_i  \end{array} \right]
\end{equation}

where the only change from Equation~\eqref{eq:unk-eb-res} is that we no longer have the $\psi_{\mu_x}$ score conditions, and $J_i$ treats the sample mean as a fixed quantity.

\subsubsection{Average Treatment Effect of the Treated}\label{app:mest_att}

In general, the estimating equations take a similar form when estimating the population average treatment effect on the treated (PATT), or the sample based analogs (S/ATT).  There are two important differences.  The first is in weights, which take the form:

\begin{align}
W_i = \begin{cases}
			 \frac{Z_i}{n_t} + \frac{1 - Z_i}{n_t}\,\frac{\operatorname{expit}(G_i^\top \alpha)}{1 - \operatorname{expit}(G_i^\top \alpha)}, & \text{if IPW}\\[6pt]
            \frac{Z_i}{n_t} + \frac{1 - Z_i}{n_c}\exp(G_i^\top \eta_c), & \text{if Entropy Balancing}
		 \end{cases} \label{eq:mest_weights_app}
\end{align}

This means that, for entropy balancing weights there will be no $\psi_{\eta_t}$ score conditions since the weights are all $1/n_t$ for treated units.

The second major difference is that in the residualizing models, covariates are centered on the mean of the $X$ taken over the \emph{treated} units.  This means that for the superpopulation PATT, we let $J_i = [1, Z_i, X_i - \mu_{X, 1}, Z_i(X_i - \mu_{X, 1})]^\top$ where $\mu_{X, 1}$ is $\E[X \mid Z = 1]$, and we update any score conditions to reflect this change to $\mu_{X, 1}$.  When estimating fixed sample quantities (S/ATT), we let $J_i = [1, Z_i, X_i - \overline{X}_1, Z_i*(X_i - \overline{X}_1)]^\top$ where $\overline{X}_1$ is the sample mean of the $X$ among the treated units.

Additionally, for IPW+Residualized superpopulation approach, this means the $\psi_{\mu_{x_c}}$ conditions are not necessary, because all of the information about the treated means can be calculated from the $\psi_{\mu_{zg}}$ parameters.

\subsection{Simulations}

For our simulations, we use the superpopulation DGP from the main manuscript, focusing on the same outcome described in Section~\ref{sec:sims}.  Results can be found in Table~\ref{tab:mest_sim_results}.  These simulations reproduce the results from our main simulations for our proposed residualized regression standard errors.  We estimate the standard errors in the following ways:

\begin{itemize}
    \item M-Estimation: sandwich standard errors from the stacked estimating equations in Section~\ref{sec:m_est_unknown_means} (superpopulation) or Section~\ref{sec:m_est_known_means} (design- and model-based).
    \item WLS + HC0: HC0 standard errors from the weighted regression. Under the superpopulation framework, the residualized model also includes the superpopulation correction, denoted ``(+ Corr)''.
\end{itemize}

The outcome models, including the weighted bivariate regression or our proposed residualized regression, are in the columns.

\begin{table}
\centering
\resizebox*{!}{ 0.83\textheight }{%
\begin{tabular}[t]{llcc|ccc|ccc}
\toprule
\multicolumn{4}{c}{ } & \multicolumn{3}{c}{Bivariate} & \multicolumn{3}{c}{Residualized} \\
\cmidrule(l{3pt}r{3pt}){5-7} \cmidrule(l{3pt}r{3pt}){8-10}
 &  &  &  & se$_{emp}$ & Avg($\hat{se}$) & Coverage & se$_{emp}$ & Avg($\hat{se}$) & Coverage\\
\midrule
\addlinespace[0.3em]
\multicolumn{10}{l}{\textbf{Design-Based}}\\
\addlinespace[0.3em]
\multicolumn{10}{l}{\textit{SATE}}\\
\addlinespace[0.3em]
\multicolumn{10}{l}{IPW}\\
\hspace{1em}\hspace{1em}\hspace{1em} &  &  & M-Estimation & 0.308 & 0.380 & 0.984 & 0.302 & 0.315 & 0.962\\
\cmidrule{4-10}
\hspace{1em}\hspace{1em}\hspace{1em} &  &  & OLS + HC0 & 0.308 & 0.356 & 0.977 & 0.302 & 0.315 & 0.962\\
\cmidrule{3-10}
\addlinespace[0.3em]
\multicolumn{10}{l}{Entropy Balancing}\\
\hspace{1em}\hspace{1em}\hspace{1em} &  &  & M-Estimation & 0.299 & 0.310 & 0.962 & 0.299 & 0.310 & 0.962\\
\cmidrule{4-10}
\hspace{1em}\hspace{1em}\hspace{1em} &  &  & OLS + HC0 & 0.299 & 0.348 & 0.978 & 0.299 & 0.310 & 0.962\\
\midrule
\cmidrule{2-10}
\addlinespace[0.3em]
\multicolumn{10}{l}{\textit{SATT}}\\
\addlinespace[0.3em]
\multicolumn{10}{l}{IPW}\\
\hspace{1em}\hspace{1em}\hspace{1em} &  &  & M-Estimation & 0.323 & 0.362 & 0.988 & 0.317 & 0.315 & 0.970\\
\cmidrule{4-10}
\hspace{1em}\hspace{1em}\hspace{1em} &  &  & OLS + HC0 & 0.323 & 0.353 & 0.985 & 0.317 & 0.315 & 0.970\\
\cmidrule{3-10}
\addlinespace[0.3em]
\multicolumn{10}{l}{Entropy Balancing}\\
\hspace{1em}\hspace{1em}\hspace{1em} &  &  & M-Estimation & 0.317 & 0.315 & 0.970 & 0.317 & 0.315 & 0.970\\
\cmidrule{4-10}
\hspace{1em}\hspace{1em}\hspace{1em} &  &  & OLS + HC0 & 0.317 & 0.353 & 0.986 & 0.317 & 0.315 & 0.970\\
\midrule
\cmidrule{1-10}
\addlinespace[0.3em]
\multicolumn{10}{l}{\textbf{Model-Based}}\\
\addlinespace[0.3em]
\multicolumn{10}{l}{\textit{ATE}}\\
\addlinespace[0.3em]
\multicolumn{10}{l}{IPW}\\
\hspace{1em}\hspace{1em}\hspace{1em} &  &  & M-Estimation & 0.281 & 0.397 & 0.982 & 0.280 & 0.324 & 0.958\\
\cmidrule{4-10}
\hspace{1em}\hspace{1em}\hspace{1em} &  &  & OLS + HC0 & 0.281 & 0.375 & 0.978 & 0.280 & 0.325 & 0.958\\
\cmidrule{3-10}
\addlinespace[0.3em]
\multicolumn{10}{l}{Entropy Balancing}\\
\hspace{1em}\hspace{1em}\hspace{1em} &  &  & M-Estimation & 0.278 & 0.321 & 0.949 & 0.278 & 0.321 & 0.949\\
\cmidrule{4-10}
\hspace{1em}\hspace{1em}\hspace{1em} &  &  & OLS + HC0 & 0.278 & 0.364 & 0.973 & 0.278 & 0.321 & 0.949\\
\midrule
\cmidrule{2-10}
\addlinespace[0.3em]
\multicolumn{10}{l}{\textit{ATT}}\\
\addlinespace[0.3em]
\multicolumn{10}{l}{IPW}\\
\hspace{1em}\hspace{1em}\hspace{1em} &  &  & M-Estimation & 0.300 & 0.415 & 0.990 & 0.296 & 0.323 & 0.964\\
\cmidrule{4-10}
\hspace{1em}\hspace{1em}\hspace{1em} &  &  & OLS + HC0 & 0.300 & 0.398 & 0.987 & 0.296 & 0.323 & 0.963\\
\cmidrule{3-10}
\addlinespace[0.3em]
\multicolumn{10}{l}{Entropy Balancing}\\
\hspace{1em}\hspace{1em}\hspace{1em} &  &  & M-Estimation & 0.296 & 0.323 & 0.964 & 0.296 & 0.323 & 0.964\\
\cmidrule{4-10}
\hspace{1em}\hspace{1em}\hspace{1em} &  &  & OLS + HC0 & 0.296 & 0.389 & 0.989 & 0.296 & 0.323 & 0.964\\
\midrule
\cmidrule{1-10}
\addlinespace[0.3em]
\multicolumn{10}{l}{\textbf{Superpopulation}}\\
\addlinespace[0.3em]
\multicolumn{10}{l}{\textit{PATE}}\\
\addlinespace[0.3em]
\multicolumn{10}{l}{IPW}\\
\hspace{1em}\hspace{1em}\hspace{1em} &  &  & M-Estimation & 0.338 & 0.336 & 0.950 & 0.327 & 0.324 & 0.946\\
\cmidrule{4-10}
\hspace{1em}\hspace{1em}\hspace{1em} &  &  & OLS + HC0 (+ Corr) & 0.338 & 0.359 & 0.963 & 0.327 & 0.325 & 0.947\\
\cmidrule{3-10}
\addlinespace[0.3em]
\multicolumn{10}{l}{Entropy Balancing}\\
\hspace{1em}\hspace{1em}\hspace{1em} &  &  & M-Estimation & 0.321 & 0.320 & 0.945 & 0.321 & 0.320 & 0.945\\
\cmidrule{4-10}
\hspace{1em}\hspace{1em}\hspace{1em} &  &  & OLS + HC0 (+ Corr) & 0.321 & 0.347 & 0.962 & 0.321 & 0.320 & 0.945\\
\midrule
\cmidrule{2-10}
\addlinespace[0.3em]
\multicolumn{10}{l}{\textit{PATT}}\\
\addlinespace[0.3em]
\multicolumn{10}{l}{IPW}\\
\hspace{1em}\hspace{1em}\hspace{1em} &  &  & M-Estimation & 0.345 & 0.345 & 0.948 & 0.337 & 0.335 & 0.948\\
\cmidrule{4-10}
\hspace{1em}\hspace{1em}\hspace{1em} &  &  & OLS + HC0 (+ Corr) & 0.345 & 0.351 & 0.953 & 0.337 & 0.335 & 0.948\\
\cmidrule{3-10}
\addlinespace[0.3em]
\multicolumn{10}{l}{Entropy Balancing}\\
\hspace{1em}\hspace{1em}\hspace{1em} &  &  & M-Estimation & 0.337 & 0.335 & 0.948 & 0.337 & 0.335 & 0.948\\
\cmidrule{4-10}
\hspace{1em}\hspace{1em}\hspace{1em} &  &  & OLS + HC0 (+ Corr) & 0.337 & 0.351 & 0.959 & 0.337 & 0.335 & 0.948\\
\midrule
\bottomrule
\end{tabular}}
\caption{\label{tab:mest_sim_results}Results for M-estimation simulations.  The column ``se$_{emp}$'' is the empirical standard error over simulations, ``Avg($\hat{se}$)'' is the average estimated standard error over simulations, and ``Coverage'' is the empirical coverage of the target estimand over simulations.  For design-based and model-based, this is the finite sample (S)ATE/ATT and for the superpopulation it is the PATE/PATT.  The superpopulation correction is only used for the residualized model.}
\end{table}

In the superpopulation DGP we use the estimating equations that account for estimating the population means for balancing covariates described in Section~\ref{sec:m_est_unknown_means}.  We see that all weighting approaches, outcome models, and estimand combinations maintain nominal coverage.  We also see that, for IPW weights, the residualized standard error is smaller than for the bivariate model, showing the improvement in precision possible with our proposed residualizing model.  

For entropy balancing, M-estimation will return identical standard error estimates for both a bivariate and our proposed residualized outcome model.  This is because the estimating equations, even for the bivariate model in Equation~\eqref{eq:unk-eb-bi}, account for the fact that the weights exactly balance the treatment and control group, thus the standard error in M-estimation reflects this balancing property in a way that OLS + HC0 standard errors will not.  IPW, on the other hand, does not enforce exact balance between the treatment and control groups, and the standard errors reflect this additional statistical uncertainty with the average bivariate standard errors greater than the ones for the residualized model.

We turn now to the design-based and model-based DGPs, where the sample is treated as fixed, and thus the target balancing means can be treated as fixed.  We see that in these scenarios, our fixed-mean estimating equations described in Section~\ref{sec:m_est_known_means} maintain nominal or conservative coverage across all weighting approaches, outcome models, and estimand combinations.  As with the superpopulation results, M-estimation will return identical standard error estimates for both the weighted bivariate and our proposed residualized outcome model in the fixed-means case; additionally, in the fixed-means case, M-estimation standard errors for our proposed residualized model are essentially identical to the residualized OLS + HC0 standard errors.

Finally, we consider the performance of the M-estimation approach to our proposed residualized regression with HC0 standard error (plus correction when necessary).  Both of these approaches are consistent estimators for the true variance of our proposed estimators, and you can see that performance for the two is very similar.  Again, this shows that while M-estimation can be used to correctly estimate appropriate standard errors, in both scenarios where weights are estimated and where they are considered ``fixed'', it is not the only way, nor a necessary way, to appropriately account for uncertainty when estimating weighted causal effects.  When the sample is treated as fixed, regression can be used to recover appropriate, or conservative, estimators.  When the sampling from a superpopulation, the ``uncertainty in the weights'' that the M-estimation literature seeks to address can be framed as an ``uncertainty in population means'' problem; regression plus our proposed superpopulation correction can also appropriately account for this uncertainty.

\section{Bootstrapped Standard Errors} \label{app:bootstrap}

In this section, we consider the bootstrap as an alternative to our proposed analytic approach and show via simulations that it recovers the uncertainty in estimating superpopulation quantities \citep{hernan2020causal, ding2023coursecausalinference}. The bootstrap procedure we consider bootstraps the whole procedure and re-calculates weights for each bootstrap resample. We expect this procedure to perform well because it accounts for the uncertainty in the estimation of the weights. Previous work has demonstrated the validity of this approach with IPW weights \citep{austin2016variance, austin2022bootstrap, kostouraki2024variance}. If estimating weights is too computationally intensive, researchers can consider only re-scaling the original sample weights when generating weights for the bootstrap sample \citep{zhang2025improving}.

These simulations follow the same setup as in the main text, and we focus on the heterogeneity case with entropy balancing and IPW weighting. With sample size of $n = 500$, we conduct 500 simulations following the superpopulation re-sampling paradigm. We set the number of bootstrap replicates to 1,000. We compare the bootstrap procedure to our superpopulation corrected standard errors, denoted \textit{Residualized}.

Table \ref{tab:sim_results_boot} contains the simulation results. Like the main text, we report the average standard error, ratio to the empirical standard error, and coverage to the superpopulation target (PATE or PATT). Although we use a superpopulation resampling paradigm, we also include coverage to the sample target (SATE or SATT) for illustration, denoted  \textit{Coverage$_s$}. 

We find that the bootstrap procedure and our analytic approach with a correction both achieve nominal coverage rates. Our proposed approach is comparable to the bootstrap procedure while being less computationally intensive. All approaches are overly conservative for the sample target. 

For matching estimators, \citet{abadie2008failure} show that the simple bootstrap fails since the  bootstrap distribution of the number of times each unit is used as a match differs from the original distribution. A variety of other bootstrap and analytic approaches have been proposed for matching estimators \citep{otsu2017bootstrap, ding2023coursecausalinference, abadie2012martingale, abadie2022robust}. 

\begin{table}
\centering
\caption{\label{tab:sim_results_boot}Bootstrap Simulation Results}
\centering
\resizebox{\ifdim\width>\linewidth\linewidth\else\width\fi}{!}{
\begin{tabular}[t]{llllll}
\toprule
\multicolumn{2}{c}{ } & \multicolumn{2}{c}{ATE} & \multicolumn{2}{c}{ATT} \\
\cmidrule(l{3pt}r{3pt}){3-4} \cmidrule(l{3pt}r{3pt}){5-6}
 &  & Residualized & Bootstrap & Residualized & Bootstrap\\
\midrule
\multirow{4}{*}{\rotatebox[origin=c]{90}{\textbf{EBal}}} & Avg($\hat{se}$) & 0.45 (0.03) & 0.46 (0.03) & 0.48 (0.03) & 0.48 (0.03)\\
 & $\hat{se}/se_{emp}$ & 0.99 & 1.01 & 0.99 & 1.00\\
 & Coverage (\%) & 95.2 (1.0) & 95.0 (1.0) & 95.6 (0.9) & 95.8 (0.9)\\
 & Coverage$_s$ (\%) & 97.0 (0.8) & 97.4 (0.7) & 97.8 (0.7) & 98.2 (0.6)\\
\midrule
\multirow{4}{*}{\rotatebox[origin=c]{90}{\textbf{IPW}}} & Avg($\hat{se}$) & 0.46 (0.03) & 0.48 (0.06) & 0.48 (0.03) & 0.50 (0.04)\\
 & $\hat{se}/se_{emp}$ & 0.99 & 1.01 & 0.99 & 1.00\\
 & Coverage (\%) & 94.8 (1.0) & 95.6 (0.9) & 95.6 (0.9) & 95.2 (1.0)\\
 & Coverage$_s$ (\%) & 96.0 (0.9) & 97.4 (0.7) & 97.8 (0.7) & 98.2 (0.6)\\
\bottomrule
\end{tabular}}
\end{table}

\clearpage

\section{Residualization as augmented IPW: derivations}\label{app:AIPW}

Throughout, weights are normalized within arm, $\sum_{i:Z_i=z} w_i = 1$, and $\hat{m}_z(\cdot)$ are the arm-specific fits in Equation \eqref{eq:armfits}, which coincide with arm-by-arm weighted least squares because the estimation in Equation~\eqref{eq:wResModel} is fully interacted with $Z$.

\subsection{\texorpdfstring{Decomposition of $\hat{\tau}$ into $\wdim$ and slope $\times$ imbalance (Eq.~\ref{eq:decomp})}{Decomposition of tau-hat into wDIM and slope x imbalance}}

Write both fits in Equation \eqref{eq:armfits} as $\hat{m}_z(\phi) = \hat{c}_z + (\phi - \phibstar)^\top \hat{b}_z$, where $\hat{c}_0 = \hat\beta_0$ and $\hat{c}_1 = \hat\beta_0 + \hat\tau$ are the arm intercepts and $\hat{b}_0 = \hat\beta$, $\hat{b}_1 = \hat\beta + \hat\gamma$ the arm slopes; note $\hat{c}_z = \hat{m}_z(\phibstar)$, by substituting $\phi = \phibstar$, so that $\hat\tau = \hat{m}_1(\phibstar) - \hat{m}_0(\phibstar)$. Starting from the intercept first-order condition in arm $z$,

\begin{align*}
\overline{Y}^w_z &= \sum_{i:Z_i=z} w_i\, \hat{m}_z(\phi_i)
  && \text{(intercept first-order condition)} \\
&= \sum_{i:Z_i=z} w_i \left[ \hat{c}_z + \big(\phi_i - \phibstar\big)^\top \hat{b}_z \right]
  && \text{(affine form of $\hat{m}_z$)} \\
&= \hat{c}_z + \big( \overline\phi^w_z - \phibstar \big)^{\!\top} \hat{b}_z
  && \text{(weights sum to one in arm $z$)} \\
&= \hat{m}_z(\phibstar) + \hat\delta_z^\top \hat{b}_z .
  && \text{($\hat\delta_z := \overline\phi^w_z - \phibstar$)}
\end{align*}
Rearranging, $\hat{m}_z(\phibstar) = \overline{Y}^w_z - \hat\delta_z^\top \hat{b}_z$; differencing the arms gives
$\hat\tau = \wdim - \hat\delta_1^\top \hat{b}_1 + \hat\delta_0^\top \hat{b}_0$, which is Equation \eqref{eq:decomp}.

\subsection{Equivalence to AIPW}

\paragraph{ATE.} Consider Equation \eqref{eq:aipw}. The intercept first-order condition of each arm's weighted regression is $\sum_{i:Z_i=z} w_i (Y_i - \hat{m}_z(\phi_i)) = 0$, so both augmentation sums are identically zero. For the first term, linearity of the fits and centering at the full-sample mean $\phibstar = n^{-1}\sum_i \phi_i$ give
\[
\frac{1}{n}\sum_i \big[\hat{m}_1(\phi_i) - \hat{m}_0(\phi_i)\big]
= \hat{m}_1(\phibstar) - \hat{m}_0(\phibstar) = \hat\tau .
\]

\paragraph{ATT.} With ATT weights (uniform on the treated, normalized odds weights on the controls) the H\'ajek AIPW-ATT estimator is

\[
\hat\tau_{AIPW} = \frac{1}{n_t}\sum_i Z_i \big(Y_i - \hat{m}_0(\phi_i)\big) \;-\; \sum_i w_i (1-Z_i) \big(Y_i - \hat{m}_0(\phi_i)\big).
\]

The second term vanishes by the control-arm intercept condition. For the first, with $\phibstar = \overline\phi_1$ (the treated mean) the treated-arm fit satisfies $\hat{m}_1(\overline\phi_1) = \overline{Y}_1$, so
$n_t^{-1} \sum_i Z_i (Y_i - \hat{m}_0(\phi_i)) = \overline{Y}_1 - \hat{m}_0(\overline\phi_1) = \hat{m}_1(\phibstar) - \hat{m}_0(\phibstar) = \hat\tau$. \hfill$\square$

\section{Simulation Results}\label{app:sim_figures}

This section presents Tables~\ref{tab:sim_results_heterogeneous} and \ref{tab:sim_results_homogeneous} with results for simulations for two other outcomes and with homogeneous treatment effects. We also include visualizations of these results to simplify interpretation. 

The three outcomes we consider are: 

\vspace{-0.5em}
\begin{itemize}\itemsep=-0.25em
\item Highly linear ($Y_1$): $Y_{1c} = X_1 + X_2 + X_3 - X_4 + X_5 + X_6 + \nu_1$
\item Moderately non-linearity ($Y_2$): $Y_{2c} = X_1 + X_2 + 0.2 X_3 X_4 - \sqrt{X_5} + \nu_2$
\item Severe non-linearity ($Y_3$): $Y_{3c} = (X_1 + X_2 + X_5)^2 + \nu_3$
\end{itemize}

\noindent with $\nu_1 \sim N(0, 25.8), \nu_2 \sim N(0, 17.1), \nu_3 \sim N(0, 181.1)$---values chosen so as to calibrate the overall linear signal to noise ratio to around 0.25 for the potential outcomes under control. The simulation setup for these outcomes is the same as in the main text. Table \ref{tab:sim_ess} contains the average effective sample size across simulations for each estimand, weighting method, and resampling process.  

\begin{table}
\centering
\caption{\label{tab:sim_ess}Mean effective sample size of the weights across simulations}
\centering
\begin{tabular}[t]{c>{}c|c>{}c|cc}
\toprule
\multicolumn{2}{c}{ } & \multicolumn{2}{c}{ATE} & \multicolumn{2}{c}{ATT} \\
\cmidrule(l{3pt}r{3pt}){3-4} \cmidrule(l{3pt}r{3pt}){5-6}
 &  & Treated & Control & Treated & Control\\
\midrule
 & EBal & 460.9 & 452.8 & 503.9 & 347.7\\
\cmidrule{2-6}
 & IPW & 455.2 & 446.2 & 503.9 & 346.4\\
\cmidrule{2-6}
\multirow{-3}{*}{\centering\arraybackslash Design-Based} & Matching & 366.9 & 358.8 & 503.9 & 198.6\\
\cmidrule{1-6}
 & EBal & 461.3 & 438.3 & 511.0 & 317.5\\
\cmidrule{2-6}
 & IPW & 456.3 & 414.7 & 511.0 & 291.1\\
\cmidrule{2-6}
\multirow{-3}{*}{\centering\arraybackslash Model-Based} & Matching & 373.2 & 356.5 & 511.0 & 201.8\\
\cmidrule{1-6}
 & EBal & 459.0 & 459.3 & 500.0 & 357.5\\
\cmidrule{2-6}
 & IPW & 453.3 & 454.2 & 500.0 & 357.8\\
\cmidrule{2-6}
\multirow{-3}{*}{\centering\arraybackslash Superpopulation} & Matching & 365.0 & 365.0 & 500.0 & 202.3\\
\bottomrule
\end{tabular}
\end{table}

\begin{landscape}\begin{table}[ht]

\centering
\caption{\label{tab:sim_results_heterogeneous}Results for heterogeneous simulations}
\centering
\resizebox{1.25\textwidth}{!}{%
\begin{tabular}[t]{c>{}c|c>{}cc>{}cc>{}cc>{}cc>{}cc>{}c}
\toprule
\multicolumn{2}{c}{ } & \multicolumn{6}{c}{ATE} & \multicolumn{6}{c}{ATT} \\
\cmidrule(l{3pt}r{3pt}){3-8} \cmidrule(l{3pt}r{3pt}){9-14}
\multicolumn{2}{c}{ } & \multicolumn{2}{c}{Y1} & \multicolumn{2}{c}{Y2} & \multicolumn{2}{c}{Y3} & \multicolumn{2}{c}{Y1} & \multicolumn{2}{c}{Y2} & \multicolumn{2}{c}{Y3} \\
\cmidrule(l{3pt}r{3pt}){3-4} \cmidrule(l{3pt}r{3pt}){5-6} \cmidrule(l{3pt}r{3pt}){7-8} \cmidrule(l{3pt}r{3pt}){9-10} \cmidrule(l{3pt}r{3pt}){11-12} \cmidrule(l{3pt}r{3pt}){13-14}
\rotatebox{-45}{} & \rotatebox{-45}{} & \rotatebox{-45}{Bivariate} & \rotatebox{-45}{Residualized} & \rotatebox{-45}{Bivariate} & \rotatebox{-45}{Residualized} & \rotatebox{-45}{Bivariate} & \rotatebox{-45}{Residualized} & \rotatebox{-45}{Bivariate} & \rotatebox{-45}{Residualized} & \rotatebox{-45}{Bivariate} & \rotatebox{-45}{Residualized} & \rotatebox{-45}{Bivariate} & \rotatebox{-45}{Residualized}\\
\midrule
\addlinespace[0.3em]
\multicolumn{14}{l}{\textbf{}}\\
\addlinespace[0.3em]
\multicolumn{14}{l}{\textbf{Design-Based}}\\
\hspace{1em}\hspace{1em} & Avg($\hat{se}$) & 0.45 (0.02) & \cellcolor{gray!20}{0.36 (0.01)} & 0.35 (0.01) & \cellcolor{gray!20}{0.31 (0.01)} & 1.33 (0.06) & \cellcolor{gray!20}{1.07 (0.03)} & 0.44 (0.02) & \cellcolor{gray!20}{0.37 (0.01)} & 0.35 (0.01) & \cellcolor{gray!20}{0.32 (0.01)} & 1.52 (0.14) & \cellcolor{gray!20}{1.12 (0.05)}\\
\cmidrule{2-14}
\hspace{1em}\hspace{1em} & $\hat{se}/se_{emp}$ & 1.27 & \cellcolor{gray!20}{1.02} & 1.16 & \cellcolor{gray!20}{1.04} & 1.24 & \cellcolor{gray!20}{0.99} & 1.18 & \cellcolor{gray!20}{0.99} & 1.11 & \cellcolor{gray!20}{0.99} & 1.33 & \cellcolor{gray!20}{0.98}\\
\cmidrule{2-14}
\hspace{1em}\hspace{1em}\multirow{-3}{*}{\centering\arraybackslash EBal} & Coverage & 98.6 (0.2) & \cellcolor{gray!20}{95.9 (0.3)} & 97.7 (0.2) & \cellcolor{gray!20}{96.0 (0.3)} & 98.4 (0.2) & \cellcolor{gray!20}{94.8 (0.3)} & 97.9 (0.2) & \cellcolor{gray!20}{95.0 (0.3)} & 97.0 (0.3) & \cellcolor{gray!20}{95.0 (0.3)} & 99.1 (0.2) & \cellcolor{gray!20}{94.4 (0.4)}\\
\cmidrule{1-14}
\hspace{1em}\hspace{1em} & Avg($\hat{se}$) & 0.46 (0.02) & \cellcolor{gray!20}{0.37 (0.01)} & 0.36 (0.02) & \cellcolor{gray!20}{0.32 (0.01)} & 1.35 (0.08) & \cellcolor{gray!20}{1.09 (0.03)} & 0.44 (0.02) & \cellcolor{gray!20}{0.37 (0.01)} & 0.35 (0.02) & \cellcolor{gray!20}{0.32 (0.01)} & 1.53 (0.16) & \cellcolor{gray!20}{1.12 (0.05)}\\
\cmidrule{2-14}
\hspace{1em}\hspace{1em} & $\hat{se}/se_{emp}$ & 1.26 & \cellcolor{gray!20}{1.03} & 1.16 & \cellcolor{gray!20}{1.04} & 1.23 & \cellcolor{gray!20}{0.99} & 1.18 & \cellcolor{gray!20}{0.99} & 1.10 & \cellcolor{gray!20}{0.99} & 1.20 & \cellcolor{gray!20}{0.98}\\
\cmidrule{2-14}
\hspace{1em}\hspace{1em}\multirow{-3}{*}{\centering\arraybackslash IPW} & Coverage & 98.6 (0.2) & \cellcolor{gray!20}{96.1 (0.3)} & 97.7 (0.2) & \cellcolor{gray!20}{96.2 (0.3)} & 98.4 (0.2) & \cellcolor{gray!20}{94.5 (0.4)} & 98.0 (0.2) & \cellcolor{gray!20}{95.0 (0.3)} & 96.7 (0.3) & \cellcolor{gray!20}{95.1 (0.3)} & 98.0 (0.2) & \cellcolor{gray!20}{94.3 (0.4)}\\
\cmidrule{1-14}
\hspace{1em}\hspace{1em} & Avg($\hat{se}$) & 0.51 (0.04) & \cellcolor{gray!20}{0.40 (0.02)} & 0.40 (0.03) & \cellcolor{gray!20}{0.35 (0.02)} & 1.50 (0.18) & \cellcolor{gray!20}{1.19 (0.07)} & 0.52 (0.04) & \cellcolor{gray!20}{0.43 (0.03)} & 0.41 (0.04) & \cellcolor{gray!20}{0.36 (0.03)} & 1.82 (0.38) & \cellcolor{gray!20}{1.31 (0.12)}\\
\cmidrule{2-14}
\hspace{1em}\hspace{1em} & $\hat{se}/se_{emp}$ & 1.19 & \cellcolor{gray!20}{1.00} & 1.13 & \cellcolor{gray!20}{1.02} & 1.18 & \cellcolor{gray!20}{0.99} & 1.12 & \cellcolor{gray!20}{0.97} & 1.09 & \cellcolor{gray!20}{0.97} & 1.13 & \cellcolor{gray!20}{0.97}\\
\cmidrule{2-14}
\hspace{1em}\hspace{1em}\multirow{-3}{*}{\centering\arraybackslash Matching} & Coverage & 97.9 (0.2) & \cellcolor{gray!20}{95.2 (0.3)} & 97.2 (0.3) & \cellcolor{gray!20}{95.4 (0.3)} & 97.6 (0.2) & \cellcolor{gray!20}{94.3 (0.4)} & 96.8 (0.3) & \cellcolor{gray!20}{93.7 (0.4)} & 96.4 (0.3) & \cellcolor{gray!20}{94.0 (0.4)} & 96.9 (0.3) & \cellcolor{gray!20}{94.0 (0.4)}\\
\cmidrule{1-14}
\addlinespace[0.3em]
\multicolumn{14}{l}{\textbf{}}\\
\addlinespace[0.3em]
\multicolumn{14}{l}{\textbf{Model-Based}}\\
\hspace{1em}\hspace{1em} & Avg($\hat{se}$) & 0.46 (0.01) & \cellcolor{gray!20}{0.37 (0.01)} & 0.36 (0.01) & \cellcolor{gray!20}{0.32 (0.01)} & 1.35 (0.03) & \cellcolor{gray!20}{1.10 (0.03)} & 0.46 (0.02) & \cellcolor{gray!20}{0.38 (0.02)} & 0.39 (0.03) & \cellcolor{gray!20}{0.32 (0.01)} & 1.56 (0.10) & \cellcolor{gray!20}{1.16 (0.07)}\\
\cmidrule{2-14}
\hspace{1em}\hspace{1em} & $\hat{se}/se_{emp}$ & 1.34 & \cellcolor{gray!20}{1.10} & 1.31 & \cellcolor{gray!20}{1.15} & 1.49 & \cellcolor{gray!20}{1.22} & 1.26 & \cellcolor{gray!20}{1.05} & 1.32 & \cellcolor{gray!20}{1.09} & 1.62 & \cellcolor{gray!20}{1.21}\\
\cmidrule{2-14}
\hspace{1em}\hspace{1em}\multirow{-3}{*}{\centering\arraybackslash EBal} & Coverage & 99.1 (0.1) & \cellcolor{gray!20}{96.3 (0.3)} & 99.0 (0.2) & \cellcolor{gray!20}{96.9 (0.3)} & 99.6 (0.1) & \cellcolor{gray!20}{97.9 (0.2)} & 98.4 (0.2) & \cellcolor{gray!20}{95.5 (0.3)} & 98.9 (0.2) & \cellcolor{gray!20}{96.4 (0.3)} & 99.8 (0.1) & \cellcolor{gray!20}{97.8 (0.2)}\\
\cmidrule{1-14}
\hspace{1em}\hspace{1em} & Avg($\hat{se}$) & 0.46 (0.02) & \cellcolor{gray!20}{0.38 (0.01)} & 0.38 (0.02) & \cellcolor{gray!20}{0.33 (0.01)} & 1.40 (0.06) & \cellcolor{gray!20}{1.14 (0.05)} & 0.47 (0.03) & \cellcolor{gray!20}{0.38 (0.02)} & 0.41 (0.04) & \cellcolor{gray!20}{0.32 (0.01)} & 1.65 (0.15) & \cellcolor{gray!20}{1.16 (0.08)}\\
\cmidrule{2-14}
\hspace{1em}\hspace{1em} & $\hat{se}/se_{emp}$ & 1.34 & \cellcolor{gray!20}{1.10} & 1.34 & \cellcolor{gray!20}{1.16} & 1.53 & \cellcolor{gray!20}{1.25} & 1.26 & \cellcolor{gray!20}{1.05} & 1.35 & \cellcolor{gray!20}{1.09} & 1.67 & \cellcolor{gray!20}{1.20}\\
\cmidrule{2-14}
\hspace{1em}\hspace{1em}\multirow{-3}{*}{\centering\arraybackslash IPW} & Coverage & 99.1 (0.1) & \cellcolor{gray!20}{96.3 (0.3)} & 99.1 (0.1) & \cellcolor{gray!20}{97.1 (0.3)} & 99.6 (0.1) & \cellcolor{gray!20}{98.3 (0.2)} & 98.4 (0.2) & \cellcolor{gray!20}{95.4 (0.3)} & 98.9 (0.2) & \cellcolor{gray!20}{96.1 (0.3)} & 99.9 (0.1) & \cellcolor{gray!20}{97.8 (0.2)}\\
\cmidrule{1-14}
\hspace{1em}\hspace{1em} & Avg($\hat{se}$) & 0.49 (0.02) & \cellcolor{gray!20}{0.41 (0.02)} & 0.39 (0.02) & \cellcolor{gray!20}{0.35 (0.01)} & 1.40 (0.06) & \cellcolor{gray!20}{1.20 (0.05)} & 0.52 (0.03) & \cellcolor{gray!20}{0.43 (0.03)} & 0.42 (0.03) & \cellcolor{gray!20}{0.36 (0.02)} & 1.64 (0.12) & \cellcolor{gray!20}{1.27 (0.08)}\\
\cmidrule{2-14}
\hspace{1em}\hspace{1em} & $\hat{se}/se_{emp}$ & 1.30 & \cellcolor{gray!20}{1.08} & 1.27 & \cellcolor{gray!20}{1.13} & 1.39 & \cellcolor{gray!20}{1.19} & 1.23 & \cellcolor{gray!20}{1.03} & 1.23 & \cellcolor{gray!20}{1.06} & 1.48 & \cellcolor{gray!20}{1.14}\\
\cmidrule{2-14}
\hspace{1em}\hspace{1em}\multirow{-3}{*}{\centering\arraybackslash Matching} & Coverage & 98.7 (0.2) & \cellcolor{gray!20}{96.3 (0.3)} & 98.5 (0.2) & \cellcolor{gray!20}{97.2 (0.3)} & 99.2 (0.1) & \cellcolor{gray!20}{97.9 (0.2)} & 98.3 (0.2) & \cellcolor{gray!20}{95.7 (0.3)} & 98.2 (0.2) & \cellcolor{gray!20}{96.4 (0.3)} & 99.4 (0.1) & \cellcolor{gray!20}{97.5 (0.2)}\\
\cmidrule{1-14}
\addlinespace[0.3em]
\multicolumn{14}{l}{\textbf{}}\\
\addlinespace[0.3em]
\multicolumn{14}{l}{\textbf{Superpopulation}}\\
\hspace{1em}\hspace{1em} & Avg($\hat{se}$) & 0.45 (0.03) & \cellcolor{gray!20}{0.37 (0.02)} & 0.35 (0.02) & \cellcolor{gray!20}{0.32 (0.02)} & 1.30 (0.12) & \cellcolor{gray!20}{1.10 (0.06)} & 0.44 (0.02) & \cellcolor{gray!20}{0.39 (0.02)} & 0.35 (0.02) & \cellcolor{gray!20}{0.33 (0.01)} & 1.45 (0.18) & \cellcolor{gray!20}{1.14 (0.08)}\\
\cmidrule{2-14}
\hspace{1em}\hspace{1em} & $\hat{se}/se_{emp}$ & 1.20 & \cellcolor{gray!20}{1.00} & 1.08 & \cellcolor{gray!20}{1.00} & 1.17 & \cellcolor{gray!20}{0.99} & 1.11 & \cellcolor{gray!20}{0.99} & 1.04 & \cellcolor{gray!20}{0.99} & 1.24 & \cellcolor{gray!20}{0.98}\\
\cmidrule{2-14}
\hspace{1em}\hspace{1em}\multirow{-3}{*}{\centering\arraybackslash EBal} & Coverage & 98.2 (0.2) & \cellcolor{gray!20}{95.0 (0.3)} & 96.5 (0.3) & \cellcolor{gray!20}{94.6 (0.4)} & 97.7 (0.2) & \cellcolor{gray!20}{94.9 (0.3)} & 97.2 (0.3) & \cellcolor{gray!20}{94.7 (0.4)} & 95.7 (0.3) & \cellcolor{gray!20}{94.8 (0.4)} & 98.2 (0.2) & \cellcolor{gray!20}{94.6 (0.4)}\\
\cmidrule{1-14}
\hspace{1em}\hspace{1em} & Avg($\hat{se}$) & 0.46 (0.06) & \cellcolor{gray!20}{0.38 (0.02)} & 0.36 (0.05) & \cellcolor{gray!20}{0.33 (0.02)} & 1.35 (0.17) & \cellcolor{gray!20}{1.13 (0.07)} & 0.44 (0.02) & \cellcolor{gray!20}{0.39 (0.02)} & 0.35 (0.02) & \cellcolor{gray!20}{0.33 (0.02)} & 1.46 (0.23) & \cellcolor{gray!20}{1.14 (0.08)}\\
\cmidrule{2-14}
\hspace{1em}\hspace{1em} & $\hat{se}/se_{emp}$ & 1.17 & \cellcolor{gray!20}{0.99} & 1.06 & \cellcolor{gray!20}{0.99} & 1.15 & \cellcolor{gray!20}{0.99} & 1.11 & \cellcolor{gray!20}{0.99} & 1.02 & \cellcolor{gray!20}{0.99} & 1.13 & \cellcolor{gray!20}{0.98}\\
\cmidrule{2-14}
\hspace{1em}\hspace{1em}\multirow{-3}{*}{\centering\arraybackslash IPW} & Coverage & 97.9 (0.2) & \cellcolor{gray!20}{94.9 (0.3)} & 96.3 (0.3) & \cellcolor{gray!20}{94.8 (0.4)} & 97.5 (0.2) & \cellcolor{gray!20}{94.6 (0.4)} & 97.2 (0.3) & \cellcolor{gray!20}{94.7 (0.4)} & 95.3 (0.3) & \cellcolor{gray!20}{94.8 (0.4)} & 97.2 (0.3) & \cellcolor{gray!20}{94.6 (0.4)}\\
\cmidrule{1-14}
\hspace{1em}\hspace{1em} & Avg($\hat{se}$) & 0.51 (0.06) & \cellcolor{gray!20}{0.41 (0.02)} & 0.40 (0.04) & \cellcolor{gray!20}{0.35 (0.02)} & 1.47 (0.21) & \cellcolor{gray!20}{1.23 (0.09)} & 0.51 (0.04) & \cellcolor{gray!20}{0.45 (0.03)} & 0.41 (0.04) & \cellcolor{gray!20}{0.38 (0.02)} & 1.71 (0.34) & \cellcolor{gray!20}{1.32 (0.12)}\\
\cmidrule{2-14}
\hspace{1em}\hspace{1em} & $\hat{se}/se_{emp}$ & 1.15 & \cellcolor{gray!20}{0.98} & 1.08 & \cellcolor{gray!20}{0.99} & 1.17 & \cellcolor{gray!20}{1.01} & 1.09 & \cellcolor{gray!20}{0.98} & 1.06 & \cellcolor{gray!20}{0.98} & 1.15 & \cellcolor{gray!20}{1.00}\\
\cmidrule{2-14}
\hspace{1em}\hspace{1em}\multirow{-3}{*}{\centering\arraybackslash Matching} & Coverage & 97.8 (0.2) & \cellcolor{gray!20}{94.8 (0.4)} & 96.3 (0.3) & \cellcolor{gray!20}{94.2 (0.4)} & 97.8 (0.2) & \cellcolor{gray!20}{95.0 (0.3)} & 96.7 (0.3) & \cellcolor{gray!20}{94.3 (0.4)} & 96.0 (0.3) & \cellcolor{gray!20}{94.2 (0.4)} & 97.5 (0.2) & \cellcolor{gray!20}{94.8 (0.3)}\\
\bottomrule
\end{tabular}
}
\end{table}
\end{landscape}

\begin{landscape}\begin{table}[ht]

\centering
\caption{\label{tab:sim_results_homogeneous}Results for homogeneous simulations}
\centering
\resizebox{1.25\textwidth}{!}{%
\begin{tabular}[t]{c>{}c|c>{}cc>{}cc>{}cc>{}cc>{}cc>{}c}
\toprule
\multicolumn{2}{c}{ } & \multicolumn{6}{c}{ATE} & \multicolumn{6}{c}{ATT} \\
\cmidrule(l{3pt}r{3pt}){3-8} \cmidrule(l{3pt}r{3pt}){9-14}
\multicolumn{2}{c}{ } & \multicolumn{2}{c}{Y1} & \multicolumn{2}{c}{Y2} & \multicolumn{2}{c}{Y3} & \multicolumn{2}{c}{Y1} & \multicolumn{2}{c}{Y2} & \multicolumn{2}{c}{Y3} \\
\cmidrule(l{3pt}r{3pt}){3-4} \cmidrule(l{3pt}r{3pt}){5-6} \cmidrule(l{3pt}r{3pt}){7-8} \cmidrule(l{3pt}r{3pt}){9-10} \cmidrule(l{3pt}r{3pt}){11-12} \cmidrule(l{3pt}r{3pt}){13-14}
\rotatebox{-45}{} & \rotatebox{-45}{} & \rotatebox{-45}{Bivariate} & \rotatebox{-45}{Residualized} & \rotatebox{-45}{Bivariate} & \rotatebox{-45}{Residualized} & \rotatebox{-45}{Bivariate} & \rotatebox{-45}{Residualized} & \rotatebox{-45}{Bivariate} & \rotatebox{-45}{Residualized} & \rotatebox{-45}{Bivariate} & \rotatebox{-45}{Residualized} & \rotatebox{-45}{Bivariate} & \rotatebox{-45}{Residualized}\\
\midrule
\addlinespace[0.3em]
\multicolumn{14}{l}{\textbf{}}\\
\addlinespace[0.3em]
\multicolumn{14}{l}{\textbf{Design-Based}}\\
\hspace{1em}\hspace{1em} & Avg($\hat{se}$) & 0.39 (0.01) & \cellcolor{gray!20}{0.33 (0.01)} & 0.31 (0.01) & \cellcolor{gray!20}{0.27 (0.00)} & 1.34 (0.05) & \cellcolor{gray!20}{1.10 (0.03)} & 0.40 (0.01) & \cellcolor{gray!20}{0.35 (0.01)} & 0.32 (0.01) & \cellcolor{gray!20}{0.29 (0.01)} & 1.55 (0.14) & \cellcolor{gray!20}{1.14 (0.05)}\\
\cmidrule{2-14}
\hspace{1em}\hspace{1em} & $\hat{se}/se_{emp}$ & 1.15 & \cellcolor{gray!20}{0.99} & 1.12 & \cellcolor{gray!20}{0.99} & 1.20 & \cellcolor{gray!20}{0.99} & 1.13 & \cellcolor{gray!20}{0.99} & 1.11 & \cellcolor{gray!20}{0.99} & 1.33 & \cellcolor{gray!20}{0.98}\\
\cmidrule{2-14}
\hspace{1em}\hspace{1em}\multirow{-3}{*}{\centering\arraybackslash EBal} & Coverage & 97.6 (0.2) & \cellcolor{gray!20}{94.9 (0.3)} & 97.4 (0.3) & \cellcolor{gray!20}{95.1 (0.3)} & 98.2 (0.2) & \cellcolor{gray!20}{94.8 (0.4)} & 97.5 (0.2) & \cellcolor{gray!20}{94.7 (0.4)} & 97.4 (0.2) & \cellcolor{gray!20}{94.8 (0.4)} & 99.0 (0.2) & \cellcolor{gray!20}{94.4 (0.4)}\\
\cmidrule{1-14}
\hspace{1em}\hspace{1em} & Avg($\hat{se}$) & 0.39 (0.01) & \cellcolor{gray!20}{0.33 (0.01)} & 0.32 (0.01) & \cellcolor{gray!20}{0.27 (0.01)} & 1.36 (0.07) & \cellcolor{gray!20}{1.12 (0.03)} & 0.40 (0.02) & \cellcolor{gray!20}{0.35 (0.01)} & 0.32 (0.02) & \cellcolor{gray!20}{0.29 (0.01)} & 1.56 (0.16) & \cellcolor{gray!20}{1.14 (0.05)}\\
\cmidrule{2-14}
\hspace{1em}\hspace{1em} & $\hat{se}/se_{emp}$ & 1.16 & \cellcolor{gray!20}{0.99} & 1.13 & \cellcolor{gray!20}{0.99} & 1.20 & \cellcolor{gray!20}{0.99} & 1.12 & \cellcolor{gray!20}{0.99} & 1.10 & \cellcolor{gray!20}{0.99} & 1.19 & \cellcolor{gray!20}{0.98}\\
\cmidrule{2-14}
\hspace{1em}\hspace{1em}\multirow{-3}{*}{\centering\arraybackslash IPW} & Coverage & 97.8 (0.2) & \cellcolor{gray!20}{95.0 (0.3)} & 97.6 (0.2) & \cellcolor{gray!20}{95.1 (0.3)} & 98.1 (0.2) & \cellcolor{gray!20}{94.6 (0.4)} & 97.3 (0.3) & \cellcolor{gray!20}{94.6 (0.4)} & 97.4 (0.3) & \cellcolor{gray!20}{94.8 (0.4)} & 98.0 (0.2) & \cellcolor{gray!20}{94.4 (0.4)}\\
\cmidrule{1-14}
\hspace{1em}\hspace{1em} & Avg($\hat{se}$) & 0.43 (0.02) & \cellcolor{gray!20}{0.37 (0.02)} & 0.35 (0.02) & \cellcolor{gray!20}{0.30 (0.01)} & 1.50 (0.17) & \cellcolor{gray!20}{1.23 (0.07)} & 0.48 (0.04) & \cellcolor{gray!20}{0.41 (0.03)} & 0.39 (0.04) & \cellcolor{gray!20}{0.34 (0.03)} & 1.85 (0.38) & \cellcolor{gray!20}{1.33 (0.11)}\\
\cmidrule{2-14}
\hspace{1em}\hspace{1em} & $\hat{se}/se_{emp}$ & 1.12 & \cellcolor{gray!20}{0.97} & 1.12 & \cellcolor{gray!20}{0.98} & 1.15 & \cellcolor{gray!20}{0.98} & 1.07 & \cellcolor{gray!20}{0.96} & 1.08 & \cellcolor{gray!20}{0.96} & 1.12 & \cellcolor{gray!20}{0.97}\\
\cmidrule{2-14}
\hspace{1em}\hspace{1em}\multirow{-3}{*}{\centering\arraybackslash Matching} & Coverage & 97.1 (0.3) & \cellcolor{gray!20}{94.3 (0.4)} & 97.0 (0.3) & \cellcolor{gray!20}{94.5 (0.4)} & 97.2 (0.3) & \cellcolor{gray!20}{94.2 (0.4)} & 96.0 (0.3) & \cellcolor{gray!20}{93.9 (0.4)} & 96.4 (0.3) & \cellcolor{gray!20}{94.0 (0.4)} & 96.8 (0.3) & \cellcolor{gray!20}{94.1 (0.4)}\\
\cmidrule{1-14}
\addlinespace[0.3em]
\multicolumn{14}{l}{\textbf{}}\\
\addlinespace[0.3em]
\multicolumn{14}{l}{\textbf{Model-Based}}\\
\hspace{1em}\hspace{1em} & Avg($\hat{se}$) & 0.39 (0.01) & \cellcolor{gray!20}{0.34 (0.01)} & 0.32 (0.01) & \cellcolor{gray!20}{0.27 (0.01)} & 1.36 (0.03) & \cellcolor{gray!20}{1.13 (0.03)} & 0.42 (0.02) & \cellcolor{gray!20}{0.36 (0.02)} & 0.36 (0.03) & \cellcolor{gray!20}{0.29 (0.01)} & 1.59 (0.10) & \cellcolor{gray!20}{1.18 (0.07)}\\
\cmidrule{2-14}
\hspace{1em}\hspace{1em} & $\hat{se}/se_{emp}$ & 1.15 & \cellcolor{gray!20}{0.98} & 1.14 & \cellcolor{gray!20}{0.99} & 1.51 & \cellcolor{gray!20}{1.25} & 1.15 & \cellcolor{gray!20}{0.98} & 1.21 & \cellcolor{gray!20}{0.98} & 1.65 & \cellcolor{gray!20}{1.23}\\
\cmidrule{2-14}
\hspace{1em}\hspace{1em}\multirow{-3}{*}{\centering\arraybackslash EBal} & Coverage & 97.0 (0.3) & \cellcolor{gray!20}{94.5 (0.4)} & 96.8 (0.3) & \cellcolor{gray!20}{94.6 (0.4)} & 99.6 (0.1) & \cellcolor{gray!20}{98.4 (0.2)} & 97.2 (0.3) & \cellcolor{gray!20}{93.8 (0.4)} & 97.9 (0.2) & \cellcolor{gray!20}{93.8 (0.4)} & 99.9 (0.0) & \cellcolor{gray!20}{98.0 (0.2)}\\
\cmidrule{1-14}
\hspace{1em}\hspace{1em} & Avg($\hat{se}$) & 0.40 (0.02) & \cellcolor{gray!20}{0.34 (0.01)} & 0.34 (0.02) & \cellcolor{gray!20}{0.28 (0.01)} & 1.42 (0.06) & \cellcolor{gray!20}{1.18 (0.05)} & 0.43 (0.03) & \cellcolor{gray!20}{0.36 (0.02)} & 0.38 (0.04) & \cellcolor{gray!20}{0.29 (0.02)} & 1.68 (0.14) & \cellcolor{gray!20}{1.18 (0.08)}\\
\cmidrule{2-14}
\hspace{1em}\hspace{1em} & $\hat{se}/se_{emp}$ & 1.17 & \cellcolor{gray!20}{0.98} & 1.20 & \cellcolor{gray!20}{0.99} & 1.54 & \cellcolor{gray!20}{1.29} & 1.16 & \cellcolor{gray!20}{0.97} & 1.25 & \cellcolor{gray!20}{0.98} & 1.70 & \cellcolor{gray!20}{1.23}\\
\cmidrule{2-14}
\hspace{1em}\hspace{1em}\multirow{-3}{*}{\centering\arraybackslash IPW} & Coverage & 97.2 (0.3) & \cellcolor{gray!20}{94.4 (0.4)} & 97.7 (0.2) & \cellcolor{gray!20}{94.6 (0.4)} & 99.7 (0.1) & \cellcolor{gray!20}{98.6 (0.2)} & 97.4 (0.3) & \cellcolor{gray!20}{93.7 (0.4)} & 98.2 (0.2) & \cellcolor{gray!20}{93.9 (0.4)} & 99.9 (0.0) & \cellcolor{gray!20}{98.0 (0.2)}\\
\cmidrule{1-14}
\hspace{1em}\hspace{1em} & Avg($\hat{se}$) & 0.43 (0.02) & \cellcolor{gray!20}{0.37 (0.02)} & 0.35 (0.02) & \cellcolor{gray!20}{0.30 (0.01)} & 1.43 (0.06) & \cellcolor{gray!20}{1.24 (0.05)} & 0.48 (0.04) & \cellcolor{gray!20}{0.41 (0.03)} & 0.39 (0.03) & \cellcolor{gray!20}{0.33 (0.02)} & 1.67 (0.12) & \cellcolor{gray!20}{1.29 (0.07)}\\
\cmidrule{2-14}
\hspace{1em}\hspace{1em} & $\hat{se}/se_{emp}$ & 1.14 & \cellcolor{gray!20}{0.97} & 1.14 & \cellcolor{gray!20}{0.98} & 1.42 & \cellcolor{gray!20}{1.23} & 1.15 & \cellcolor{gray!20}{0.97} & 1.14 & \cellcolor{gray!20}{0.97} & 1.50 & \cellcolor{gray!20}{1.16}\\
\cmidrule{2-14}
\hspace{1em}\hspace{1em}\multirow{-3}{*}{\centering\arraybackslash Matching} & Coverage & 97.1 (0.3) & \cellcolor{gray!20}{94.1 (0.4)} & 97.0 (0.3) & \cellcolor{gray!20}{94.2 (0.4)} & 99.3 (0.1) & \cellcolor{gray!20}{98.3 (0.2)} & 97.2 (0.3) & \cellcolor{gray!20}{94.3 (0.4)} & 97.4 (0.3) & \cellcolor{gray!20}{94.4 (0.4)} & 99.5 (0.1) & \cellcolor{gray!20}{97.7 (0.2)}\\
\cmidrule{1-14}
\addlinespace[0.3em]
\multicolumn{14}{l}{\textbf{}}\\
\addlinespace[0.3em]
\multicolumn{14}{l}{\textbf{Superpopulation}}\\
\hspace{1em}\hspace{1em} & Avg($\hat{se}$) & 0.38 (0.01) & \cellcolor{gray!20}{0.33 (0.01)} & 0.31 (0.01) & \cellcolor{gray!20}{0.27 (0.01)} & 1.29 (0.11) & \cellcolor{gray!20}{1.11 (0.06)} & 0.40 (0.02) & \cellcolor{gray!20}{0.35 (0.01)} & 0.33 (0.02) & \cellcolor{gray!20}{0.29 (0.01)} & 1.46 (0.18) & \cellcolor{gray!20}{1.15 (0.08)}\\
\cmidrule{2-14}
\hspace{1em}\hspace{1em} & $\hat{se}/se_{emp}$ & 1.15 & \cellcolor{gray!20}{0.99} & 1.14 & \cellcolor{gray!20}{1.00} & 1.14 & \cellcolor{gray!20}{0.99} & 1.14 & \cellcolor{gray!20}{0.99} & 1.15 & \cellcolor{gray!20}{0.99} & 1.25 & \cellcolor{gray!20}{0.98}\\
\cmidrule{2-14}
\hspace{1em}\hspace{1em}\multirow{-3}{*}{\centering\arraybackslash EBal} & Coverage & 97.8 (0.2) & \cellcolor{gray!20}{95.1 (0.3)} & 97.4 (0.3) & \cellcolor{gray!20}{95.0 (0.3)} & 97.4 (0.3) & \cellcolor{gray!20}{94.5 (0.4)} & 97.9 (0.2) & \cellcolor{gray!20}{95.0 (0.3)} & 97.9 (0.2) & \cellcolor{gray!20}{95.1 (0.3)} & 98.2 (0.2) & \cellcolor{gray!20}{94.5 (0.4)}\\
\cmidrule{1-14}
\hspace{1em}\hspace{1em} & Avg($\hat{se}$) & 0.39 (0.01) & \cellcolor{gray!20}{0.33 (0.01)} & 0.32 (0.01) & \cellcolor{gray!20}{0.27 (0.01)} & 1.32 (0.15) & \cellcolor{gray!20}{1.15 (0.07)} & 0.40 (0.02) & \cellcolor{gray!20}{0.35 (0.01)} & 0.33 (0.02) & \cellcolor{gray!20}{0.29 (0.01)} & 1.47 (0.23) & \cellcolor{gray!20}{1.15 (0.08)}\\
\cmidrule{2-14}
\hspace{1em}\hspace{1em} & $\hat{se}/se_{emp}$ & 1.15 & \cellcolor{gray!20}{0.99} & 1.16 & \cellcolor{gray!20}{1.00} & 1.13 & \cellcolor{gray!20}{0.99} & 1.12 & \cellcolor{gray!20}{0.99} & 1.12 & \cellcolor{gray!20}{0.99} & 1.13 & \cellcolor{gray!20}{0.98}\\
\cmidrule{2-14}
\hspace{1em}\hspace{1em}\multirow{-3}{*}{\centering\arraybackslash IPW} & Coverage & 97.6 (0.2) & \cellcolor{gray!20}{95.0 (0.3)} & 97.6 (0.2) & \cellcolor{gray!20}{95.0 (0.3)} & 97.3 (0.3) & \cellcolor{gray!20}{94.5 (0.4)} & 97.6 (0.2) & \cellcolor{gray!20}{94.9 (0.3)} & 97.4 (0.3) & \cellcolor{gray!20}{95.1 (0.3)} & 97.1 (0.3) & \cellcolor{gray!20}{94.4 (0.4)}\\
\cmidrule{1-14}
\hspace{1em}\hspace{1em} & Avg($\hat{se}$) & 0.43 (0.02) & \cellcolor{gray!20}{0.37 (0.02)} & 0.36 (0.02) & \cellcolor{gray!20}{0.30 (0.01)} & 1.45 (0.20) & \cellcolor{gray!20}{1.25 (0.09)} & 0.48 (0.04) & \cellcolor{gray!20}{0.41 (0.03)} & 0.40 (0.04) & \cellcolor{gray!20}{0.34 (0.03)} & 1.72 (0.34) & \cellcolor{gray!20}{1.32 (0.12)}\\
\cmidrule{2-14}
\hspace{1em}\hspace{1em} & $\hat{se}/se_{emp}$ & 1.12 & \cellcolor{gray!20}{0.97} & 1.14 & \cellcolor{gray!20}{0.97} & 1.14 & \cellcolor{gray!20}{1.00} & 1.10 & \cellcolor{gray!20}{0.97} & 1.14 & \cellcolor{gray!20}{0.97} & 1.15 & \cellcolor{gray!20}{1.00}\\
\cmidrule{2-14}
\hspace{1em}\hspace{1em}\multirow{-3}{*}{\centering\arraybackslash Matching} & Coverage & 97.2 (0.3) & \cellcolor{gray!20}{94.5 (0.4)} & 97.4 (0.3) & \cellcolor{gray!20}{94.5 (0.4)} & 97.6 (0.2) & \cellcolor{gray!20}{94.8 (0.4)} & 96.8 (0.3) & \cellcolor{gray!20}{94.1 (0.4)} & 97.4 (0.3) & \cellcolor{gray!20}{94.3 (0.4)} & 97.5 (0.2) & \cellcolor{gray!20}{94.5 (0.4)}\\
\bottomrule
\end{tabular}
}
\end{table}
\end{landscape}

\begin{figure}[!ht]
    \centering
    \includegraphics[width=\textwidth]{"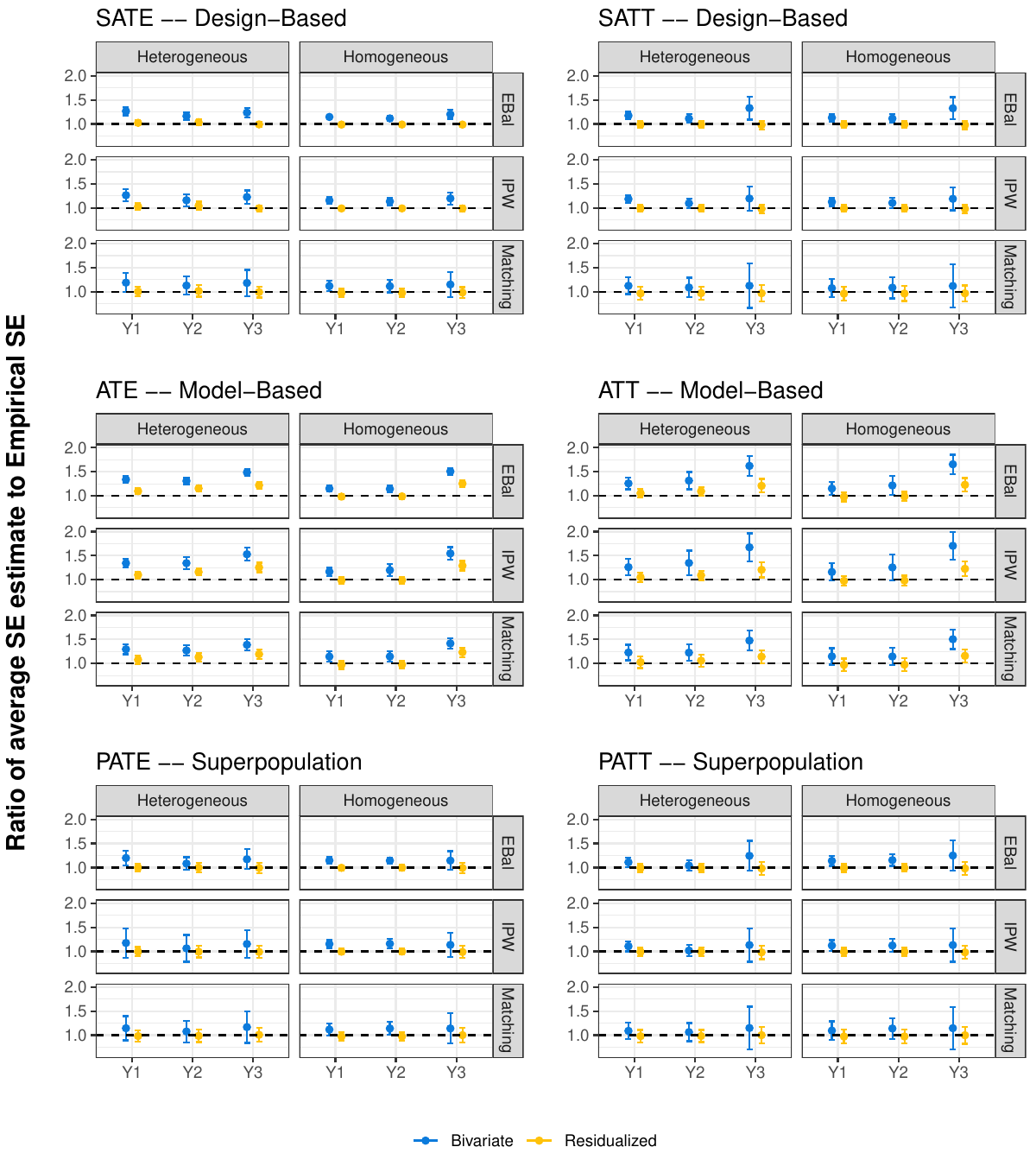"}%
    \caption{Precision gains for residualized standard errors under different resampling paradigms.  Intervals represent 95\% confidence intervals for simulation uncertainty.  Target estimand and sampling frame listed by panel.}
    \label{fig:sims_SEs}
\end{figure}

\begin{figure}[!ht]
    \centering
    \includegraphics[width=\textwidth]{"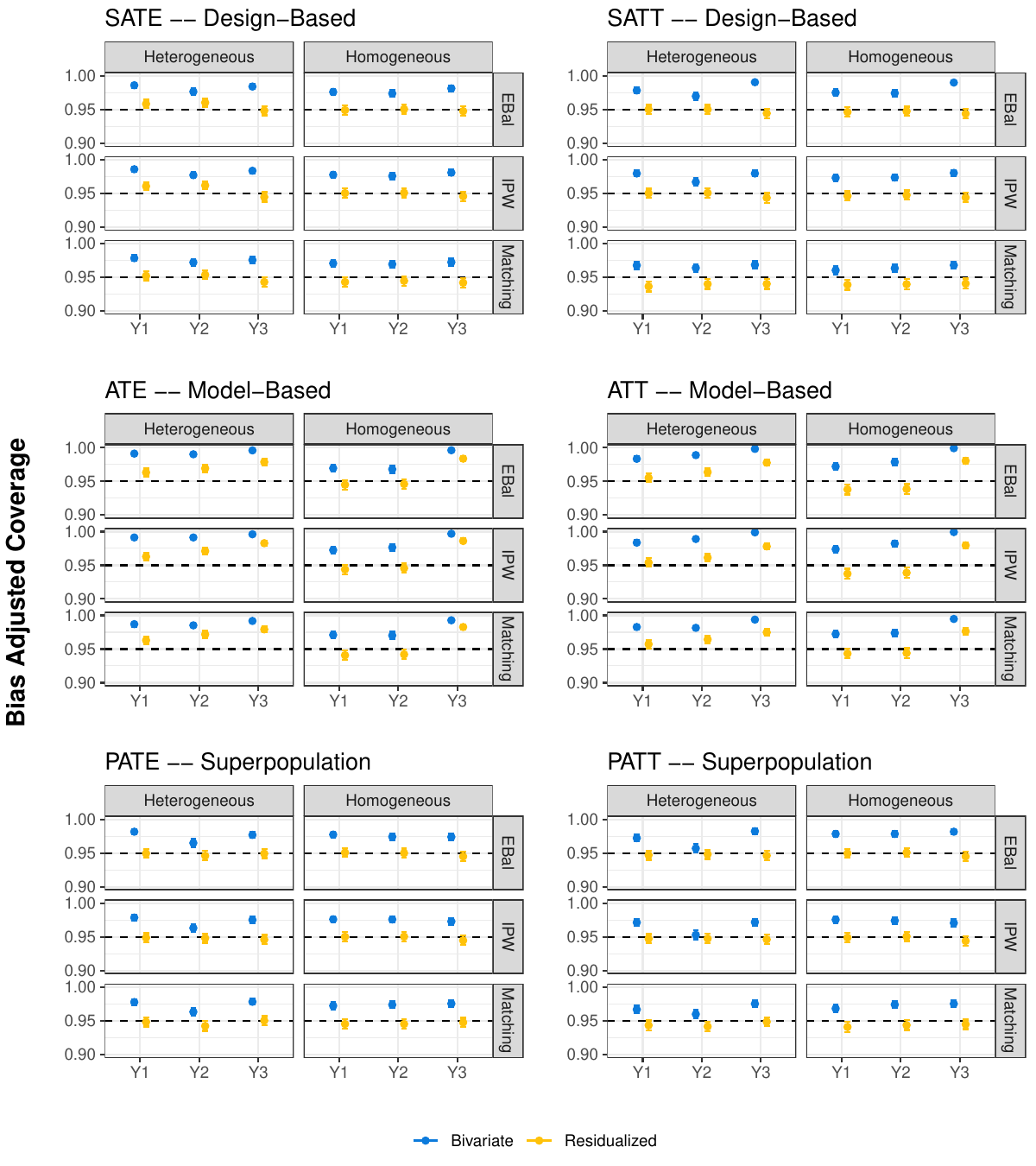"}%
    \caption{Bias-adjusted coverage for estimated standard errors under different resampling paradigms.  Intervals represent 95\% confidence intervals for simulation uncertainty. Target estimand and sampling frame listed by panel.}
    \label{fig:ba_cov}
\end{figure}

\FloatBarrier

\section{Empirical Applications}\label{app:application}

\subsection{Original Results}

In Table \ref{tab:replication_summary} we present results from existing analyses of the three data sets in Section \ref{sec:app}. 

\citet{ladd2009exploiting} first study the persuasive power of the news media in the 1997 British election using matching with two different specifications. The first is genetic matching with all 39 covariates, while the second only uses 9 categorical variables in exact matching. Table \ref{tab:replication_summary} contains their reported point estimates and standard errors from both analyses on all readers (found in Table 2 in Ladd and Lenz (2009)). A re-analysis by \citet{hainmueller_entropy_2012} uses entropy balancing on all 39 covariates and reports the results on page 43 of the main text. 

\citet{chan2016globally} analyze the effect of school lunch on BMI using various weighting approaches. Below we include their reported results from calibration weighting with four different loss functions. Cal (ET), Cal (EL), Cal (Q) and Cal (IL) are exponential tilting, empirical likelihood, quadratic, and inverse logistic, respectively. The results can be found in Table 6 of the manuscript.

\citet{black2016courting} originally study the effect of promotion prospects to the Supreme Court on the behavior of circuit court judges separately for contender and non-contender judges. They use coarsened exact matching on seven covariates. For contender judges, they report an absolute difference of 0.09 on page 38. Using the replication package, we calculate the standard error and confidence interval presented in Figure 1. For non-contender judges, we use the replication package to report point estimates, standard errors, and confidence intervals. \citet{xu2023hierarchically} uses hierarchically regularized entropy balancing to re-analyze the data. We use the replication code with the specific commit hash of the \texttt{hbal} package to calculate the point estimates and confidence intervals represented in their Figure 2.

\begin{table}[!ht]
\centering
 \resizebox{\columnwidth}{!}{
\begin{tabular}{llllllll}
\toprule
\textbf{Data} & \textbf{Estimand} & \textbf{Analysis} & \textbf{Cov.} & \textbf{Weighting} & \textbf{Point Est.} & \textbf{SE} & \textbf{CI} \\ 
\midrule
\multirow{3}{*}{News} & \multirow{3}{*}{ATT} & Ladd \& Lenz (2009) & 39 & GenMatch & 0.10 & 0.04 & (0.02, 0.19) \\
 &  & Ladd \& Lenz (2009) & 9 & Exact Match & 0.11 & 0.04 & (0.03, 0.19) \\
 &  & Hainmueller (2012) & 39 & EBal & 0.12 & 0.04 & (0.04, 0.20) \\
\midrule
\multirow{4}{*}{School lunch} & \multirow{4}{*}{ATE} & \multirow{4}{*}{Chan et al. (2016)} & 11 & Cal (ET) & -0.04 & 0.22 & (-0.48, 0.40) \\
 & & & 11 & Cal (EL) & -0.06 & 0.22 & (-0.50, 0.38) \\
 & & & 11 & Cal (Q) & 0.00 & 0.22 & (-0.44, 0.44) \\
 & & & 11 & Cal (IL) & -0.08 & 0.22 & (-0.52, 0.36) \\
\midrule
\multirow{2}{*}{\shortstack[l]{Promotion \\ Contender}} & \multirow{2}{*}{ATT} & Black \& Owens (2016) & 7 & CEM & 0.09 & 0.03 & (0.03, 0.14) \\
\addlinespace
 &  & Xu \& Yang (2023) & 7 & HBal & 0.03 & 0.01 & (0.00, 0.05) \\
\midrule
\multirow{2}{*}{\shortstack[l]{Promotion \\ Non-contender}} & \multirow{2}{*}{ATT} & Black \& Owens (2016) & 6 & CEM & 0.01 & 0.01 & (-0.02, 0.03)\\
\addlinespace
 &  & Xu \& Yang (2023) & 6 & HBal & 0.01 & 0.01 & (-0.01, 0.03) \\
\bottomrule
\end{tabular}
}
\caption{Results from existing analyses of empirical applications. Confidence intervals are at the 95\% confidence level. }
\label{tab:replication_summary}
\end{table}

\subsection{\texorpdfstring{Covariate-level balance and $R^2$}{Covariate-level balance and R2}}\label{app:application-diagnostics}
Tables \ref{tab:ladd_lenz_app_tab}, \ref{tab:chan_app_tab}, \ref{tab:black_app_tab_con}, and \ref{tab:black_app_tab_noncon} present covariate balance and partial $R^2$ values for the three application studies. For each study, in the below tables we present the initial imbalance and remaining imbalance after IPW and matching in terms of the standardized mean difference. Note that entropy balancing is an exact balancing procedure, so there is no imbalance. We also include the partial $R^2$ of each covariate on the outcome from a weighted regression. Since we employ entropy balancing (EB), IPW, and matching, we include the partial $R^2$ from each of these models, referred to as Partial $R^2_{Y\sim X}$-EB, Partial $R^2_{Y\sim X}$-IPW, and Partial $R^2_{Y\sim X}$-Mat respectively. The last two columns contain the partial $R^2$ of each covariate from a weighted regression of treatment on covariates with the IPW and matching weights, denoted $R^2_{Z\sim X}-$IPW and $R^2_{Z\sim X}$-Mat. These partial $R^2$ values help explain which covariates contribute most to the reduction in standard errors through residualization. For IPW and matching, they can also reveal how the point estimate might change between the unadjusted and adjusted regressions.  

\begin{table}[!ht]
\centering
 \resizebox{\columnwidth}{!}{
\begin{tabular}{lcccccccc}
\toprule
Covariate & Initial Imbalance & Post-IPW Imbalance & Post-Match Imbalance & $R^2_{Y\sim X}$-EB & $R^2_{Y\sim X}$-IPW & $R^2_{Y\sim X}$-Mat & $R^2_{Z\sim X}$-IPW & $R^2_{Z\sim X}$-Mat\\
\midrule
conservative & -0.013 & 0.035 & 0.027 & 0.002 & 0.003 & 0.003 & 0.00014 & 0.00044\\
labor & 0.046 & -0.050 & -0.090 & 0.012 & 0.011 & 0.013 & 0.00058 & 0.00179\\
liberal & -0.063 & -0.011 & 0.025 & 0.006 & 0.002 & 0.011 & 0.00034 & 0.00014\\
white & 0.082 & 0.016 & -0.010 & 0.002 & 0.000 & 0.002 & 0.00007 & 0.00000\\
wkclass & 0.297 & -0.007 & -0.032 & 0.006 & 0.008 & 0.003 & 0.00000 & 0.00013\\
parent labor & 0.166 & -0.006 & -0.035 & 0.000 & 0.000 & 0.001 & 0.00012 & 0.00004\\
f ideo92 & -0.011 & 0.001 & 0.032 & 0.000 & 0.003 & 0.000 & 0.00009 & 0.00000\\
vote l 92 & 0.135 & -0.030 & -0.028 & 0.018 & 0.008 & 0.008 & 0.00001 & 0.00055\\
vote c 92 & -0.031 & 0.037 & 0.053 & 0.000 & 0.002 & 0.001 & 0.00002 & 0.00035\\
vote lib 92 & -0.087 & -0.007 & -0.030 & 0.001 & 0.004 & 0.003 & 0.00003 & 0.00003\\
labfel92 & 0.076 & -0.032 & -0.041 & 0.006 & 0.005 & 0.010 & 0.00001 & 0.00008\\
confel92 & 0.009 & 0.042 & 0.050 & 0.000 & 0.000 & 0.001 & 0.00021 & 0.00013\\
know 3 & -0.329 & -0.018 & 0.080 & 0.001 & 0.009 & 0.002 & 0.00007 & 0.00018\\
TVnewseither & -0.173 & 0.027 & 0.009 & 0.000 & 0.004 & 0.001 & 0.00026 & 0.00009\\
read paper & 1.038 & 0.023 & 0.000 & 0.002 & 0.001 & 0.001 & 0.00009 & 0.00003\\
ideo92 & 0.082 & -0.004 & -0.035 & 0.001 & 0.000 & 0.000 & 0.00000 & 0.00003\\
auth92 & 0.057 & 0.006 & 0.051 & 0.000 & 0.000 & 0.000 & 0.00006 & 0.00029\\
tusa92 & -0.054 & 0.015 & 0.035 & 0.000 & 0.000 & 0.003 & 0.00009 & 0.00018\\
copemg92 & 0.077 & -0.004 & 0.017 & 0.006 & 0.009 & 0.011 & 0.00003 & 0.00005\\
hedqul92 & 0.328 & -0.007 & -0.077 & 0.000 & 0.000 & 0.000 & 0.00016 & 0.00009\\
hhincq92 & -0.018 & -0.010 & 0.051 & 0.000 & 0.001 & 0.000 & 0.00002 & 0.00025\\
ragect92 & -0.152 & -0.013 & -0.015 & 0.001 & 0.001 & 0.001 & 0.00005 & 0.00002\\
rsex92 & -0.202 & 0.013 & -0.046 & 0.000 & 0.002 & 0.000 & 0.00007 & 0.00001\\
region2 & 0.102 & -0.036 & -0.134 & 0.001 & 0.001 & 0.002 & 0.00001 & 0.00022\\
region3 & 0.022 & 0.035 & 0.053 & 0.006 & 0.000 & 0.019 & 0.00066 & 0.00049\\
region4 & -0.022 & 0.024 & 0.024 & 0.000 & 0.000 & 0.001 & 0.00041 & 0.00018\\
region5 & -0.010 & -0.011 & 0.019 & 0.000 & 0.000 & 0.005 & 0.00009 & 0.00014\\
region6 & -0.161 & 0.000 & -0.035 & 0.000 & 0.000 & 0.002 & 0.00008 & 0.00008\\
region7 & -0.036 & -0.005 & -0.016 & 0.002 & 0.000 & 0.016 & 0.00016 & 0.00006\\
region8 & 0.000 & 0.035 & 0.109 & 0.003 & 0.000 & 0.016 & 0.00063 & 0.00095\\
region9 & 0.136 & -0.026 & 0.060 & 0.001 & 0.001 & 0.013 & 0.00007 & 0.00032\\
region10 & 0.099 & 0.043 & -0.051 & 0.002 & 0.002 & 0.007 & 0.00091 & 0.00001\\
region11 & -0.239 & -0.011 & -0.027 & 0.001 & 0.001 & 0.014 & 0.00008 & 0.00000\\
occupation2 & -0.139 & -0.006 & -0.021 & 0.006 & 0.001 & 0.002 & 0.00015 & 0.00001\\
occupation3 & -0.168 & -0.012 & 0.024 & 0.005 & 0.000 & 0.003 & 0.00004 & 0.00003\\
occupation4 & -0.254 & 0.006 & -0.012 & 0.008 & 0.001 & 0.002 & 0.00014 & 0.00007\\
occupation5 & 0.039 & 0.008 & -0.002 & 0.003 & 0.001 & 0.002 & 0.00011 & 0.00005\\
occupation6 & 0.356 & 0.015 & 0.012 & 0.007 & 0.001 & 0.003 & 0.00023 & 0.00026\\
occupation7 & -0.063 & -0.029 & 0.035 & 0.012 & 0.006 & 0.006 & 0.00001 & 0.00018\\
\bottomrule
\end{tabular}

}
\caption{Balance and Partial $R^2$ for Ladd and Lenz (2009)}
\label{tab:ladd_lenz_app_tab}
\end{table}

\begin{table}[!ht]
\centering
 \resizebox{\columnwidth}{!}{
\begin{tabular}{lcccccccc}
\toprule
Covariate & Initial Imbalance & Post-IPW Imbalance & Post-Match Imbalance & $R^2_{Y\sim X}$-EB & $R^2_{Y\sim X}$-IPW & $R^2_{Y\sim X}$-Mat & $R^2_{Z\sim X}$-IPW & $R^2_{Z\sim X}$-Mat\\
\midrule
age & 0.055 & -0.035 & -0.030 & 0.273 & 0.279 & 0.269 & 0.00036 & 0.00000\\
ChildSex & -0.026 & -0.003 & 0.011 & 0.000 & 0.000 & 0.000 & 0.00001 & 0.00002\\
black & 0.254 & -0.010 & -0.021 & 0.002 & 0.002 & 0.003 & 0.00001 & 0.00071\\
mexam & 0.351 & -0.013 & -0.042 & 0.005 & 0.005 & 0.004 & 0.00002 & 0.00067\\
pir200 plus & -0.916 & 0.021 & 0.020 & 0.002 & 0.002 & 0.003 & 0.00002 & 0.00092\\
WIC & 0.383 & -0.038 & 0.041 & 0.000 & 0.000 & 0.000 & 0.00014 & 0.00011\\
Food Stamp & 0.775 & -0.059 & 0.088 & 0.000 & 0.000 & 0.000 & 0.00032 & 0.00154\\
fsdchbi & 0.427 & -0.050 & 0.035 & 0.000 & 0.000 & 0.000 & 0.00024 & 0.00009\\
AnyIns & -0.140 & -0.010 & 0.025 & 0.000 & 0.000 & 0.000 & 0.00006 & 0.00000\\
RefSex & -0.220 & 0.003 & -0.031 & 0.002 & 0.002 & 0.002 & 0.00003 & 0.00006\\
RefAge & -0.173 & -0.002 & -0.073 & 0.003 & 0.003 & 0.000 & 0.00000 & 0.00077\\
\bottomrule
\end{tabular}
}
\caption{Balance and Partial $R^2$ for Chan et al. (2016)}
\label{tab:chan_app_tab}
\end{table}

\begin{table}[!ht]
\centering
\resizebox{\columnwidth}{!}{
\begin{tabular}{lcccccccc}
\toprule
Covariate & Initial Imbalance & Post-IPW Imbalance & Post-Match Imbalance & $R^2_{Y\sim X}$-EB & $R^2_{Y\sim X}$-IPW & $R^2_{Y\sim X}$-Mat & $R^2_{Z\sim X}$-IPW & $R^2_{Z\sim X}$-Mat\\
\midrule
judgeJCS & 0.289 & -0.039 & -0.044 & 0.068 & 0.070 & 0.084 & 0.00009 & 0.00158\\
presDist & -1.522 & 0.079 & 0.079 & 0.000 & 0.000 & 0.000 & 0.00172 & 0.00110\\
panelDistJCS & 0.223 & -0.012 & -0.080 & 0.002 & 0.002 & 0.000 & 0.00000 & 0.00121\\
circmed & 0.562 & -0.011 & 0.082 & 0.000 & 0.000 & 0.001 & 0.00001 & 0.00242\\
sctmed & -0.060 & 0.033 & 0.016 & 0.010 & 0.008 & 0.006 & 0.00054 & 0.00000\\
coarevtc & -0.067 & -0.004 & 0.049 & 0.017 & 0.020 & 0.018 & 0.00001 & 0.00144\\
casepub & -0.481 & -0.045 & -0.077 & 0.003 & 0.004 & 0.001 & 0.00070 & 0.00167\\
\bottomrule
\end{tabular}
}
\caption{Balance and Partial $R^2$ for Black and Owens (2016) for the contender subgroup}
\label{tab:black_app_tab_con}
\end{table}

\begin{table}[!ht]
\centering
 \resizebox{\columnwidth}{!}{
\begin{tabular}{lcccccccc}
\toprule
Covariate & Initial Imbalance & Post-IPW Imbalance & Post-Match Imbalance & $R^2_{Y\sim X}$-EB & $R^2_{Y\sim X}$-IPW & $R^2_{Y\sim X}$-Mat & $R^2_{Z\sim X}$-IPW & $R^2_{Z\sim X}$-Mat\\
\midrule
judgeJCS & -0.054 & -0.042 & -0.189 & 0.001 & 0.000 & 0.000 & 0.00016 & 0.00225\\
presDist & 0.170 & 0.020 & 0.039 & 0.000 & 0.000 & 0.001 & 0.00007 & 0.00022\\
panelDistJCS & 0.039 & -0.011 & -0.085 & 0.000 & 0.000 & 0.000 & 0.00002 & 0.00037\\
circmed & -0.002 & -0.037 & -0.113 & 0.001 & 0.000 & 0.001 & 0.00004 & 0.00023\\
sctmed & -0.247 & -0.035 & 0.002 & 0.006 & 0.002 & 0.012 & 0.00020 & 0.00002\\
coarevtc & -0.011 & -0.001 & -0.023 & 0.000 & 0.003 & 0.000 & 0.00000 & 0.00009\\
\bottomrule
\end{tabular}
}
\caption{Balance and Partial $R^2$ for Black and Owens (2016) for the non-contender subgroup}
\label{tab:black_app_tab_noncon}
\end{table}

\subsection{Superpopulation Corrected Standard Errors}\label{app:application-superpop-correction}

In Table \ref{tab:sp_corr} we apply the correction described in Equations~\eqref{eqn:sp_correction} and \eqref{eqn:sp_correction_ATT} to generate standard errors under the superpopulation framework which target the PATE and PATT.

\begin{table}[tp]
\caption{Point estimates and Superpopulation Corrected Standard Errors by Method}\label{tab:sp_corr}
\centering
\begin{tabular}[ht]{lcccccc}
\toprule
\multicolumn{1}{c}{ } & \multicolumn{2}{c}{Entropy Balancing} & \multicolumn{2}{c}{IPW Weighting} & \multicolumn{2}{c}{Matching} \\
\cmidrule(l{3pt}r{3pt}){2-3} \cmidrule(l{3pt}r{3pt}){4-5} \cmidrule(l{3pt}r{3pt}){6-7}
Study & $\hat{\tau}$ & SE & $\hat{\tau}$ & SE & $\hat{\tau}$ & SE\\
\midrule
Ladd and Lenz (2009) & 0.119 & 0.031 & 0.119 & 0.033 & 0.116 & 0.033\\
Chan et al. (2016) & -0.046 & 0.225 & -0.054 & 0.227 & -0.199 & 0.258\\
\addlinespace[0.3em]
\multicolumn{7}{l}{Black and Owens (2016)}\\
\hspace{1em}\quad Contender & 0.028 & 0.012 & 0.029 & 0.012 & 0.029 & 0.024\\
\hspace{1em} \quad Non-contender & 0.042 & 0.009 & 0.042 & 0.009 & 0.044 & 0.010\\
\bottomrule
\end{tabular}
\end{table}

\subsection{Point estimate changes}\label{app:application-disc}

In the \citet{chan2016globally} study, the IPW and matching point estimates are substantially different once we control for covariates. Although the difference in the bivariate and residualized point estimates is statistically insignificant, we highlight potential reasons for the difference. 

For IPW, age is highly predictive of the outcome with a partial-$R^2$ of 0.28 (see Table \ref{tab:chan_app_tab}), but it remains relatively imbalanced. Since IPW does not balance age well, including this control in the regression may clean up this residual imbalance, moving it closer to the  entropy balancing point estimates.\footnote{With IPW, controlling for all variables except for age results in a point estimate of -0.164 with a standard error of 0.276.} Some  other covariates also remain as or more imbalanced than age after IPW weighting and matching, but including them in the adjusted regression does not change the point estimate since they do not explain much variation in the outcome.\footnote{WIC program remains imbalanced around the same level as age. However, a regression controlling for all variables except for WIC program results in a point estimate of -0.06 with a standard error of 0.227. Since WIC program is not prognostic, even though it remains imbalanced, omitting it from the residualized regression results in the same point estimate as the residualized regression with all variables.}

For matching, age is still prognostic, but the partial-$R^2_{Z\sim X}$ is very close to zero, meaning it is not as conditionally imbalanced as with IPW.\footnote{With matching, removing age from the residualized regression does not change the point estimate in the same way it does for IPW. The residualized regression without age results in a point estimate of -0.198 with a standard error of 0.313.} Other covariates are either balanced well or are not prognostic.  Equation~\eqref{eq:decomp} nonetheless accounts for the change exactly, as the sum over covariates of residual imbalance times the fitted slope. The per-covariate partial $R^2$ values summarize these two ingredients but can understate a change that accumulates across many covariates with modest imbalance and modest slopes. This is more likely in the matched sample, where the effective sample size falls from 1,046 control and 1,284 treated units to 519 and 629, so that the weighted imbalances and slopes are correspondingly noisy.

\clearpage
\printbibliography

@article{buja2019models,
  title={Models as approximations I},
  author={Buja, Andreas and Brown, Lawrence and Berk, Richard and George, Edward and Pitkin, Emil and Traskin, Mikhail and Zhang, Kai and Zhao, Linda},
  journal={Statistical Science},
  volume={34},
  number={4},
  pages={523--544},
  year={2019},
  publisher={JSTOR}
}

@article{wainstein2025sensitivity,
  title={Sensitivity of weighted least squares estimators to omitted variables},
  author={Wainstein, Leonard and Hazlett, Chad},
  journal={arXiv preprint arXiv:2508.02954},
  year={2025}
}

@article{robins1994estimation,
  title={Estimation of regression coefficients when some regressors are not always observed},
  author={Robins, James M and Rotnitzky, Andrea and Zhao, Lue Ping},
  journal={Journal of the American statistical Association},
  volume={89},
  number={427},
  pages={846--866},
  year={1994},
  publisher={Taylor \& Francis}
}

@article{black2016courting,
  title={Courting the president: how circuit court judges alter their behavior for promotion to the Supreme Court},
  author={Black, Ryan C and Owens, Ryan J},
  journal={American Journal of Political Science},
  volume={60},
  number={1},
  pages={30--43},
  year={2016},
  publisher={Wiley Online Library}, 
doi = {10.1111/ajps.12132}
}

@article{lovell1963seasonal,
  title={Seasonal adjustment of economic time series and multiple regression analysis},
  author={Lovell, Michael C},
  journal={Journal of the American Statistical Association},
  volume={58},
  number={304},
  pages={993--1010},
  year={1963},
  publisher={Taylor \& Francis}
}

@article{frisch1933partial,
  title={Partial time regressions as compared with individual trends},
  author={Frisch, Ragnar and Waugh, Frederick V},
  journal={Econometrica: Journal of the Econometric Society},
  pages={387--401},
  year={1933},
  publisher={JSTOR}
}

@article{hirano2003efficient,
  title={Efficient estimation of average treatment effects using the estimated propensity score},
  author={Hirano, Keisuke and Imbens, Guido W and Ridder, Geert},
  journal={Econometrica},
  volume={71},
  number={4},
  pages={1161--1189},
  year={2003},
  publisher={Wiley Online Library}
}

@article{chan2016globally,
  title={Globally efficient non-parametric inference of average treatment effects by empirical balancing calibration weighting},
  author={Chan, Kwun Chuen Gary and Yam, Sheung Chi Phillip and Zhang, Zheng},
  journal={Journal of the Royal Statistical Society Series B: Statistical Methodology},
  volume={78},
  number={3},
  pages={673--700},
  year={2016},
  publisher={Oxford University Press}, 
    doi = {10.1111/rssb.12129}
}

@article{lunceford2004stratification,
  title={Stratification and weighting via the propensity score in estimation of causal treatment effects: a comparative study},
  author={Lunceford, Jared K and Davidian, Marie},
  journal={Statistics in medicine},
  volume={23},
  number={19},
  pages={2937--2960},
  year={2004},
  publisher={Wiley Online Library}
}

@article{glynn2010introduction,
  title={An introduction to the augmented inverse propensity weighted estimator},
  author={Glynn, Adam N and Quinn, Kevin M},
  journal={Political analysis},
  volume={18},
  number={1},
  pages={36--56},
  year={2010},
  publisher={Cambridge University Press}
}

@article{hahn1998role,
  title={On the role of the propensity score in efficient semiparametric estimation of average treatment effects},
  author={Hahn, Jinyong},
  journal={Econometrica},
  pages={315--331},
  year={1998},
  publisher={JSTOR}
}

@article{shinkrehazlettdemystifying,
  title={Demystifying and avoiding the OLS" weighting problem": Unmodeled heterogeneity and straightforward solutions},
  author={Hazlett, Chad and Shinkre, Tanvi},
  journal={arXiv preprint arXiv:2403.03299},
  year={2024}
}

@article{ho2007matching,
  title={Matching as nonparametric preprocessing for reducing model dependence in parametric causal inference},
  author={Ho, Daniel E and Imai, Kosuke and King, Gary and Stuart, Elizabeth A},
  journal={Political analysis},
  volume={15},
  number={3},
  pages={199--236},
  year={2007},
  publisher={Cambridge University Press}
}

@article{hazlett2020kernel,
  title={Kernel balancing},
  author={Hazlett, Chad},
  journal={Statistica Sinica},
  volume={30},
  number={3},
  pages={1155--1189},
  year={2020},
  publisher={JSTOR}
}

@article{zhao2017entropy,
  title={Entropy balancing is doubly robust},
  author={Zhao, Qingyuan and Percival, Daniel},
  journal={Journal of Causal Inference},
  volume={5},
  number={1},
  pages={20160010},
  year={2017},
  publisher={De Gruyter}
}

@article{lin_agnostic_2013,
	title = {Agnostic notes on regression adjustments to experimental data: {Reexamining} {Freedman}’s critique},
	volume = {7},
	issn = {1932-6157, 1941-7330},
	shorttitle = {Agnostic notes on regression adjustments to experimental data},
	url = {https://projecteuclid.org/journals/annals-of-applied-statistics/volume-7/issue-1/Agnostic-notes-on-regression-adjustments-to-experimental-data--Reexamining/10.1214/12-AOAS583.full},
	doi = {10.1214/12-AOAS583},
	number = {1},
	urldate = {2022-04-06},
	journal = {The Annals of Applied Statistics},
	author = {Lin, Winston},
	month = mar,
	year = {2013},
	note = {Publisher: Institute of Mathematical Statistics},
	pages = {295--318},
}

@article{abadie_sampling-based_2020,
	title = {Sampling-{Based} versus {Design}-{Based} {Uncertainty} in {Regression} {Analysis}},
	volume = {88},
	issn = {1468-0262},
	url = {https://onlinelibrary.wiley.com/doi/abs/10.3982/ECTA12675},
	doi = {10.3982/ECTA12675},
	language = {en},
	number = {1},
	urldate = {2022-04-08},
	journal = {Econometrica},
	author = {Abadie, Alberto and Athey, Susan and Imbens, Guido W. and Wooldridge, Jeffrey M.},
	year = {2020},
	pages = {265--296},
}

@article{white_using_1980,
	title = {Using {Least} {Squares} to {Approximate} {Unknown} {Regression} {Functions}},
	volume = {21},
	issn = {0020-6598},
	url = {https://www.jstor.org/stable/2526245},
	doi = {10.2307/2526245},
	number = {1},
	urldate = {2022-07-15},
	journal = {International Economic Review},
	author = {White, Halbert},
	year = {1980},
	note = {Publisher: [Economics Department of the University of Pennsylvania, Wiley, Institute of Social and Economic Research, Osaka University]},
	pages = {149--170},
}

@article{hainmueller_entropy_2012,
	title = {Entropy {Balancing} for {Causal} {Effects}: {A} {Multivariate} {Reweighting} {Method} to {Produce} {Balanced} {Samples} in {Observational} {Studies}},
	volume = {20},
	issn = {1047-1987, 1476-4989},
	shorttitle = {Entropy {Balancing} for {Causal} {Effects}},
url = {https://www.cambridge.org/core/journals/political-analysis/article/entropy-balancing-for-causal-effects-a-multivariate-reweighting-method-to-produce-balanced-samples-in-observational-studies/220E4FC838066552B53128E647E4FAA7},
	doi = {10.1093/pan/mpr025},
	language = {en},
	number = {1},
	urldate = {2022-04-13},
	journal = {Political Analysis},
	author = {Hainmueller, Jens},
	year = {2012},
	note = {Publisher: Cambridge University Press},
	pages = {25--46},
}

@article{rubin1974estimating,
  title={Estimating causal effects of treatments in randomized and nonrandomized studies.},
  author={Rubin, Donald B},
  journal={Journal of educational Psychology},
  volume={66},
  number={5},
  pages={688},
  year={1974},
  publisher={American Psychological Association}
}

@article{rosenbaum_central_1983,
	title = {The central role of the propensity score in observational studies for causal effects},
	volume = {70},
	issn = {0006-3444},
	url = {https://doi.org/10.1093/biomet/70.1.41},
	doi = {10.1093/biomet/70.1.41},
	number = {1},
	urldate = {2022-11-09},
	journal = {Biometrika},
	author = {Rosenbaum, Paul R. and Rubin, Donald B.},
	month = apr,
	year = {1983},
	pages = {41--55},
}

@article{chattopadhyay2023implied,
  title={On the implied weights of linear regression for causal inference},
  author={Chattopadhyay, Ambarish and Zubizarreta, Jos{\'e} R},
  journal={Biometrika},
  volume={110},
  number={3},
  pages={615--629},
  year={2023},
  publisher={Oxford University Press}
}

@article{ladd2009exploiting,
  title={Exploiting a rare communication shift to document the persuasive power of the news media},
  author={Ladd, Jonathan McDonald and Lenz, Gabriel S},
  journal={American Journal of Political Science},
  volume={53},
  number={2},
  pages={394--410},
  year={2009},
  publisher={Wiley Online Library}, 
doi = {10.1111/j.1540-5907.2009.00377.x}
}

@book{hernan2020causal,
  title={Causal inference},
  author={Hern{\'a}n, Miguel A and Robins, James M},
  year={2020},
  publisher={Boca Raton: Chapman \& Hall/CRC}
}

@article{reifeis_hudgens_2022,
    author = {Reifeis, Sarah A and Hudgens, Michael G},
    title = {On Variance of the Treatment Effect in the Treated When Estimated by Inverse Probability Weighting},
    journal = {American Journal of Epidemiology},
    volume = {191},
    number = {6},
    pages = {1092-1097},
    year = {2022},
    month = {02},
    issn = {0002-9262},
    doi = {10.1093/aje/kwac014},
    url = {https://doi.org/10.1093/aje/kwac014},
    eprint = {https://academic.oup.com/aje/article-pdf/191/6/1092/44799742/kwac014.pdf},
}

@article{berk_etal_2013,
author = {Richard Berk and Emil Pitkin and Lawrence Brown and Andreas Buja and Edward George and Linda Zhao},
title ={Covariance Adjustments for the Analysis of Randomized Field Experiments},
journal = {Evaluation Review},
volume = {37},
number = {3-4},
pages = {170-196},
year = {2013},
doi = {10.1177/0193841X13513025},
    note ={PMID: 24647925},
URL = { 
    
        https://doi.org/10.1177/0193841X13513025
    
    

},
eprint = { 
    
        https://doi.org/10.1177/0193841X13513025
    
    

}
}

@article{negi_woolridge_2021,
author = {Negi, Akanksha and Wooldridge, Jeffrey},
title = {Revisiting regression adjustment in experiments with heterogeneous treatment effects},
journal = {Econometric Reviews},
volume = {40},
number = {5},
pages = {504--534},
year = {2021},
publisher = {Taylor \& Francis},
doi = {10.1080/07474938.2020.1824732},
URL = { 
        https://doi.org/10.1080/07474938.2020.1824732
},
eprint = { 
        https://doi.org/10.1080/07474938.2020.1824732
}

}

@misc{ding2023coursecausalinference,
      title={A First Course in Causal Inference}, 
      author={Peng Ding},
      year={2023},
      eprint={2305.18793},
      archivePrefix={arXiv},
      primaryClass={stat.ME},
      url={https://arxiv.org/abs/2305.18793}, 
}

@article{rubin1980randomization,
  title={Randomization analysis of experimental data: The Fisher randomization test comment},
  author={Rubin, Donald B},
  journal={Journal of the American statistical association},
  volume={75},
  number={371},
  pages={591--593},
  year={1980},
  publisher={JSTOR}
}

@incollection{cohn2023balancing,
  title={Balancing weights for causal inference},
  author={Cohn, Eric R and Ben-Michael, Eli and Feller, Avi and Zubizarreta, Jos{\'e} R},
  booktitle={Handbook of Matching and Weighting Adjustments for Causal Inference},
  pages={293--312},
  year={2023},
  publisher={Chapman and Hall/CRC}
}

@article{deville1992calibration,
  title={Calibration estimators in survey sampling},
  author={Deville, Jean-Claude and S{\"a}rndal, Carl-Erik},
  journal={Journal of the American statistical Association},
  volume={87},
  number={418},
  pages={376--382},
  year={1992},
  publisher={Taylor \& Francis}, 
  doi = {10.2307/2290268}
}

@article{kallberg2023large,
  title={Large Sample Properties of Entropy Balancing Estimators of Average Causal Effects},
  author={K{\"a}llberg, David and Waernbaum, Ingeborg},
  journal={Econometrics and Statistics},
  year={2023},
  publisher={Elsevier}
}

@book{sarndal2005estimation,
  title={Estimation in surveys with nonresponse},
  author={S{\"a}rndal, Carl-Erik and Lundstr{\"o}m, Sixten},
  year={2005},
  publisher={John Wiley \& Sons}
}

@article{d2010linearization,
  title={Linearization variance estimation for generalized raking estimators in the presence of nonresponse},
  author={D'Arrigo, Julia and Skinner, Chris J},
  journal={Survey Methodology},
  volume={36},
  number={2},
  pages={181--192},
  year={2010},
  publisher={Statistics Canada}
}

@article{hartman2024kpop,
  title={Kpop: A kernel balancing approach for reducing specification assumptions in survey weighting},
  author={Hartman, Erin and Hazlett, Chad and Sterbenz, Ciara},
  journal={Journal of the Royal Statistical Society Series A: Statistics in Society},
  pages={qnae082},
  year={2024},
  publisher={Oxford University Press UK}
}

@article{breidt_opsomer_2017,
author = {F. Jay Breidt and Jean D. Opsomer},
title = {{Model-Assisted Survey Estimation with Modern Prediction Techniques}},
volume = {32},
journal = {Statistical Science},
number = {2},
publisher = {Institute of Mathematical Statistics},
pages = {190 -- 205},
year = {2017},
doi = {10.1214/16-STS589},
URL = {https://doi.org/10.1214/16-STS589}
}

@book{fuller_2009,
author = {Wayne Fuller},
publisher = {John Wiley \& Sons, Ltd},
isbn = {9780470523551},
title = {Frontmatter},
booktitle = {Sampling Statistics},
chapter = {},
doi = {https://doi.org/10.1002/9780470523551.fmatter},
url = {https://onlinelibrary.wiley.com/doi/abs/10.1002/9780470523551.fmatter},
eprint = {https://onlinelibrary.wiley.com/doi/pdf/10.1002/9780470523551.fmatter},
year = {2009}}

@article{lunceford_davidian_2004,
author = {Lunceford, Jared K. and Davidian, Marie},
title = {Stratification and weighting via the propensity score in estimation of causal treatment effects: a comparative study},
journal = {Statistics in Medicine},
volume = {23},
number = {19},
pages = {2937-2960},
doi = {https://doi.org/10.1002/sim.1903},
url = {https://onlinelibrary.wiley.com/doi/abs/10.1002/sim.1903},
eprint = {https://onlinelibrary.wiley.com/doi/pdf/10.1002/sim.1903},
year = {2004}
}

@article{gabriel2024inverse,
  title={Inverse probability of treatment weighting with generalized linear outcome models for doubly robust estimation},
  author={Gabriel, Erin E and Sachs, Michael C and Martinussen, Torben and Waernbaum, Ingeborg and Goetghebeur, Els and Vansteelandt, Stijn and Sj{\"o}lander, Arvid},
  journal={Statistics in Medicine},
  volume={43},
  number={3},
  pages={534--547},
  year={2024},
  publisher={Wiley Online Library}
}

@misc{benmichael2021balancingactcausalinference,
      title={The Balancing Act in Causal Inference}, 
      author={Eli Ben-Michael and Avi Feller and David A. Hirshberg and José R. Zubizarreta},
      year={2021},
      eprint={2110.14831},
      archivePrefix={arXiv},
      primaryClass={stat.ME},
      url={https://arxiv.org/abs/2110.14831}, 
}

@article{xu2023hierarchically,
  title={Hierarchically regularized entropy balancing},
  author={Xu, Yiqing and Yang, Eddie},
  journal={Political Analysis},
  volume={31},
  number={3},
  pages={457--464},
  year={2023},
  publisher={Cambridge University Press}
}

@article{stefanski2002calculus,
  title={The calculus of M-estimation},
  author={Stefanski, Leonard A and Boos, Dennis D},
  journal={The American Statistician},
  volume={56},
  number={1},
  pages={29--38},
  year={2002},
  publisher={Taylor \& Francis}
}

@Manual{greifer2025,
  title = {WeightIt: Weighting for Covariate Balance in Observational Studies},
  author = {Noah Greifer},
  year = {2025},
  note = {R package version 1.4.0.9002},
  url = {https://ngreifer.github.io/WeightIt/},
}

@article{greifer2021matching,
    author = {Greifer, Noah and Stuart, Elizabeth A},
    title = {Matching Methods for Confounder Adjustment: An Addition to the Epidemiologist’s Toolbox},
    journal = {Epidemiologic Reviews},
    volume = {43},
    number = {1},
    pages = {118-129},
    year = {2021},
    month = {06},
    issn = {1478-6729},
    doi = {10.1093/epirev/mxab003},
    url = {https://doi.org/10.1093/epirev/mxab003},
    eprint = {https://academic.oup.com/epirev/article-pdf/43/1/118/48355081/mxab003.pdf},
}

@article{zubizarreta2015,
author = {José R. Zubizarreta},
title = {Stable Weights that Balance Covariates for Estimation With Incomplete Outcome Data},
journal = {Journal of the American Statistical Association},
volume = {110},
number = {511},
pages = {910--922},
year = {2015},
publisher = {ASA Website},
doi = {10.1080/01621459.2015.1023805},
URL = { https://doi.org/10.1080/01621459.2015.1023805},
eprint = { https://doi.org/10.1080/01621459.2015.1023805}
}

@article{savje2021generalized,
  title={Generalized full matching},
  author={S{\"a}vje, Fredrik and Higgins, Michael J and Sekhon, Jasjeet S},
  journal={Political Analysis},
  volume={29},
  number={4},
  pages={423--447},
  year={2021},
  publisher={Cambridge University Press}
}

@Article{hoetal_matchit_2011,
  title = {{MatchIt}: Nonparametric Preprocessing for Parametric Causal Inference},
  author = {Daniel E. Ho and Kosuke Imai and Gary King and Elizabeth A. Stuart},
  year = {2011},
  journal = {Journal of Statistical Software},
  volume = {42},
  number = {8},
  pages = {1--28},
  doi = {10.18637/jss.v042.i08},
}

@book{wooldridge2010econometric,
  title={Econometric analysis of cross section and panel data},
  author={Wooldridge, Jeffrey M},
  year={2010},
  publisher={MIT press}
}

@article{ross2024m,
  title={M-estimation for common epidemiological measures: introduction and applied examples},
  author={Ross, Rachael K and Zivich, Paul N and Stringer, Jeffrey SA and Cole, Stephen R},
  journal={International Journal of Epidemiology},
  volume={53},
  number={2},
  pages={dyae030},
  year={2024},
  publisher={Oxford University Press}
}

@article{shook2025double,
  title={Double robust variance estimation with parametric working models},
  author={Shook-Sa, Bonnie E and Zivich, Paul N and Lee, Chanhwa and Xue, Keyi and Ross, Rachael K and Edwards, Jessie K and Stringer, Jeffrey SA and Cole, Stephen R},
  journal={Biometrics},
  volume={81},
  number={2},
  pages={ujaf054},
  year={2025},
  publisher={Oxford University Press}
}

@article{imai2014covariate,
  title={Covariate balancing propensity score},
  author={Imai, Kosuke and Ratkovic, Marc},
  journal={Journal of the Royal Statistical Society Series B: Statistical Methodology},
  volume={76},
  number={1},
  pages={243--263},
  year={2014},
  publisher={Oxford University Press}
}

@article{austin2022bootstrap,
  title={Bootstrap vs asymptotic variance estimation when using propensity score weighting with continuous and binary outcomes},
  author={Austin, Peter C},
  journal={Statistics in Medicine},
  volume={41},
  number={22},
  pages={4426--4443},
  year={2022},
  publisher={Wiley Online Library}
}

@article{austin2016variance,
  title={Variance estimation when using inverse probability of treatment weighting (IPTW) with survival analysis},
  author={Austin, Peter C},
  journal={Statistics in medicine},
  volume={35},
  number={30},
  pages={5642--5655},
  year={2016},
  publisher={Wiley Online Library}
}

@article{kostouraki2024variance,
  title={On variance estimation of the inverse probability-of-treatment weighting estimator: A tutorial for different types of propensity score weights},
  author={Kostouraki, Andriana and Hajage, David and Rachet, Bernard and Williamson, Elizabeth J and Chauvet, Guillaume and Belot, Aur{\'e}lien and Leyrat, Cl{\'e}mence},
  journal={Statistics in Medicine},
  volume={43},
  number={13},
  pages={2672--2694},
  year={2024},
  publisher={Wiley Online Library}
}

@article{abadie2008failure,
  title={On the failure of the bootstrap for matching estimators},
  author={Abadie, Alberto and Imbens, Guido W},
  journal={Econometrica},
  volume={76},
  number={6},
  pages={1537--1557},
  year={2008},
  publisher={Wiley Online Library}
}

@article{otsu2017bootstrap,
  title={Bootstrap inference of matching estimators for average treatment effects},
  author={Otsu, Taisuke and Rai, Yoshiyasu},
  journal={Journal of the American Statistical Association},
  volume={112},
  number={520},
  pages={1720--1732},
  year={2017},
  publisher={Taylor \& Francis}
}

@article{abadie2012martingale,
  title={A martingale representation for matching estimators},
  author={Abadie, Alberto and Imbens, Guido W},
  journal={Journal of the American Statistical Association},
  volume={107},
  number={498},
  pages={833--843},
  year={2012},
  publisher={Taylor \& Francis}
}

@article{abadie2022robust,
  title={Robust post-matching inference},
  author={Abadie, Alberto and Spiess, Jann},
  journal={Journal of the American Statistical Association},
  volume={117},
  number={538},
  pages={983--995},
  year={2022},
  publisher={Taylor \& Francis}
}

@article{zhang2025improving,
  title={Improving Variance and Confidence Interval Estimation in Small-Sample Propensity Score Analyses: Bootstrap vs. Asymptotic Methods},
  author={Zhang, Baoshan and O'Brien, Sean M and Wu, Yuan and Thomas, Laine E},
  journal={arXiv preprint arXiv:2511.10911},
  year={2025}
}

@article{aronow2013conservative,
  title={Conservative variance estimation for sampling designs with zero pairwise inclusion probabilities},
  author={Aronow, Peter M and Samii, Cyrus},
  journal={Survey Methodology},
  volume={39},
  number={1},
  pages={231--241},
  year={2013},
  publisher={Statistics Canada}
}
\end{refsection}
\end{document}